\documentclass[11pt]{article}
\usepackage{eurosym}
\usepackage{amsfonts}
\usepackage{amssymb}
\usepackage{graphicx}
\usepackage{amsmath}
\usepackage{makeidx}
\usepackage{indentfirst}
\usepackage[T1]{fontenc}
\usepackage[utf8]{inputenc}

\newcounter{resultnum}[section]
\newcounter{conclusionnum}[section]
\newcounter{conditionnum}[section]
\newcounter{conjecturenum}[section]
\newcounter{examplenum}[section]
\newcounter{exercisenum}[section]
\newcounter{lemmanum}[section]
\newcounter{notationnum}[section]
\newcounter{theoremnum}[section]
\newcounter{definitionnum}[section]
\newcounter{corollarynum}[section]
\newcounter{remarknum}[section]
\newcounter{propositionnum}[section]
\newcounter{acknowledgementnum}[section]
\newcounter{algorithmnum}[section]
\newcounter{axiomnum}[section]
\newcounter{casenum}[section]
\newcounter{claimnum}[section]
\newcounter{summarynum}[section]
\newcounter{problemnum}[section]
\begin{document}

\title{Nonassociative black holes in R-flux deformed phase spaces\\
and relativistic models of Perelman thermodynamics}
\date{March 27, 2023}
\author{ {\textbf{Lauren\c{t}iu Bubuianu}\thanks{%
email: laurentiu.bubuianu@tvr.ro and laurfb@gmail.com}} \and {\small \textit{%
SRTV - Studioul TVR Ia\c{s}i} and \textit{University Appolonia}, 2 Muzicii
street, Ia\c{s}i, 700399, Romania} \vspace{.1 in} \and {\textbf{Douglas
Singleton }} \thanks{%
email: dougs@mail.fresnostate.edu} \and {\small \textit{Department of
Physics,\ California State University Fresno, Fresno, CA 93740-8031, USA}}
\vspace{.1 in} \and \textbf{Sergiu I. Vacaru} \thanks{%
emails: sergiu.vacaru@fulbrightmail.org ; sergiuvacaru@mail.fresnostate.edu
; sergiu.vacaru@gmail.com ; \newline \textit{Address for post correspondence in
2022-2023 as a visitor senior researcher at YF CNU Ukraine:\ }  \newline
Vokzal'na (former Yu. Gagarin) street, 37-3, Chernivtsi, Ukraine, 58008} \and {\small \textit{Department of
Physics, California State University at Fresno, Fresno, CA 93740, USA; }}
\and {\small \textit{Institute of Applied-Physics and Computer Sciences,
Kotsyubinsky 2, Chernivtsi, 58012, Ukraine; }} \and {\small \textit{Faculty
of Physics, Babes-Bolyai University, 1 Mihail Kogalniceanu Street,
Cluj-Napoca, 400084, Romania}} }
\maketitle

\begin{abstract}
This paper explores new classes of black hole (BH) solutions
in nonassociative and noncommutative gravity, focusing on features that
generalize to higher dimensions. The theories we study are modelled on (co)
tangent Lorentz bundles with
a star product structure determined by R-flux deformations in string theory. For the nonassociative vacuum Einstein equations we consider both real and complex
effective sources. In order to analyze the nonassociative vacuum Einstein equations we develop the anholonomic frame and
connection deformation methods, which allows one to decoupled  and solve these equations. The metric coefficients can depend on both space-time coordinates and energy-momentum. By imposing conditions on the
integration functions and effective sources we find physically important, exact solutions: (1) 6-d Tangherlini BHs, which are star product and R-flux
distorted to 8-d black ellipsoids (BEs)
and BHs; (2) nonassocitative space-time and co-fiber space double BH and/or
BE configurations generalizing Schwarzschild - de Sitter metrics. We also investigate the concept of Bekenstein-Hawking entropy and find it applicable only for very
special classes of nonassociative BHs with conventional horizons and
(anti) de Sitter configurations. Finally, we show how analogs of the
relativistic Perelman W-entropy and related geometric thermodynamic
variables can be defined and computed for general classes of off-diagonal
solutions with nonassociative R-flux deformations.

\vskip3pt

\textbf{Keywords:}\ nonassociative geometry and gravity; R-flux
non-geometric background; nonholonomic star deformations; non-symmetric
metrics; non-associative black holes; generalized Bekenstein-Hawking
entropy; Perelman entropy; relativistic geometric flows and statistical
thermodynamics.
\end{abstract}

\newpage 

\tableofcontents

\section{Introduction}

Nonassociative and noncommutative geometric and physical models have been
subjects of great interest for almost 90 years, since the first works of
nonassociative quantum mechanics \cite{jordan32,jordan34} and further
developments in mathematical particle physics \cite%
{kurdgelaidze,drinf,okubo,castro1,mylonas13,kupriyanov15,gunaydin}. Much is
known about nonassociative gravity, nonassociative gauge theories, nonassociative membrane theory and double
field theory, from string theory \cite%
{alvarez06,luest10, blumenhagen10,condeescu13,blumenhagen13,kupriyanov19a}.
Nonassociative structures also arise in the world volume of a D-brane (for open strings) and for flux
compactification (for
closed strings). Reviews of various aspects of nonassociative physics and
gravity and a comprehensive list of references can be found in \cite%
{szabo19,blumenhagen16,aschieri17, partner01,partner02,partner03}.

Self-consistent theories of nonassociative vacuum gravity were formulated
for $\star $-product ({\it i.e.} star-product) deformations determined by
R-flux backgrounds in string gravity \cite{blumenhagen16,aschieri17}. Such
nonassociative and noncommutative geometric and gravity theories are modelled
on a conventional phase space $\mathcal{M}=T^{\ast }\mathbf{V,}$ which for
theories including general relativity (GR) is a
cotangent bundle, $T^{\ast }\mathbf{V}$, on a spacetime Lorentz manifold, $%
\mathbf{V}$. The nonassociative, vacuum, gravitational equations, $\
Ric^{\star }[\nabla ^{\star }]=0,$ were postulated as phase space $\star $%
-deformations of the standard Ricci tensor $Ric$ in GR. The nonassociative
tensor $Ric^{\star }$ was constructed for a unique nonassociative
Levi-Civita (LC) connection, $\nabla ^{\star },$ which is torsionless and
compatible with the respective $\star $-deformed symmetric, $\ _{\star
}^{\shortparallel }\mathbf{g,}$ and nonsymmetric, $\ _{\star
}^{\shortparallel }\mathbf{q},$ metric structures.\footnote{%
Our conventions can be found in \cite{partner01,partner02,partner03} and are explained
in the next section and Appendix \ref{appendixa}. In this work, the phase
space dimension is $\dim \mathcal{M}=8,$ with local coordinates labeled like $\ ^{\shortparallel }u^{\alpha _{s}}=(x^{i_{s}},\
^{\shortparallel }p_{a_{s}})$. Here $^{\shortparallel }$ on the left indicates a phase space with spacetime coordinates, $%
x^{j_{s}}=(x^{j},t)$, and {\it complex} momentum coordinates, $\
^{\shortparallel }p_{a_{s}}=\ ip_{a_{s}}=(ip_{a},iE$), with $i^{2}=-1$.
}

Nonassociative and noncommutative modified gravity theories (MGTs)
have attracted attention as a type of bimetric gravity theory \cite%
{rosen40,rosen77} (see \cite{gheor14,luest21} for reviews), 
with the second metric structure being nonsymmetric \cite%
{einstein25,einstein45,eisenhart51,eisenhart52}. Earlier works on the
development of nonassociative gravity are given in \cite{moffat95,partner01}. There are also
geometric models on a phase space $\mathcal{M}^{\star }$
enabled with $\star$-product structure which involves nonassociative generalizatons
of relativistic and supersymmetric/(non) commutative
Finsler-Lagrange-Hamilton spaces \cite{vacaru96b,vacaru09a,bubuianu19}.  In this paper, we do not use the explicit
form of Finsler-like geometric objects and variables, but re-define the formulas for nonlinear
quadratic elements in a form so that nonassociative phase space BH solutions
can be generated as real configurations coming from star R-flux
deformations. We also compute nontrivial phase space components of
the nonsymmetric parts of the metric which can occur in
nonassociative gravity.

The works \cite{blumenhagen16,aschieri17,partner01,partner02} raised two important questions in nonassociative gravity: (1) How to formulate and analyze fundamental properties of nonassociative
modified Einstein equations? (2) How to
construct in explicit form classes of exact, physically
important solutions in nonassociative gravity and determine the physical
meaning of such solutions? In regard to question (2) nonassociative versions of gravitational and
matter field equations cannot be solved by standard methods. It
is difficult to construct and study the physical properties of
nonassociative BH and cosmological solutions (see \cite{partner02,partner03}).

The main \textbf{hypothesis} in our works on nonassociative and
noncommutative gravity is that such \textit{viable physical theories can be
formulated as nonholonomic real and/or (almost) complex deformations of GR to geometric 
models with extra-dimensional coordinates and generalized (almost) complex/
symplectic structures}. We carry out constructions with a generalized metric,
frame, and (non) linear connection structures on nonholonomic manifolds
and/or (co) tangent bundles.\footnote{%
Briefly, a nonholonomic manifold is enabled with a non-integrable
distribution, with a prescribed system of anholonomic frames,
or certain Whitney/ direct sums defining some (non) linear connection
structure, see references \cite{partner01,partner02} for details.} The works 
\cite{vacaru96b,vacaru09a,bubuianu19} further laid out various extensions/generalizations involving supersymmetry, noncommutativity, and nonassociativity.

This is the fourth paper in a series of papers \cite%
{partner01,partner02,partner03} devoted to constructing exact,
physically important solutions in
nonassociative geometry and (modified) gravity models determined by $\star$%
-products induced by R-flux deformations in string theory. In \cite%
{partner02}, we developed ,in nonassociative geometric form, the anholonomic
frame and connection deformation method, AFCDM. Using geometric and analytic methods, we can
decouple and solve in a general form, vacuum and non-vacuum gravitational
and matter field equations and geometric flow evolution equations. For such
solutions, the coefficients of the symmetric and nonsymmetric generic
off-diagonal metrics and (non) linear connections depend, in general, on all
phase space coordinates.

The AFCDM is different from the usual approach of constructing exact
solutions which uses a diagonal ansatz for the metric, and has metric components
that only depend on spacetime coordinates. In the usual approach the systems of nonlinear PDEs
are transformed into systems of ODEs (ordinary differential equations),
which can be integrated in some exact, parametric forms depending on
integration constants. The physical meaning of these constants is defined by
symmetry and asymptotic conditions (for the well known, exact solutions in GR, details can
be found in \cite{misner,hawking73,wald82,kramer03}). Applying the AFCDM, we
can integrate directly, in exact, parametric form, physically
important systems of nonlinear PDEs. In \cite{partner03}, we constructed
exact, parametric solutions, in $\kappa $ and $\hbar $, describing 4-d
$\star$ R-flux distortions of Schwarzschild BHs, ellipsoid
configurations defining black ellipsoids (BEs) and BHs with ellipsoidal thin
accretion disks. We also used 4-d nonholonomic constraints of the quadratic
elements for 8-d quasi-stationary nonassociative vacuum solutions found in
\cite{partner02}.

In this paper we further develop our program of constructing BH and BE solutions in
nonassociative gravity \cite{partner02,partner03}. In addition to previous solutions
\cite{partner03}, we construct two new classes of BH configurations in 8-d nonassociative phase
spaces. For this purpose, we use modified quasi-stationary integrals for
metrics and (non) linear connections with an energy Killing symmetry. 
The main goal of this article is to generalize
and apply the AFCDM in such a form which allows one to generate, in
nonassociative form, classes of 8-d phase space BH solutions. We find that for our 
generic off-diagonal solutions it is only possible, in very special
cases, to define the concept of a generalized Bekenstein--Hawking entropy. However, we show that one can define the concept of an entropy for these solutions using Perelman's geometric flow thermodynamic variables \cite{perelman1}. Previous work on applying Perelman's thermodynamics in relativistic theories can be found in \cite{bubuianu19}. 

These constructions of entropy for relativistic and noncommutative geometric flow models using Perelman thermodynamics were given in detail in the works \cite{vacaru08,ibubuianu20,ibubuianu21}. 
In this paper, we consider Perelman-type geometric, entropic and thermodynamic
values determined by real R-flux effective sources as in \cite%
{aschieri17,partner02,partner03}.

The paper is organized as follows: In section \ref{sec2}, we review the
necessary geometric concepts, methods, and formulas of \cite%
{partner01,partner02,partner03}. The nonassociative vacuum gravitational
equations are provided in nonholonomic, dyadic variables with a shell by
shell parametric decompositions of the 8-d phase space. We also present the
quadratic elements for quasi-stationary phase spaces, in terms of
gravitational polarizations and generating functions, which will be used for
constructing BH solutions in later sections.

Section \ref{sec3} is devoted to nonassociative $\star$-product R-flux
deformations of Schwarzschild - de Sitter metrics with the metric
coefficients having additional dependencies on energy and momentum. We
provide explicit formulas for $\star$ $\kappa$--parametric deformations of
Tangherlini's BHs \cite{tangherlini63} into quasi-stationary configurations
on nonassociative phase spaces. We construct examples of double BHs, BEs,
configurations both on 4-d phase spacetime and in 4-d (co) fiber space with
explicit energy dependence and related nonassociative and nonlinear
symmetries. We then give two classes of exact solutions describing
quasi-stationary BHs and BEs with an energy-like Killing symmetry. The first
solution describes nonassociative R-flux distortions of 6-d Tangherlini like
BHs to 8-d phase spaces. The second solution is for nonassociative,
nonholonomic deformations of BHs into BEs with a fixed energy parameter.

Section \ref{sec4} focuses on the entropy and geometric
thermodynamics of BHs and BEs as quasi-stationary Ricci solitons encoding
nonassociative R-flux data. We speculate on the phase space generalizations
of the Bekenstein-Hawking entropy for nonassociative BH and BE solutions
with hypersurface horizons. For more general classes of off-diagonal
solutions, with nonassociative deformations of relativistic and phase space
Ricci solitons, we conclude that the Bekenstein-Hawking approach is not
applicable. However we show how relativistic generalizations of Grigori Perelman's
entropy and statistical/ geometric thermodynamics can be applied in those cases where the Bekenstein-Hawking approach to entropy does not exist.  We
give two examples of how to compute the generalized Perelman thermodynamic
variables for quasi-stationary phase space R-flux deformed Tangherlini BHs
and double 4-d BHs.

A summary and discussion of results, with conclusions, are given in
section \ref{sec5}. The Appendix contains a brief overview of coordinate and
index conventions and useful formulas for nonassociative AFCDM
and generating functions for various solutions.

\section{Nonassociative phase space geometry of quasi-stationary star-deformed Einstein spaces}

\label{sec2} We are primarily interested in constructing BH solutions in
nonassociative vacuum gravity defined on phase spaces. Such metrics may
depend both on spacetime and co-fiber coordinates and encode $\star $%
--product and R--flux contributions. In this section we review a
nonassociative geometric formalism \cite{blumenhagen16,aschieri17}
reformulated in nonholonomic dyadic variables \cite{partner01}, which will
allow us to construct exact and parametric solutions in the following
sections. We follow the conventions and definitions on (non) associative
phase space geometry and the AFCDM given in \cite{partner02}. We
consider extensions of generic off-diagonal ansatz which allow us to
generalize quasi-stationary solutions for 4-d spacetimes \cite{partner03} and
generate new classes of vacuum 8-d nonassociative phase spaces.

\subsection{Nonholonomic dyadic shell coordinates, s-frames, and star
products}

In this work, nonassociative geometric and physical theories are constructed
on nonholonomic phase space modeled as a cotangent Lorentz bundle $\mathcal{M%
}=T_{s\shortparallel }^{\ast }\mathbf{V}$ on a pseudo-Euclidean manifold of
signature $(+++-)$. Such a phase space is enabled with a quasi-Hopf
structure \cite{drinf,aschieri17} adapted to a nonholonomic dyadic
decomposition into four oriented shells $s=1,2,3,4$ with conventional
(2+2)+(2+2) splitting of dimensions - briefly, an s-decomposition. Boldface and not boldface symbols carry the same meaning
as in \cite{partner01,partner02} and footnote 4 from \cite{partner03}, and are discussed more in the following paragraph. Further necessary conventions and coordinate formulas are given in
Appendix \ref{appendixa}.

The (2+2)+(2+2) nonholonomic shell splitting of a phase spaces can be defined by
a nonlinear connection ({\it i.e.} N--connection) structure. The s-splitting is given as:
\begin{eqnarray}
\ _{s}^{\shortmid }\mathbf{N}:\ \ _{s}T\mathbf{T}^{\ast }\mathbf{V} &=&\
^{1}hT^{\ast }V\oplus \ ^{2}vT^{\ast }V\oplus \ ^{3}cT^{\ast }V\oplus \
^{4}cT^{\ast }V,\mbox{ and }  \notag \\
\ _{s}^{\shortparallel }\mathbf{N}:\ \ _{s}T\mathbf{T}_{\shortparallel
}^{\ast }\mathbf{V} &=&\ ^{1}hT_{\shortparallel }^{\ast }V\oplus \
^{2}vT_{\shortparallel }^{\ast }V\oplus \ ^{3}cT_{\shortparallel }^{\ast
}V\oplus \ ^{4}cT_{\shortparallel }^{\ast }V,\mbox{  for }s=1,2,3,4.
\label{ncon}
\end{eqnarray}%
The double dash on the left again means phase space coordinates with complex momentum, while the single dash means phase space coordinates with real momentum. More details can be found in Appendix A.1. 
The nonlinear s-connection (\ref{ncon}) is characterized by a
set of coefficients $\ _{s}^{\shortparallel }\mathbf{N}=\{\ ^{\shortparallel
}N_{\ i_{s}a_{s}}(u)\}$ which can be used for constructing N-elongated bases
(N-/ s-adapted bases),
\begin{equation}
\ ^{\shortparallel }\mathbf{e}_{\alpha _{s}}=(\ \ ^{\shortparallel }\mathbf{e%
}_{i_{s}}=\ \frac{\partial }{\partial x^{i_{s}}}-\ ^{\shortparallel }N_{\
i_{s}a_{s}}\frac{\partial }{\partial \ ^{\shortparallel }p_{a_{s}}},\ \
^{\shortparallel }e^{b_{s}}=\frac{\partial }{\partial \ ^{\shortparallel
}p_{b_{s}}})\mbox{ on }\ _{s}T\mathbf{T}_{\shortparallel }^{\ast }\mathbf{V,}
\label{nadapbdsc}
\end{equation}%
and, dual s-adapted bases (also called s-cobases)%
\begin{equation}
\ ^{\shortparallel }\mathbf{e}^{\alpha _{s}}=(\ ^{\shortparallel }\mathbf{e}%
^{i_{s}}=dx^{i_{s}},\ ^{\shortparallel }\mathbf{e}_{a_{s}}=d\
^{\shortparallel }p_{a_{s}}+\ ^{\shortparallel }N_{\ i_{s}a_{s}}dx^{i_{s}})%
\mbox{ on }\ \ _{s}T^{\ast }\mathbf{T}_{\shortparallel }^{\ast }\mathbf{V.}
\label{nadapbdss}
\end{equation}%
In similar form, we can construct s-bases $\ ^{\shortmid }\mathbf{e}%
_{\alpha _{s}}[\ ^{\shortmid }N_{\ i_{s}a_{s}}]$ and s-cobases $\
^{\shortmid }\mathbf{e}^{\alpha _{s}}[\ ^{\shortmid }N_{\ i_{s}a_{s}}]$ as
linear N-operators. Such bases are not integrable, {\it i.e.} nonholonomic
(equivalently, anholonomic) \cite{partner01,partner02}.

We now consider the action of $\ ^{\shortmid }\mathbf{e}_{i_{s}}$ (here the single dash on the left means an N-elongated base like \eqref{nadapbdsc} depending on real momenta) on some functions $ f(x,p)$ and $b(x,p),$ we can define the nonholonomic s-adapted star
product $\star _{s}$:
\begin{eqnarray}
f\star _{s}b &:= &\cdot \lbrack \mathcal{F}_{s}^{-1}(f,b)]  \label{starpn} \\
&=&\cdot \lbrack \exp (-\frac{1}{2}i\hbar (\ ^{\shortmid }\mathbf{e}%
_{i_{s}}\otimes \ ^{\shortmid }e^{i_{s}}-\ ^{\shortmid }e^{i_{s}}\otimes \
^{\shortmid }\mathbf{e}_{i_{s}})+\frac{i\mathit{\ell }_{s}^{4}}{12\hbar }%
R^{i_{s}j_{s}a_{s}}(p_{a_{s}}\ ^{\shortmid }\mathbf{e}_{i_{s}}\otimes \
^{\shortmid }\mathbf{e}_{j_{a}}-\ ^{\shortmid }\mathbf{e}_{j_{s}}\otimes
p_{a_{s}}\ ^{\shortmid }\mathbf{e}_{i_{s}}))]f\otimes b  \notag \\
&=&f\cdot b-\frac{i}{2}\hbar \lbrack (\ ^{\shortmid }\mathbf{e}_{i_{s}}f)(\
^{\shortmid }e^{i_{s}}b)-(\ ^{\shortmid }e^{i_{s}}f)(\ ^{\shortmid }\mathbf{e%
}_{i_{s}}b)]+\frac{i\mathit{\ell }_{s}^{4}}{6\hbar }%
R^{i_{s}j_{s}a_{s}}p_{a_{s}}(\ ^{\shortmid }\mathbf{e}_{i_{s}}f)(\
^{\shortmid }\mathbf{e}_{j_{s}}b)+\ldots .  \notag
\end{eqnarray}%
This nonassociative operator involves a constant $\mathit{\ell }$
characterizing the R-flux contributions determined by an antisymmetric $%
R^{i_{s}j_{s}a_{s}}$ background in string theory. The tensor product $\otimes $ can also be written in a
s-adapted form $\otimes _{s}.$ For a decomposition in parameters $\hbar $ and $\kappa =%
\mathit{\ell }_{s}^{3}/6\hbar ,$ the tensor products turn into usual
multiplications as in the third line of (\ref{starpn}).

To investigate associative geometric flows, commutative geometric flows, and physical models
on phase space we work with (pseudo) Riemannian symmetric metrics on
cotangent Lorentz bundle $T^{\ast }V$. Such a metric field is a tensor $\
^{\shortparallel }g=\{\ ^{\shortparallel }g_{\alpha \beta }\}\in $ $TT^{\ast
}V\otimes TT^{\ast }V$ of local signature $(+,+,+,-;+,+,+,-).$ We can write $%
\ ^{\shortparallel }g$ in equivalent form as an s-metric $\
_{s}^{\shortparallel }\mathbf{g}=\{\ ^{\shortparallel }\mathbf{g}_{\alpha
_{s}\beta _{s}}\}$ with coefficients computed with respect to a nonholonomic
s-base and using cofiber coordinates multiplied to complex identity $i$ if
such an s-metric is on a phase space $\mathcal{M}=\mathbf{T}_{\shortparallel
}^{\ast }\mathbf{V.}$ Considering symmetric tensor products of $\
^{\shortparallel }\mathbf{\ e}^{\alpha _{s}}\in T^{\ast }\mathbf{T}%
_{s\shortparallel }^{\ast }\mathbf{V}$ (\ref{nadapbdss}), we can obtain an
s-decomposition of the form
\begin{eqnarray}
g=\ _{s}^{\shortparallel }\mathbf{g} &=&(h_{1}\ ^{\shortparallel }\mathbf{g}%
,~v_{2}\ ^{\shortparallel }\mathbf{g},\ c_{3}\ ^{\shortparallel }\mathbf{g,}%
c_{4}\ ^{\shortparallel }\mathbf{g})\in T\mathbf{T}_{\shortparallel }^{\ast }%
\mathbf{V}\otimes _{\star }T\mathbf{T}_{\shortparallel }^{\ast }\mathbf{V}
\label{sdm} 
\\
&=&\ ^{\shortparallel }\mathbf{g}_{\alpha _{s}\beta _{s}}(\ _{s}^{\shortparallel }u)\ ^{\shortparallel }\mathbf{e}^{\alpha _{s}}\otimes
 \ ^{\shortparallel }\mathbf{e}^{\beta _{s}}=\{\ ^{\shortparallel }\mathbf{g}_{\alpha _{s}\beta _{s}}=(\ ^{\shortparallel }\mathbf{g}%
_{i_{1}j_{1}},\ ^{\shortparallel }\mathbf{g}_{a_{2}b_{2}},\ ^{\shortparallel
}\mathbf{g}^{a_{3}b_{3}},\ ^{\shortparallel }\mathbf{g}^{a_{4}b_{4}})\}.
\notag
\end{eqnarray}%
A star product and R-flux deformation in nonassociative geometry transforms
a symmetric metric $\ _{s}^{\shortparallel }\mathbf{g}$ into a general
nonsymmetric one \cite{blumenhagen16,aschieri17} with symmetric and nonsymmetric components, $%
\ _{\star }^{\shortparallel }\mathbf{g,}$ and $\ _{\star
}^{\shortparallel }\mathbf{q}\mathfrak{,}$. Nonholonomic
s-adapted constructions \cite{partner01,partner02} can be used to split  $%
\ _{\star }^{\shortparallel }\mathbf{g,}$ and $\ _{\star
}^{\shortparallel }\mathbf{q}\mathfrak{,}$ as 
\begin{eqnarray}
&& \mbox{ symmetric: } \ _{\star s}^{\shortparallel }\mathbf{g}=(h_{1}\
_{\star }^{\shortparallel }\mathbf{g},v_{2}\ _{\star }^{\shortparallel }%
\mathbf{g,}c_{3}\ _{\star }^{\shortparallel }\mathbf{g,}c_{4}\ _{\star
}^{\shortparallel }\mathbf{g})  \label{ssdm} \\
&&=\{\ _{\star }^{\shortparallel }\mathbf{g}_{\alpha _{s}\beta _{s}}=\
_{\star }^{\shortparallel }\mathbf{g}_{\beta _{s}\alpha _{s}}=(\ _{\star
}^{\shortparallel }\mathbf{g}_{i_{1}j_{1}}=\ _{\star }^{\shortparallel }%
\mathbf{g}_{j_{1}i_{1}},\ _{\star }^{\shortparallel }\mathbf{g}%
_{a_{2}b_{2}}=\ _{\star }^{\shortparallel }\mathbf{g}_{b_{2}a_{2}},\ _{\star
}^{\shortparallel }\mathbf{g}^{a_{3}b_{3}}=\ \ _{\star }^{\shortparallel }%
\mathbf{g}^{b_{3}a_{3}},\ _{\star }^{\shortparallel }\mathbf{g}%
^{a_{4}b_{4}}=\ _{\star }^{\shortparallel }\mathbf{g}^{b_{4}a_{4}})\}  \notag
\\
&&  \notag \\
&&\mbox{ and nonsymmetric: }\ _{\star }^{\shortparallel }\mathbf{q}=(h_{1}\
_{\star }^{\shortparallel }\mathbf{q},v_{2}\ _{\star }^{\shortparallel }%
\mathbf{q},c_{3}\ _{\star }^{\shortparallel }\mathbf{q}, c_{4}\
_{\star }^{\shortparallel }\mathbf{q})  \label{nssdm} \\
&&=\{\ _{\star }^{\shortparallel }\mathbf{q}_{\alpha _{s}\beta _{s}}=(\
_{\star }^{\shortparallel }\mathbf{q}_{i_{1}j_{1}}\neq \ _{\star
}^{\shortparallel }\mathbf{q}_{j_{1}i_{1}},\ \ _{\star }^{\shortparallel }%
\mathbf{q}_{a_{2}b_{2}}\neq \ _{\star }^{\shortparallel }\mathbf{q}%
_{b_{2}a_{2}}\ _{\star }^{\shortparallel }\mathbf{q}^{a_{3}b_{3}}\neq \
_{\star }^{\shortparallel }\mathbf{q}^{b_{3}a_{3}},\ _{\star
}^{\shortparallel }\mathbf{q}^{a_{4}b_{4}}\neq \ _{\star }^{\shortparallel }%
\mathbf{q}^{b_{4}a_{4}})\neq \ _{\star }^{\shortparallel }\mathbf{q}_{\beta
_{s}\alpha _{s}}\}.  \notag
\end{eqnarray}

For any associative and commutative phase space s-metrics, $\
_{s}^{\shortparallel }\mathbf{g}$ , we can construct two
linear connection structures - the Levi-Civita (LC) connection and
the canonical s-connection: 
\begin{equation}
(\ _{s}^{\shortparallel }\mathbf{g,\ }_{s}^{\shortparallel }\mathbf{N}%
)\rightarrow \left\{
\begin{array}{cc}
\ ^{\shortparallel }\mathbf{\nabla :} & \ ^{\shortparallel }\mathbf{\nabla }%
\ \ _{s}^{\shortparallel }\mathbf{g}=0;\ _{\nabla }^{\shortparallel }%
\mathcal{T}=0,\ \mbox{\  LC--connection }; \\
\ _{s}^{\shortparallel }\widehat{\mathbf{D}}: &
\begin{array}{c}
\ _{s}^{\shortparallel }\widehat{\mathbf{D}}\ _{s}^{\shortparallel }\mathbf{g%
}=0;\ h_{1}\ ^{\shortparallel }\widehat{\mathcal{T}}=0,v_{2}\
^{\shortparallel }\widehat{\mathcal{T}}=0,c_{3}\ ^{\shortparallel }\widehat{%
\mathcal{T}}=0,c_{4}\ ^{\shortparallel }\widehat{\mathcal{T}}=0, \\
h_{1}v_{2}\ ^{\shortparallel }\widehat{\mathcal{T}}\neq 0,h_{1}c_{s}\
^{\shortparallel }\widehat{\mathcal{T}}\neq 0,v_{2}c_{s}\ ^{\shortparallel }%
\widehat{\mathcal{T}}\neq 0,c_{3}c_{4}\ ^{\shortparallel }\widehat{\mathcal{T%
}}\neq 0,%
\end{array}%
\begin{array}{c}
\mbox{ canonical } \\
\mbox{ s-connection  }.%
\end{array}%
\end{array}%
\right.  \label{twocon}
\end{equation}%
We use hat labels for canonical geometric objects determined by $\
_{s}^{\shortparallel }\widehat{\mathbf{D}}=(h_{1}\ ^{\shortparallel }%
\widehat{\mathbf{D}},\ v_{2}\ ^{\shortparallel }\widehat{\mathbf{D}},\
c_{3}\ ^{\shortparallel }\widehat{\mathbf{D}},\ c_{4}\ ^{\shortparallel }%
\widehat{\mathbf{D}})$ acting on tangent spaces of phase space, {\it i.e.} on $T%
\mathbf{T}_{\shortparallel }^{\ast }\mathbf{V.}$ Fundamental geometric
objects like torsions, $\ _{\nabla }^{\shortparallel }\mathcal{T}=0$ and $\
\ ^{\shortparallel }\widehat{\mathcal{T}}\neq 0,$ and curvature, $\ _{\nabla
}^{\shortparallel }\mathcal{R}=\{\ _{\nabla }^{\shortparallel }R_{\ \beta
_{s}\gamma _{s}\delta _{s}}^{\alpha _{s}}\}$ and $\ _{s}^{\shortparallel }%
\widehat{\mathcal{R}}=\{\ ^{\shortparallel }\widehat{\mathbf{R}}_{\ \beta
_{s}\gamma _{s}\delta _{s}}^{\alpha _{s}}\},$ and other s-tensor objects can be defined and
computed (see details in \cite{partner01,partner02}).

In \cite{partner02}, we studied in detail the nonassociative (non)
symmetric and generalized connection structures on 8-d phase spaces. A nonassociative symmetric metric
s-tensor, as in (\ref{ssdm}), on a phase space with R-flux induced terms can be represented as
\begin{equation*}
\ _{\star }^{\shortparallel }\mathbf{g}=\ _{\star }^{\shortparallel }\mathbf{%
g}_{\alpha _{s}\beta _{s}}\star (\ ^{\shortparallel }\mathbf{e}^{\alpha
_{s}}\otimes _{\star }\ ^{\shortparallel }\mathbf{e}^{\beta _{s}}),%
\mbox{
where }\ _{\star }^{\shortparallel }\mathbf{g}(\ ^{\shortparallel }\mathbf{e}%
_{\alpha _{s}},\ ^{\shortparallel }\mathbf{e}_{\beta _{s}})=\ _{\star
}^{\shortparallel }\mathbf{g}_{\alpha _{s}\beta _{s}}=\ _{\star
}^{\shortparallel }\mathbf{g}_{\beta _{s}\alpha _{s}}\in \mathcal{A}%
_{s}^{\star }.
\end{equation*}%
Such a nonassociatve s-metric is compatible with star R-flux
deformations of an s-connection $\ _{s}^{\shortparallel }\mathbf{D}%
\rightarrow \ _{s}^{\shortparallel }\mathbf{D}^{\star }$ if the condition $%
\ _{s}^{\shortparallel }\mathbf{D}^{\star }\ _{\star }^{\shortparallel }%
\mathbf{g}=0$ is satisfied. Here, we note that with
respect to s-adapted bases, $\ ^{\shortparallel }\mathbf{e}_{\xi _{s}}$, and
tensor products of their dual s-bases, a nonsymmetric s-metric structure (%
\ref{nssdm}) can be parameterized in a $[(2\times 2)+(2\times 2)]+$ $%
[(2\times 2)+(2\times 2)]$ block form,
\begin{equation}
\ _{\star }^{\shortparallel }\mathbf{q}_{\alpha _{s}\beta _{s}}=\ _{\star
}^{\shortparallel }\mathbf{g}_{\alpha _{s}\beta _{s}}-i\kappa \overline{%
\mathcal{R}}_{\quad \alpha _{s}}^{\tau _{s}\xi _{s}}\ \mathbf{%
^{\shortparallel }e}_{\xi _{s}}\ _{\star }^{\shortparallel }\mathbf{g}%
_{\beta _{s}\tau _{s}}.  \label{dmss1}
\end{equation}%
Formulas (\ref{aux40bb}), (\ref{aux40b}) and (\ref{aux40aa}) from Appendix \ref{appendixa}
show how to split $
\ _{\star }^{\shortparallel }\mathbf{q}_{\alpha _{s}\beta _{s}}$ into
symmetric and anti-symmetric parts.

Here we use the convention \cite{partner01,partner02} that on phase spaces, the star product (\ref{starpn}) can
be defined using nonholonomic dyadic decompositions with $\ ^{\shortparallel
}\mathbf{e}_{\alpha _{s}}$ (\ref{nadapbdsc}) and R-flux terms. This can be used to 
compute star deformations of canonical, s-adapted, geometric
objects into nonassociative metric of the form:%
\begin{equation}
\begin{array}{ccc}
\fbox{$(\star _{N},\ \ \mathcal{A}_{N}^{\star },\ _{\star }^{\shortparallel }%
\mathbf{g,\ _{\star }^{\shortparallel }q\mathfrak{,}\ \ ^{\shortparallel }N},%
\mathbf{\ \ ^{\shortparallel }e}_{\alpha }\mathbf{,\ \mathbf{\mathbf{\mathbf{%
\ ^{\shortparallel }}}}D}^{\star })$} & \Leftrightarrow & \fbox{$(\star
_{s},\ \ \mathcal{A}_{s}^{\star },\ _{\star s}^{\shortparallel }\mathbf{g,\
_{\star s}^{\shortparallel }q\mathfrak{,}}\ \ _{s}^{\shortparallel }\mathbf{N%
},\mathbf{\ \ ^{\shortparallel }e}_{\alpha _{s}}\mathbf{,\ \mathbf{\mathbf{%
\mathbf{\ _{s}^{\shortparallel }}}}D}^{\star })$} \\
& \Uparrow &  \\
\fbox{$(\ \ ^{\shortparallel }\mathbf{g,\ \ ^{\shortparallel }N},\mathbf{\ \
^{\shortparallel }e}_{\alpha }\mathbf{,}\ ^{\shortparallel }\widehat{\mathbf{%
D}})$} & \Leftrightarrow & \fbox{$(\ \ _{s}^{\shortparallel }\mathbf{g,\ \
_{s}^{\shortparallel }N},\mathbf{\ \ ^{\shortparallel }e}_{\alpha _{s}}%
\mathbf{,}\ _{s}^{\shortparallel }\widehat{\mathbf{D}}),$}%
\end{array}
\label{conv2s}
\end{equation}%
where $\ ^{\shortparallel }\mathbf{D}^{\star }=\ ^{\shortparallel }\nabla
^{\star }+\ ^{\shortparallel }\widehat{\mathbf{Z}}^{\star },$ for a
nonholonomic splitting 4+4, and $\ _{s}^{\shortparallel }\mathbf{D}^{\star
}=\ ^{\shortparallel }\nabla ^{\star }+\ _{s}^{\shortparallel }\widehat{%
\mathbf{Z}}^{\star },$ for an s-splitting.

Following the convention of (\ref{conv2s}), we can compute
nonassociative star deformations of LC- and canonical
s-connections from (\ref{twocon}) as follows: {\small
\begin{equation}
(\ _{\star s}^{\shortparallel }\mathbf{g,\ _{s}^{\shortparallel }N}%
)\rightarrow \left\{
\begin{array}{cc}
_{\star }^{\shortparallel }\mathbf{\nabla :}\  &
\begin{array}{c}
\fbox{\ $_{\star }^{\shortparallel }\mathbf{\nabla }$\ $_{\star
}^{\shortparallel }\mathbf{g}=0$;\ $_{\nabla }^{\shortparallel }\mathcal{T}%
^{\star }=0,$}\mbox{\ star
LC-connection}; \\
\end{array}
\\
\ _{s}^{\shortparallel }\widehat{\mathbf{D}}^{\star }: & \fbox{$%
\begin{array}{c}
\ _{s}^{\shortparallel }\widehat{\mathbf{D}}^{\star }\ _{\star
s}^{\shortparallel }\mathbf{g}=0;\ h_{1}\ ^{\shortparallel }\widehat{%
\mathcal{T}}^{\star }=0,v_{2}\ ^{\shortparallel }\widehat{\mathcal{T}}%
^{\star }=0,c_{3}\ ^{\shortparallel }\widehat{\mathcal{T}}^{\star }=0,c_{4}\
^{\shortparallel }\widehat{\mathcal{T}}^{\star }=0, \\
h_{1}v_{2}\ ^{\shortparallel }\widehat{\mathcal{T}}^{\star }\neq
0,h_{1}c_{s}\ ^{\shortparallel }\widehat{\mathcal{T}}^{\star }\neq
0,v_{2}c_{s}\ ^{\shortparallel }\widehat{\mathcal{T}}^{\star }\neq
0,c_{3}c_{4}\ ^{\shortparallel }\widehat{\mathcal{T}}^{\star }\neq 0,%
\end{array}%
$}\mbox{ canonical  s-connection }.%
\end{array}%
\right.  \label{twoconsstar}
\end{equation}%
} In these formulas, we can consider nonholonomic, dyadic horizontal and (co)
vertical decompositions of the form
$\ _{s}^{\shortparallel }\widehat{\mathbf{D}}^{\star }=(h_{1}\
^{\shortparallel }\widehat{\mathbf{D}}^{\star },\ v_{2}\ ^{\shortparallel }%
\widehat{\mathbf{D}}^{\star },\ c_{3}\ ^{\shortparallel }\widehat{\mathbf{D}}%
^{\star },\ c_{4}\ ^{\shortparallel }\widehat{\mathbf{D}}^{\star })$, and adapted
to nonlinear s--connection structures $\ _{s}^{\shortparallel }%
\mathbf{N.}$

There is a canonical distortion relation between two linear connections in (%
\ref{twoconsstar}), given by
\begin{equation}
\ _{s}^{\shortparallel }\widehat{\mathbf{D}}^{\star }=\ ^{\shortparallel
}\nabla ^{\star }+\ _{\star }^{\shortparallel }\widehat{\mathbf{Z}}.
\label{candistrnas}
\end{equation}%
Without the star labels \eqref{candistrnas} takes the associative and
commutative form in (\ref{twocon}). In order to apply the AFCDM to
construct physically interesting solutions it is more convenient 
to work with (non) associative nonholonomic
dyadic canonical geometric quantities $(\ _{\star }^{\shortparallel }\mathbf{q},\
_{s}^{\shortparallel }\widehat{\mathbf{D}}^{\star }).$ After constructing classes
of nonholonomic solutions in explicit form, we can
redefine the constructions in terms of star deformed LC-configurations using (\ref{candistrnas}). Nonassociative
LC-configurations can be extracted imposing zero s-torsion conditions,%
\begin{equation}
\ _{\star s}^{\shortparallel }\widehat{\mathbf{Z}}=0,%
\mbox{ which is
equivalent to }\ _{s}^{\shortparallel }\widehat{\mathbf{D}}^{\star }=\
^{\shortparallel }\nabla ^{\star }\mbox{ for }\mid \ _{s}^{\shortparallel }%
\widehat{\mathbf{T}}=0.  \label{lccondnonass}
\end{equation}%
In general, all type of solutions subjected, or not, to some conditions of
type (\ref{lccondnonass}) contain certain nonzero anholonomy coefficients of
frame structures. 

\subsection{Nonassociative vacuum Einstein equations and dyadic $\kappa$--parametric splitting}

The nonassociative Riemannian, $\ ^{\shortparallel }\widehat{\mathcal{\Re }}%
_{\quad }^{\star }=\{\ ^{\shortparallel }\widehat{\mathcal{\Re }}_{\quad
\alpha _{s}\beta _{s}\gamma _{s}}^{\star \mu _{s}}\},$ and Ricci, $\
_{s}^{\shortparallel }\widehat{\mathcal{R}}ic^{\star }=\{\ ^{\shortparallel }%
\widehat{\mathbf{R}}_{\ \beta _{s}\gamma _{s}}^{\star }\},$ s-tensors for
the canonical s-connection $\ _{s}^{\shortparallel }\widehat{\mathbf{D}}%
^{\star }=\{\ ^{\shortparallel }\widehat{\mathbf{\Gamma }}_{\star \alpha
_{s}\beta _{s}}^{\gamma _{s}}\}$ (\ref{candistrnas}) are defined
respectively by formulas (\ref{nadriemhopfcan}) and (\ref{driccicanonstar1})
in Appendix \ref{appendixa}. Such fundamental geometric objects can be
decomposed in parametric form (\ref{paramscon}) with $[01,10,11]:=\left%
\lceil \hbar ,\kappa \right\rceil $ components containing nonassociative and
noncommutative contributions from star product deformations which can be
real or complex ones. 

We can consider nonholonomic distributions on phase
space when $\ _{[00]}^{\shortparallel }\widehat{\mathbf{R}}ic_{\alpha
_{s}\beta _{s}}^{\star }=\ ^{\shortparallel }\widehat{\mathbf{R}}_{\ \alpha
_{s}\beta _{s}}$ \ are determined by associative and commutative s-adapted
canonical s-connection \cite{bubuianu19}. As a result, the
star s-deformed Ricci s-tensor can be expressed in parametric form,
\begin{eqnarray}
\ _{s}^{\shortparallel }\widehat{\mathcal{R}}ic^{\star } &=&\{\
^{\shortparallel }\widehat{\mathbf{R}}_{\ \beta _{s}\gamma _{s}}^{\star
}\}=\ _{s}^{\shortparallel }\widehat{\mathcal{R}}ic+\ _{s}^{\shortparallel }%
\widehat{\mathcal{K}}ic\left\lceil \hbar ,\kappa \right\rceil =\{\
^{\shortparallel }\widehat{\mathbf{R}}_{\ \beta _{s}\gamma _{s}}+\
^{\shortparallel }\widehat{\mathbf{K}}_{\ \beta _{s}\gamma _{s}}\left\lceil
\hbar ,\kappa \right\rceil \},  \label{paramsricci} \\
&& \mbox{ where }\ _{s}^{\shortparallel }\widehat{\mathcal{K}}ic =\{\
^{\shortparallel }\widehat{\mathbf{K}}_{\ \beta _{s}\gamma _{s}}\left\lceil
\hbar ,\kappa \right\rceil =\ _{[01]}^{\shortparallel }\widehat{\mathbf{%
\mathbf{\mathbf{\mathbf{R}}}}}ic_{\beta _{s}\gamma _{s}}^{\star }+\
_{[10]}^{\shortparallel }\widehat{\mathbf{R}}ic_{\beta _{s}\gamma
_{s}}^{\star }+\ _{[11]}^{\shortparallel }\widehat{\mathbf{R}}ic_{\beta
_{s}\gamma _{s}}^{\star }\} ~. \notag
\end{eqnarray}%
the above gives nonassociative deformations of the canonical Ricci s-tenor.

For the canonical s-connection $\ _{s}^{\shortparallel }\widehat{\mathbf{D}}%
^{\star },$ the nonassociative phase space vacuum Einstein equations for (\ref{paramsricci}), with a
nonzero cosmological constant on at least one shell  ($\ _{s}^{\shortparallel
}\lambda \neq 0$ for some $s,$), can be written in the form
\begin{equation}
\ ^{\shortparallel }\widehat{\mathbf{R}}ic_{\alpha _{s}\beta _{s}}^{\star }-%
\frac{1}{2}\ _{\star }^{\shortparallel }\mathbf{g}_{\alpha _{s}\beta
_{s}}\ _{s}^{\shortparallel }\widehat{\mathbf{R}}sc^{\star }=\
_{s}^{\shortparallel }\lambda \ _{\star }^{\shortparallel }\mathbf{g}%
_{\alpha _{s}\beta _{s}},  \label{nonassocdeinst1}
\end{equation}%
where the nonassociative scalar curvature $_{s}^{\shortparallel }\widehat{%
\mathbf{R}}sc^{\star }$ is defined by formulas (\ref{ricciscsymnonsym}). We
can consider nonzero values of $\ ^{\shortparallel }\widehat{\mathbf{R}}%
sc^{\star }=\ ^{1}\Lambda (\ _{1}^{\shortparallel }u)+\ ^{2}\Lambda (\
_{2}^{\shortparallel }u)+\ ^{3}\Lambda (\ _{3}^{\shortparallel }u)+\
^{4}\Lambda (\ _{4}^{\shortparallel }u)$, given as a sum of effective cosmological constants on different shells $\ ^{s}\Lambda (\ _{s}^{\shortparallel }u)$. It is possible to choose
effective $\ ^{s}\Lambda (\ _{s}^{\shortparallel }u)$ and $\
_{s}^{\shortparallel }\lambda =\ ^{\shortparallel }\lambda $ in a form that $%
\ ^{\shortparallel }\lambda +\frac{1}{2}\ _{\star }^{\shortparallel }\mathbf{%
q}_{\alpha _{s}\beta _{s}}\ _{s}^{\shortparallel }\widehat{\mathbf{R}}%
sc^{\star }=0,$ when $\ _{\star }^{\shortparallel }\mathbf{q}_{\alpha
_{s}\beta _{s}}=\ _{\star }^{\shortparallel }\mathbf{\check{q}}_{\alpha
_{s}\beta _{s}}+\ _{\star }^{\shortparallel }\mathbf{a}_{\alpha _{s}\beta
_{s}}.$ The nonassociative symmetric and nonsymmetric components $\ _{\star
}^{\shortparallel }\mathbf{\check{q}}_{\alpha _{s}\beta _{s}}$ and $\
_{\star }^{\shortparallel }\mathbf{a}_{\alpha _{s}\beta _{s}}$ are computed
following formulas (\ref{aux40bb}), (\ref{aux40b}) and (\ref{aux40aa}) being
determined by $\ _{\star }^{\shortparallel }\mathbf{\check{q}}_{\alpha
_{s}\beta _{s}}^{[0]}=\ _{\star }^{\shortparallel }\mathbf{g}_{\alpha
_{s}\beta _{s}}$ and $i\kappa \overline{\mathcal{R}}_{\quad \alpha
_{s}}^{\tau _{s}\xi _{s}}\ \mathbf{^{\shortparallel }e}_{\xi _{s}}$ in (\ref%
{dmss1}).

Using (\ref{paramsricci}) for the canonical Ricci s-tensor, we express the nonassociative vacuum gravitational field equations (\ref{nonassocdeinst1}) in the form
\begin{eqnarray}
\ ^{\shortparallel }\widehat{\mathbf{R}}_{\ \beta _{s}\gamma _{s}}~ &=&\
^{\shortparallel }\mathbf{K}_{_{\beta _{s}\gamma _{s}}},%
\mbox{ for effective
nonassociative sources }  \label{cannonsymparamc2} \\
\ ^{\shortparallel }\mathbf{K}_{_{\beta _{s}\gamma _{s}}} &=&\
_{[0]}^{\shortparallel }\Upsilon _{_{\beta _{s}\gamma _{s}}}+\
_{[1]}^{\shortparallel }\mathbf{K}_{_{\beta _{s}\gamma _{s}}}\left\lceil
\hbar ,\kappa \right\rceil ,\mbox{ where }  \notag \\
&&\ _{[0]}^{\shortparallel }\Upsilon _{_{\beta _{s}\gamma _{s}}}=\
^{s}\Lambda (\ ^{\shortparallel }u^{\gamma _{s}})\ _{\star }^{\shortparallel
}\mathbf{g}_{\beta _{s}\gamma _{s}}\ \mbox{and }  \label{assocsourc2} \\
&&\ _{[1]}^{\shortparallel }\mathbf{K}_{_{\beta _{s}\gamma _{s}}}\left\lceil
\hbar ,\kappa \right\rceil =\ ^{s}\Lambda (\ ^{\shortparallel }u^{\gamma
_{s}})\ _{\star }^{\shortparallel }\mathbf{\check{q}}_{\beta _{s}\gamma
_{s}}^{[1]}(\kappa )-\ ^{\shortparallel }\widehat{\mathbf{K}}_{\ \beta
_{s}\gamma _{s}}\left\lceil \hbar ,\kappa \right\rceil .
\label{nassocsourc2}
\end{eqnarray}%
$_{[1]}^{\shortparallel }\mathbf{K}_{_{\beta _{s}\gamma _{s}}}\left\lceil
\hbar ,\kappa \right\rceil$ are effective parametric sources with coefficients proportional to $\hbar ,\kappa $ and $\hbar \kappa .$

The effective sources in (\ref{cannonsymparamc2}) can be parameterized for
nontrivial real \textsf{quasi-stationary} 8-d configurations\footnote{\label%
{fnqs} An s-metric is \textsf{quasi-stationary} if the corresponding (non)
associative phase spacetime geometric s-objects possess a Killing symmetry with respect to $t$ on shell $s=2$ and with respect to $\ ^{\shortparallel
}p _{7},$ or $\ ^{\shortparallel }p _{8},$ for all shells up
to $s=4.$} using coordinates $(x^{k_{3}},\ ^{\shortparallel }p_{8}),$ for $\
^{\shortmid }p_{8}=E,$ with $\ _{\star }^{\shortparallel }\mathbf{g}_{\beta
_{s}\gamma _{s}\mid \hbar ,\kappa =0}=\ ^{\shortparallel }\mathbf{g}_{\beta
_{s}\gamma _{s}},$ in such forms: {\small
\begin{eqnarray}
\ ^{\shortparallel }\mathbf{K}_{\ \beta _{s}}^{\alpha _{s}} &=&\{\
^{\shortparallel }\mathcal{K}_{\ j_{1}}^{i_{1}}(\kappa ,x^{k_{1}})=[\ \
_{1}^{\shortparallel }\Upsilon (x^{k_{1}})+\ _{1}^{\shortparallel }\mathbf{K}%
(\kappa ,x^{k_{1}})]\delta _{j_{1}}^{i_{1}},\ ^{\shortparallel }\mathcal{K}%
_{\ j_{2}}^{i_{2}}(\kappa ,x^{k_{1}},x^{3})=[\ _{2}^{\shortparallel
}\Upsilon (x^{k_{1}},x^{3})+\ _{2}^{\shortparallel }\mathbf{K}(\kappa
,x^{k_{1}},x^{3})]\delta _{b_{2}}^{a_{2}},  \notag \\
&&\ ^{\shortparallel }\mathcal{K}_{\ a_{3}}^{b_{3}}(\kappa ,x^{k_{2}},\
^{\shortparallel }p_{6})=[\ \ _{3}^{\shortparallel }\Upsilon (x^{k_{2}},\
^{\shortparallel }p_{6})+\ _{3}^{\shortparallel }\mathbf{K}(x^{k_{2}},\
^{\shortparallel }p_{6})]\ \delta _{a_{3}}^{b_{3}},  \label{ansatzsourchv} \\
&&\ ^{\shortparallel }\mathcal{K}_{\ a_{4}}^{b_{4}}(\kappa ,x^{k_{3}},\
^{\shortparallel }p_{8})=[\ \ _{4}^{\shortparallel }\Upsilon (x^{k_{3}},\
^{\shortparallel }p_{8})+\ _{4}^{\shortparallel }\mathbf{K}(x^{k_{3}},\
^{\shortparallel }p_{8})]\delta _{a_{4}}^{b_{4}}\},\mbox{ where }  \notag \\
\ ^{\shortparallel }\mathbf{K}_{\ j_{s}k_{s}} &=&-\ _{[11]}^{\shortparallel }%
\widehat{\mathbf{R}}ic_{j_{s}k_{s}}^{\star }(\ ^{\shortparallel
}u^{k_{s-2}},\ ^{\shortparallel }u^{k_{s}})\mbox{ as in }(\ref%
{driccicanonstar1}),\mbox{ for }  \notag \\
\mathbf{g}_{j_{s}k_{s}}
&=&%
\{g_{1}(x^{k_{1}}),g_{2}(x^{k_{1}}),g_{3}(x^{k_{1}},x^{3}),g_{4}(x^{k_{1}},x^{3}),\ ^{\shortparallel }g^{5}(x^{k_{2}},\ ^{\shortparallel }p_{6}),\ ^{\shortparallel }g^{6}(x^{k_{2}},\ ^{\shortparallel }p_{6}),\ ^{\shortparallel }g^{7}(x^{k_{3}},\ ^{\shortparallel }p_{8}),\ ^{\shortparallel }g^{8}(x^{k_{3}},\ ^{\shortparallel }p_{8})\}.
\notag
\end{eqnarray}%
} To apply the AFCDM for generating exact/ parametric solutions on (non) associative/ commutative 8-d phase spaces with nonholonomic 2+2+2+2 splitting \cite{bubuianu19,partner02} it is important to parameterize the effective sources (\ref{ansatzsourchv}) by using quasi-stationary shell by shell
adapted data
\begin{equation}
\ ^{\shortparallel }\mathbf{K}_{\ \beta _{s}}^{\alpha
_{s}}~=[~_{1}^{\shortparallel }\mathcal{K}(\kappa ,x^{k_{1}})\delta
_{i_{1}}^{j_{1}},~_{2}^{\shortparallel }\mathcal{K}(\kappa
,x^{k_{1}},x^{3})\delta _{b_{2}}^{a_{2}},~_{3}^{\shortparallel }\mathcal{K}%
(\kappa ,x^{k_{2}},\ ^{\shortparallel }p_{6})\delta
_{a_{3}}^{b_{3}},~_{4}^{\shortparallel }\mathcal{K}(\kappa ,x^{k_{3}},\
^{\shortparallel }p_{8})\delta _{a_{4}}^{b_{4}}].  \label{cannonsymparamc2a}
\end{equation}%
Here we note that $\ ^{\shortparallel }\mathbf{K}_{\ \beta _{s}}^{\alpha
_{s}}$ in (\ref{ansatzsourchv}) can be related to a general non-diagonal
source $\ ^{\shortparallel }\widehat{\Upsilon }_{\alpha _{s}^{\prime }\beta
_{s}^{\prime }}$ via frame transforms $\ \ ^{\shortparallel }\widehat{%
\Upsilon }_{\alpha _{s}^{\prime }\beta _{s}^{\prime }}=e_{\ \alpha
_{s}^{\prime }}^{\alpha _{s}}e_{\ \beta _{s}^{\prime }}^{\beta _{s}}\ \
^{\shortparallel }\mathcal{K}_{\alpha _{s}\beta _{s}}.$ Sources of type (\ref%
{ansatzsourchv}) \ and (\ref{cannonsymparamc2a}), with Killing symmetry on $%
\ ^{\shortparallel }p ^{7}$ are used for constructing
exact/parametric quasi-stationary solutions with such symmetries. To
generate alternative classes of solutions with Killing symmetry on $\
^{\shortparallel }p ^{8}$, we may change $\ ^{\shortparallel
}p_{8}\rightarrow \ ^{\shortparallel }p_{7}$ into coefficients of s-metrics
and effective sources. In this work, we study such configurations for a
fixed energy type parameter $\ p_{8}=E=E_{0}.$

Prescribing four effective sources $\ _{s}^{\shortparallel }%
\mathcal{K}$ (\ref{cannonsymparamc2a}) as \textbf{generating sources}, we
constrain the nonholonomic
gravitational dynamics. Such generating sources are related to conventional cosmological constants
via nonlinear symmetries, when the nonsymmetric parts of the s-metrics and
canonical Ricci s-tensors can be computed as R-flux deformations of some
off-diagonal symmetric metric configurations.

\subsection{Nonassociative parametric vacuum quasi-stationary s-metrics}

\label{ssecqs}In nonassociative nonholonomic phase space gravity, the system
in (\ref{cannonsymparamc2}) with a generating source (\ref%
{cannonsymparamc2a}) can be decoupled and solved in exact 
forms on 8-d quasi-stationary phase spaces. This can be performed if the
AFCDM is applied \cite%
{partner02}. In this subsection, we modify the constructions and formulas to be able to generate quasi-stationary and BH solutions
with nonassociative star R-flux configurations.

\subsubsection{Nonholonomic deformations encoding star R-flux sources}

We consider a commutative \textbf{prime } s-metric $\ _{s}^{\shortmid }\mathbf{\mathring{g}}$ and s-connection structures of type (\ref{sdm}) and (\ref{nadapbdss})
\begin{eqnarray}
\ ^{\shortparallel }\mathbf{\mathring{g}} &=&\ _{s}^{\shortmid }\mathbf{%
\mathring{g}}=\ ^{\shortmid }\mathring{g}_{\alpha _{s}\beta
_{s}}(x^{i_{s}},p_{a_{s}})d\ ~^{\shortparallel }u^{\alpha _{s}}\otimes d\
~^{\shortparallel }u^{\beta _{s}}=\ ~^{\shortparallel }\mathbf{\mathring{g}}%
_{\alpha _{s}\beta _{s}}(\ _{s}^{\shortmid }u)\ ~^{\shortparallel }\mathbf{%
\mathbf{\mathring{e}}}^{\alpha _{s}}\mathbf{\otimes \ ~^{\shortparallel }%
\mathbf{\mathring{e}}}^{\beta _{s}},  \label{primedm} \\
&&\ ^{\shortparallel }\mathbf{\mathring{e}}^{\alpha _{s}}=[d\
^{\shortparallel }x^{i_{s}},\ ^{\shortparallel }\mathbf{\mathring{e}}%
^{a_{s}}=d\ ^{\shortparallel }y^{a_{s}}+\ ^{\shortparallel }\mathring{N}%
_{i_{s}}^{a_{s}}(~^{\shortparallel }u)d~^{\shortparallel }x^{i_{s}}]\in
T_{s}^{\ast }\mathbf{T}_{\shortparallel }^{\ast }\mathbf{V}.  \notag
\end{eqnarray}%
By prime we mean the metric from which we begin when we make various deformations. These prime metrics may be some well known solution, such as Schwarzschild. Here, we shall consider prime metrics which are exact
solutions in 8-d nonholonomic phase space gravity \cite{bubuianu19} subjected
to nonassociative star R-flux extensions into another classes of solutions
determined, respectively, by symmetric and nonsymmetric metrics, $_{\star
}^{\shortparallel }\mathbf{\check{q}}_{\mu _{s}\nu _{s}}^{\circ }=(\ _{\star
}^{\shortparallel }\mathbf{\check{q}}_{i_{1}j_{1}}^{\circ },\ _{\star
}^{\shortparallel }\mathbf{\check{q}}_{a_{2}b_{2}}^{\circ },\ _{\star
}^{\shortparallel }\mathbf{\check{q}}_{\circ }^{a_{3}b_{3}},\ _{\star
}^{\shortparallel }\mathbf{\check{q}}_{\circ }^{a_{4}b_{4}})$ of type (\ref%
{aux40b}), and $\ _{\star }^{\shortparallel }\mathbf{a}_{\mu _{s}\nu
_{s}}^{\circ }=(0,0,\ _{\star }^{\shortparallel }\mathbf{a}%
_{c_{3}b_{3}}^{\circ },\ _{\star }^{\shortparallel }\mathbf{a}%
_{c_{4}b_{4}}^{\circ })$ of type (\ref{aux40aa}).\footnote{%
We label prime s-metrics and related geometric s-objects with a small circle
on the left/right/up of corresponding symbols.}

The main goal of this article is to study nonassociative, nonholonomic deformations
\begin{equation}
\ _{s}^{\shortparallel }\mathbf{\mathring{g}}\rightarrow \
_{s}^{\shortparallel }\mathbf{g}=[\ ^{\shortparallel }g_{\alpha _{s}}= \
^{\shortparallel }\eta _{\alpha _{s}}\ ^{\shortparallel }\mathring{g}
_{\alpha _{s}},\ ^{\shortparallel }N_{i_{s-1}}^{a_{s}}= \
^{\shortparallel}\eta _{i_{s-1}}^{a_{s}}\ ^{\shortparallel }\mathring{N}%
_{i_{s-1}}^{a_{s}}]  \label{offdiagdef}
\end{equation}%
to \textbf{target} quasi-stationary s-metrics of type $\
_{s}^{\shortparallel }\mathbf{g}$ (\ref{ssdm}). We shall use a "hat" on an
s-metric, $\ _{s}^{\shortparallel }\widehat{\mathbf{g}}$, if it defines an
exact/ parametric solution of nonassociative vacuum gravitational equations (%
\ref{cannonsymparamc2}). Such nonholonomic star s-deformations can be
described in terms of so-called gravitational polarization ($\eta $%
-polarization) functions, when the target s-metrics are parameterized as
\begin{eqnarray}
\ _{s}^{\shortparallel }\mathbf{g} &=&~^{\shortparallel }g_{i_{s}}(\hbar
,\kappa ,~^{\shortparallel }x^{k_{s}})d~^{\shortparallel }x^{i_{s}}\otimes
d~^{\shortparallel }x^{i_{s}}+~^{\shortparallel }g_{a_{s}}(\hbar ,\kappa
,~^{\shortparallel }x^{i_{s}},~^{\shortparallel }p_{b_{s}})\
~^{\shortparallel }\mathbf{e}^{a_{s}}\otimes ~^{\shortparallel }\mathbf{e}%
^{a_{s}}  \label{dmpolariz} \\
&=&~^{\shortparallel }\eta _{i_{k}}(\hbar ,\kappa
,x^{i_{1}},y^{a_{2}},~^{\shortparallel }p_{a_{3}},~^{\shortparallel
}p_{a_{4}})\ ~^{\shortparallel }\mathring{g}_{i_{s}}(\hbar ,\kappa
,x^{i_{1}},y^{a_{2}},~^{\shortparallel }p_{a_{3}},~^{\shortparallel
}p_{a_{4}})d~^{\shortparallel }x^{i_{s}}\otimes d~^{\shortparallel }x^{i_{s}}
\notag \\
&&+\ ~^{\shortparallel }\eta _{b_{s}}(\hbar ,\kappa
,x^{i_{1}},y^{a_{2}},~^{\shortparallel }p_{a_{3}},~^{\shortparallel
}p_{a_{4}})\ ~^{\shortparallel }\mathring{g}_{b_{s}}(\hbar ,\kappa
,x^{i_{1}},y^{a_{2}},~^{\shortparallel }p_{a_{3}},~^{\shortparallel
}p_{a_{4}})~^{\shortparallel }\mathbf{e}^{b_{s}}[\eta ]\otimes \
~^{\shortparallel }\mathbf{e}^{b_{s}}[\eta ],  \notag \\
~^{\shortparallel }\mathbf{e}^{\alpha _{s}}[\eta ] &=&(d~^{\shortparallel
}x^{i_{s}},\ ~^{\shortparallel }\mathbf{e}^{a_{s}}=d~^{\shortparallel
}y^{a_{s}}+~^{\shortparallel }\eta _{i_{s}}^{a_{s}}(\hbar ,\kappa
,x^{i_{1}},y^{a_{2}},~^{\shortparallel }p_{a_{3}},~^{\shortparallel
}p_{a_{4}})~^{\shortparallel }\mathring{N}_{i_{s}}^{a_{s}}(\hbar ,\kappa
,x^{i_{1}},y^{a_{2}},~^{\shortparallel }p_{a_{3}},~^{\shortparallel
}p_{a_{4}})d~^{\shortparallel }x^{i_{s}}).  \notag
\end{eqnarray}

The AFCDM can be applied for generating quasi-stationary s-metric ansatz (%
\ref{dmpolariz}) with a small parameter (string constant) $\kappa$ with $0\leq
\kappa <1.$ Also we can use decompositions on a second small parameter, $%
\hbar ,$ in order to distinguish between possible noncommutative and
nonassociative effects. For simplicity,
we only give formulas for linear approximations on $\kappa $
encoding $\hbar $ into generation functions and sources and respective
integration functions. Parametric $\kappa $--decompositions of the $\eta $%
-polarization functions (\ref{offdiagdef}) result in s-adapted coefficients of target metrics of the form:
\begin{eqnarray}
\ ^{\shortparallel }g_{i_{1}}(\kappa ,x^{k_{1}}) &=&~^{\shortparallel }\eta
_{i_{i}}\ ~^{\shortparallel }\mathring{g}_{i_{1}}=~^{\shortparallel }\zeta
_{i_{1}}(1+\kappa ~^{\shortparallel }\chi _{i_{1}})\ ~^{\shortparallel }%
\mathring{g}_{i_{1}}=  \label{aux02} \\
&&\{~^{\shortparallel }\zeta _{i_{1}}(x^{i_{1}},y^{a_{2}},~^{\shortparallel
}p_{a_{3}},~^{\shortparallel }p_{a_{4}})[1+\kappa ~^{\shortparallel }\chi
_{i_{1}}(x^{i_{1}},y^{a_{2}},~^{\shortparallel }p_{a_{3}},~^{\shortparallel
}p_{a_{4}})]\}\ ~^{\shortparallel }\mathring{g}%
_{i_{1}}(x^{i_{1}},y^{a_{2}},~^{\shortparallel }p_{a_{3}},~^{\shortparallel
}p_{a_{4}}),  \notag \\
\ ~^{\shortparallel }g_{b_{2}}(\kappa ,x^{i_{1}},y^{3}) &=&\
~^{\shortparallel }\eta _{b_{2}}\ ~^{\shortparallel }\mathring{g}%
_{b_{1}}=~^{\shortparallel }\zeta _{b_{2}}(1+\kappa \ ~^{\shortparallel
}\chi _{b_{2}})\ ~^{\shortparallel }\mathring{g}_{b_{1}}=  \notag \\
&&\{~^{\shortparallel }\zeta _{b_{2}}(x^{i_{1}},y^{a_{2}},~^{\shortparallel
}p_{a_{3}},~^{\shortparallel }p_{a_{4}})[1+\kappa \ ~^{\shortparallel }\chi
_{b_{2}}(x^{i_{1}},y^{a_{2}},~^{\shortparallel }p_{a_{3}},~^{\shortparallel
}p_{a_{4}})]\}\ ~^{\shortparallel }\mathring{g}%
_{b_{1}}(x^{i_{1}},y^{a_{2}},~^{\shortparallel }p_{a_{3}},~^{\shortparallel
}p_{a_{4}}),  \notag \\
\ ~^{\shortparallel }g^{a_{3}}(\kappa ,x^{i_{2}},~^{\shortparallel }p_{6})
&=&\ ~^{\shortparallel }\eta ^{a_{3}}\ ~^{\shortparallel }\mathring{g}%
^{a_{3}}=~^{\shortparallel }\zeta ^{a_{3}}(1+\kappa \ ~^{\shortparallel
}\chi ^{a_{3}})\ ~^{\shortparallel }\mathring{g}^{a_{3}}=  \notag \\
&&\{~^{\shortparallel }\zeta ^{a_{3}}(x^{i_{1}},y^{b_{2}},~^{\shortparallel
}p_{b_{3}},~^{\shortparallel }p_{b_{4}})\ [1+\kappa \ ~^{\shortparallel
}\chi ^{a_{3}}(x^{i_{1}},y^{b_{2}},~^{\shortparallel
}p_{b_{3}},~^{\shortparallel }p_{b_{4}})]\}\ ^{\shortmid }\mathring{g}%
^{a_{3}}(x^{i_{1}},y^{b_{2}},~^{\shortparallel }p_{b_{3}},~^{\shortparallel
}p_{b_{4}}),  \notag
\end{eqnarray}%
For N-connection s-coefficients, we use the $\kappa $--decompositions giving:
\begin{eqnarray*}
\ ^{\shortparallel }N_{i_{1}}^{a_{2}}(\kappa ,x^{k_{1}},y^{3}) &=&\
^{\shortparallel }\eta _{i_{1}}^{a_{2}}\ ~^{\shortparallel }\mathring{N}%
_{i_{1}}^{a_{2}}=~^{\shortparallel }\zeta _{i_{1}}^{a_{2}}(1+\kappa \
^{\shortparallel }\chi _{i_{1}}^{a_{2}})\ ~^{\shortparallel }\mathring{N}%
_{i_{1}}^{a_{2}} \\
&=&\{\ ^{\shortparallel }\zeta _{i_{1}}^{a_{2}}(x^{i_{1}},y^{b_{2}},\
^{\shortparallel }p_{b_{3}},~^{\shortparallel }p_{b_{4}}) [1+\kappa \ ^{\shortparallel }\chi _{i_{1}}^{a_{2}}(x^{i_{1}},y^{b_{2}},\
^{\shortparallel }p_{b_{3}},~^{\shortparallel }p_{b_{4}})]\} \\
&&\ ^{\shortparallel }\mathring{N}_{i_{1}}^{a_{2}}(x^{i_{1}},y^{b_{2}},\
^{\shortparallel }p_{b_{3}},\ ^{\shortparallel }p_{b_{4}}), \\
\ ^{\shortparallel }N_{i_{2}a_{3}}(\kappa ,x^{k_{1}},y^{b_{2}},~^{\shortparallel }p_{6}) &=&\ ^{\shortparallel }\eta
_{i_{2}a_{3}}\ ~^{\shortparallel }\mathring{N}_{i_{2}a_{3}}=~^{%
\shortparallel }\zeta _{i_{2}a_{3}}(1+\kappa ~^{\shortparallel }\chi
_{i_{2}a_{3}})\ ^{\shortparallel }\mathring{N}_{i_{2}a_{3}} \\
&=& \{\ ^{\shortparallel }\zeta
_{i_{2}a_{3}}(x^{i_{1}},y^{b_{2}},~^{\shortparallel }p_{b_{3}},\
^{\shortparallel }p_{b_{4}}) [1+\kappa ~^{\shortparallel }\chi
_{i_{2}a_{3}}(x^{i_{1}},y^{b_{2}},\ ^{\shortparallel }p_{b_{3}},\
^{\shortparallel }p_{b_{4}})]\} \\
&&\ ^{\shortmid }\mathring{N}%
_{i_{2}a_{3}}(x^{i_{1}},y^{b_{2}},~^{\shortparallel
}p_{b_{3}},~^{\shortparallel }p_{b_{4}}), 
\end{eqnarray*}
\begin{eqnarray*}
\ ^{\shortparallel }N_{i_{3}a_{4}}(\kappa ,x^{k_{1}},y^{b_{2}},\
^{\shortparallel }p_{a_{3}},\ ^{\shortparallel }p_{7}) &=&...,\mbox{ or } \\
\ ^{\shortparallel }N_{i_{3}a_{4}}(\kappa ,x^{k_{1}},y^{b_{2}},\
^{\shortparallel }p_{a_{3}},\ ^{\shortparallel }E) &=&\ ^{\shortparallel
}\eta _{i_{3}a_{4}}\ ^{\shortparallel }\mathring{N}_{i_{3}a_{4}}=\
^{\shortparallel }\zeta _{i_{3}a_{4}}(1+\kappa \ ^{\shortparallel }\chi
_{i_{3}a_{4}})\ ^{\shortparallel }\mathring{N}_{i_{3}a_{4}} \\
&=&\{\ ^{\shortparallel }\zeta _{i_{3}a_{4}}(x^{i_{1}},y^{b_{2}},\
^{\shortparallel }p_{b_{3}},\ ^{\shortparallel }p_{b_{4}}) [1+\kappa \chi
_{i_{3}a_{4}}(x^{i_{1}},y^{b_{2}},\ ^{\shortparallel }p_{b_{3}},\
^{\shortparallel }p_{b_{4}})]\} \\
&&\ ^{\shortparallel }\mathring{N}%
_{i_{3}a_{4}}(x^{i_{1}},y^{b_{2}},\ ^{\shortparallel }p_{b_{3}},\
^{\shortparallel }p_{b_{4}}).
\end{eqnarray*}

Briefly, we can write formulas for the coefficients of target s-metrics and
related N-connection structure on $_{s}\mathbf{T}_{\shortparallel }^{\ast }%
\mathbf{V}$ in symbolic form:
\begin{equation}
\ _{s}^{\shortparallel }\mathbf{\mathring{g}}\rightarrow \ _{\varepsilon
}^{\shortparallel }\mathbf{g}=[\ ^{\shortparallel }g_{\alpha _{s}}=\
^{\shortparallel }\zeta _{\alpha _{s}}(1+\kappa \ ^{\shortparallel }\chi
_{\alpha _{s}})\ \ ^{\shortparallel }\mathring{g}_{\alpha _{s}},\ \
^{\shortparallel }N_{i_{s}}^{a_{s}}=\ ^{\shortparallel }\zeta
_{i_{s-1}}^{a_{s}}(1+\kappa \ \ ^{\shortparallel }\chi _{i_{s-1}}^{a_{s}})\
\ ^{\shortparallel }\mathring{N}_{i_{s-1}}^{a_{s}}].  \label{epstargsm}
\end{equation}%
Explicit formulas for parametric generating functions defining
quasi-stationary solutions of $E$-dependent configurations, were given in Appendix
B.3 of \cite{partner02}. Here we consider a different class of
configurations for the case $E_{0}=const$ on the shell $s=4$, with explicit
dependence on $\ ^{\shortparallel }p_{7}$ of generating functions, $\
^{\shortparallel }\zeta ^{8},\ ^{\shortparallel }\chi ^{8};$ generating
source and cosmological constant $,\ ~_{4}^{\shortparallel }\mathcal{K},\ \
_{4}^{\shortparallel }\Lambda _{0};$ integration functions $,\
_{1}^{\shortparallel }n_{k_{4}},\ _{2}^{\shortparallel }n_{k_{4}};$
prescribed data for a prime s-metric, $(\ ^{\shortparallel }\mathring{g}%
^{7},\ ^{\shortparallel }\mathring{g}^{8}; \ ^{\shortparallel }\mathring{N}%
_{k_{3}7},\ ^{\shortparallel }\mathring{N}_{i_{3}8}):$%
\begin{eqnarray*}
\ ^{\shortparallel }\zeta ^{7} &=&-\frac{4}{\ \ ^{\shortparallel }\mathring{g%
}^{7}}\frac{[\ ^{\shortparallel }\partial ^{7}(|\ ^{\shortparallel }\zeta
^{8}\ ^{\shortparallel }\mathring{g}^{8}|^{1/2})]^{2}}{|\int d\
^{\shortparallel }p_{7}\{(~_{4}^{\shortparallel }\mathcal{K})\
^{\shortparallel }\partial ^{7}(\ ^{\shortparallel }\zeta ^{8}\
^{\shortparallel }\mathring{g}^{8})\}|}\mbox{ and } \\
\ ^{\shortparallel }\chi ^{7} &=&\frac{\ ^{\shortparallel }\partial
^{7}(~^{\shortparallel }\chi ^{8}|\ ^{\shortparallel }\zeta ^{8}\
^{\shortparallel }\mathring{g}^{8}|^{1/2})}{4\ ^{\shortparallel }\partial
^{7}(|\ ^{\shortparallel }\zeta ^{8}\ ^{\shortparallel }\mathring{g}%
^{8}|^{1/2})}-\frac{\int d\ ^{\shortparallel }p_{7}\{\ ^{\shortparallel
}\partial ^{7}[(~_{4}^{\shortparallel }\mathcal{K})\ (~^{\shortparallel
}\zeta ^{8}\ ^{\shortparallel }\mathring{g}^{8})\ ^{\shortparallel }\chi
^{8}]\}}{\int d\ ^{\shortparallel }p_{7}\{(~_{4}^{\shortparallel }\mathcal{K}%
)\ ^{\shortparallel }\partial ^{7}(\ ^{\shortparallel }\zeta ^{8}\
^{\shortparallel }\mathring{g}^{8})\}},
\end{eqnarray*}%
\begin{eqnarray*}
\ ^{\shortparallel }\zeta _{i_{3}7} &=&\frac{\ ^{\shortparallel }\partial
_{i_{3}}\int d\ ^{\shortparallel }p_{7}(~_{4}^{\shortparallel }\mathcal{K})\
\ ^{\shortparallel }\partial ^{7}(~^{\shortparallel }\zeta ^{8})}{(\
^{\shortparallel }\mathring{N}_{i_{3}7})(~_{4}^{\shortparallel }\mathcal{K}%
)\ ^{\shortparallel }\partial ^{7}(\ ^{\shortparallel }\zeta ^{8})}%
\mbox{and
}\ ^{\shortparallel }\chi _{i_{3}7}=\frac{\ ^{\shortparallel }\partial
_{i_{3}}[\int d\ ^{\shortparallel }p_{7}(~_{4}^{\shortparallel }\mathcal{K}%
)\ ^{\shortparallel }\partial ^{7}(\ ^{\shortparallel }\zeta ^{8}\
^{\shortparallel }\mathring{g}^{8})]}{\ ^{\shortparallel }\partial _{i_{3}}\
[\int d\ ^{\shortparallel }p_{7}(~_{4}^{\shortparallel }\mathcal{K})\
^{\shortparallel }\partial ^{7}(\ ^{\shortparallel }\zeta ^{8})]}-\frac{\
^{\shortparallel }\partial ^{7}(\ ^{\shortparallel }\zeta ^{8}\
^{\shortparallel }\mathring{g}^{8})}{\ ^{\shortparallel }\partial ^{7}(\
^{\shortparallel }\zeta ^{8})}, \\
\ ^{\shortparallel }\zeta _{i_{3}8} &=&\ (\ ^{\shortparallel }\mathring{N}%
_{i_{3}8})^{-1}[\ _{1}^{\shortparallel }n_{i_{3}}+16\ _{2}^{\shortparallel
}n_{i_{3}}[\int d\ \ ^{\shortparallel }p_{7}\{\frac{\left( \
^{\shortparallel }\partial ^{7}[(~^{\shortparallel }\zeta ^{8}\
^{\shortparallel }\mathring{g}^{8})^{-1/4}]\right) ^{2}}{|\int d\
^{\shortparallel }\ ^{\shortparallel }p_{7}\ (~_{4}^{\shortparallel }%
\mathcal{K})\ ^{\shortparallel }\partial ^{7}(\ ^{\shortparallel }\zeta
^{8}\ ^{\shortparallel }\mathring{g}^{8})\ ^{\shortparallel }\partial ^{7}|}]%
\mbox{ and } \\
\ ^{\shortparallel }\chi _{i_{3}8} &=&-\frac{16\ _{2}^{\shortparallel
}n_{i_{3}}\int d\ ^{\shortparallel }p_{7}\frac{\left( \ ^{\shortparallel
}\partial ^{7}[(\ ^{\shortparallel }\zeta ^{8}\ \ ^{\shortparallel }%
\mathring{g}^{8})^{-1/4}]\right) ^{2}}{|\int d\ ^{\shortparallel }p_{7}\ (\
_{4}^{\shortparallel }\mathcal{K})\ ^{\shortparallel }\partial ^{7}(\
^{\shortparallel }\zeta ^{8}\ ^{\shortparallel }\mathring{g}^{8})|}(\frac{\
^{\shortparallel }\partial ^{7}[(\ ^{\shortparallel }\zeta ^{8}\ \
^{\shortparallel }\mathring{g}^{8})^{-1/4}\ ^{\shortparallel }\chi ^{8})]}{%
2\ ^{\shortparallel }\partial ^{7}[(\ ^{\shortparallel }\zeta ^{8}\
^{\shortparallel }\mathring{g}^{8})^{-1/4}]}+\frac{\int d\ ^{\shortparallel
}p_{7}(\ _{4}^{\shortparallel }\mathcal{K})\ ^{\shortparallel }\partial
^{7}[(\ ^{\shortparallel }\zeta ^{8}\ \ ^{\shortparallel }\mathring{g}^{8})\
^{\shortparallel }\chi ^{8}]}{\int d\ ^{\shortparallel }p_{7}\ (\
_{4}^{\shortparallel }\mathcal{K})\ ^{\shortparallel }\partial ^{7}(\
^{\shortparallel }\zeta ^{8}\ ^{\shortparallel }\mathring{g}^{8})})}{\
_{1}^{\shortparallel }n_{i_{3}}+16\ _{2}^{\shortparallel }n_{i_{3}}[\int d\
^{\shortparallel }p_{7}\frac{\left( \ ^{\shortparallel }\partial ^{7}[(\
^{\shortparallel }\zeta ^{8}\ ^{\shortparallel }\mathring{g}%
^{8})^{-1/4}]\right) ^{2}}{|\int d\ ^{\shortparallel }p_{7}(\
_{4}^{\shortparallel }\mathcal{K})\ ^{\shortparallel }\partial ^{7}(\
^{\shortparallel }\zeta ^{8}\ ^{\shortparallel }\mathring{g}^{8})|}]}.
\end{eqnarray*}%
These formulas transform into corresponding ones with $E$-dependencies if we
change the indices $7\leftrightarrow 8$, $\ ^{\shortparallel
}p_{7}\rightarrow \ ^{\shortparallel }p_{8}=\ ^{\shortparallel }E,\
^{\shortparallel }\partial ^{7}(...)\rightarrow \ ^{\shortparallel }\partial
^{8}(...)=(...)^{\ast },$ see details in section 5 of \cite{partner02} on how
to derive analogous formulas for shells $s=1,2,3.$

The $\zeta $- and $\chi $-coefficients for deformations (\ref{epstargsm}) are of the form 
\begin{eqnarray}
\ \ \ ^{\shortparallel }\eta _{2}(\kappa ,x^{k_{1}}) &=&\ ^{\shortparallel
}\zeta _{2}(1+\kappa \ ^{\shortparallel }\chi _{2}(x^{k_{1}}));\ \ \
^{\shortparallel }\eta _{4}(\kappa ,x^{k_{1}},y^{3})=\ ^{\shortparallel
}\zeta _{4}(x^{k_{1}},y^{3})(1+\kappa \ ^{\shortparallel }\chi
_{4}(x^{k_{1}},y^{3}));  \notag \\
\ ^{\shortparallel }\eta ^{5}(\kappa ,x^{k_{1}},y^{a_{2}},~^{\shortparallel
}p_{6}) &=&\ ^{\shortparallel }\zeta
^{5}(x^{k_{1}},y^{a_{2}},~^{\shortparallel }p_{6})(1+\kappa \ \
^{\shortparallel }\chi ^{5}(x^{k_{1}},y^{a_{2}},~^{\shortparallel }p_{6}));
\label{epsilongenfdecomp} \\
\ \ ^{\shortparallel }\eta ^{8}(\kappa
,x^{k_{1}},y^{a_{2}},~^{\shortparallel }p_{a_{3}},~^{\shortparallel }p_{7})
&=&\ ^{\shortparallel }\zeta ^{8}(x^{k_{1}},y^{a_{2}},~^{\shortparallel
}p_{a_{3}},~^{\shortparallel }p_{7})(1+\kappa \ \ ^{\shortparallel }\chi
^{8}(x^{k_{1}},y^{a_{2}},~^{\shortparallel }p_{a_{3}},~^{\shortparallel
}p_{7})),\ \mbox{ for fixed }\ ^{\shortparallel }E_{0},\mbox{ or }  \notag \\
\ ^{\shortparallel }\eta ^{7}(\kappa ,x^{k_{1}},y^{a_{2}},~^{\shortparallel
}p_{a_{3}},~^{\shortparallel }E) &=&\ ^{\shortparallel }\zeta
^{7}(x^{k_{1}},y^{a_{2}}, ^{\shortparallel }p_{a_{3}},~^{\shortparallel
}E)(1+\kappa \ ^{\shortparallel }\chi
^{7}(x^{k_{1}},y^{a_{2}},~^{\shortparallel }p_{a_{3}},\ ^{\shortparallel
}E)).  \notag
\end{eqnarray}
We can generate similar solutions when
$\ ^{\shortparallel }\eta ^{6}(\kappa ,x^{i_{2}},~^{\shortparallel }p_{5})=\
^{\shortparallel }\zeta ^{6}(\kappa ,x^{i_{2}},~^{\shortparallel
}p_{5})(1+\kappa \ \ ^{\shortparallel }\chi ^{6}(\kappa
,x^{i_{2}},~^{\shortparallel }p_{5}))$ is chosen as a generating function
instead of $
\ ^{\shortparallel }\eta ^{5}(\kappa ,x^{i_{2}},~^{\shortparallel }p_{6})=\
^{\shortparallel }\zeta ^{5}(\kappa ,x^{i_{2}},~^{\shortparallel
}p_{6})(1+\kappa \ \ ^{\shortparallel }\chi ^{5}(\kappa ,x^{i_{2}}, \
^{\shortparallel }p_{6})).$ The resulting new classes of parametric solutions
posses a Killing symmetry on $p_{6}$ which is different
from off--diagonal configurations (\ref{epsilongenfdecomp}) with Killing
symmetry on $p_{5}$.

\subsubsection{Off-diagonal ansatz in radial coordinates with
fixed/or running energy dependence}

Introducing $\kappa $-coefficients (\ref{epstargsm}) instead of $\eta $%
-coefficients \ in (\ref{dmpolariz}), we generate nonlinear quadratic
elements for quasi-stationary s-metrics,
\begin{eqnarray}
d\ ^{\shortparallel }\widehat{s}^{2} &=&\ ^{\shortparallel }\widehat{g}%
_{\alpha _{s}\beta _{s}}(x^{k},y^{3},\ ^{\shortparallel }p_{a_{3}},\
^{\shortparallel }p_{a_{4}};g_{4},\ ^{\shortparallel }g^{a_{3}},\
^{\shortparallel }g^{a_{4}},\ ~_{s}^{\shortparallel }\mathcal{K},\ \
_{s}^{\shortparallel }\Lambda _{0},\kappa )d\ ^{\shortparallel }u^{\alpha
_{s}}d\ ^{\shortparallel }u^{\beta _{s}}=  \notag \\
&=&d\ _{1}\widehat{s}^{2}(x^{k},~_{1}^{\shortparallel }\mathcal{K},\ \
_{1}^{\shortparallel }\Lambda _{0},\kappa )+d\ _{2}\widehat{s}%
^{2}(x^{k},y^{3},~~_{2}^{\shortparallel }\mathcal{K},\ \
_{2}^{\shortparallel }\Lambda _{0},\kappa )+  \label{ansatz1} \\
&&d\ _{3}^{\shortparallel }\widehat{s}^{2}(x^{k},y^{3},\ ^{\shortparallel
}p_{a_{3}},~_{3}^{\shortparallel }\mathcal{K},\ _{3}^{\shortparallel
}\Lambda _{0},\kappa )+d\ _{4}^{\shortparallel }\widehat{s}%
^{2}(x^{k},y^{3},\ ^{\shortparallel }p_{a_{3}},\ ^{\shortparallel
}p_{a_{4}},\ ~_{4}^{\shortparallel }\mathcal{K},\ \ _{4}^{\shortparallel
}\Lambda _{0},\kappa ).  \notag
\end{eqnarray}%
Here we study the total phase space quasi-stationary s-metric
structures (\ref{ansatz1}) describing extensions of the BH stationary
solutions in nonassociative vacuum gravity with Killing symmetry on $%
y^{4}= t$ on the shell $s=2.$\footnote{%
Such 8-d s-metrics have a Killing symmetry on $\ ^{\shortparallel }p_{5},$ or on 
$ \ ^{\shortparallel}p_{6}, $ on $s=3$.}

The shell components of the quadratic nonlinear elements (\ref{ansatz1}) define solutions of (\ref{cannonsymparamc2})
if the coefficients of the s-metric are defined and parameterized as:

On shell $s=1$,
\begin{eqnarray}
d\ _{1}\widehat{s}^{2} &=&g_{i_{1}}(x^{k_{1}})[(dx^{i_{1}})^{2}]=e^{\psi
_{0}(x^{k_{1}})}(1+\kappa \ ^{\psi }\ ^{\shortparallel }\chi
)[(dx^{1})^{2}+(dx^{2})^{2}],\mbox{ where }  \notag \\
&&\psi (x^{k_{1}})=\psi _{0}(x^{k_{1}})+\kappa \ ^{\psi }\ ^{\shortparallel
}\chi (x^{k_{1}})\mbox{ is a solution of }\psi ^{\bullet \bullet }+\psi
^{\prime \prime }=2\ \ _{1}^{\shortparallel }\mathcal{K}.  \label{aux01}
\end{eqnarray}%
On shell $s=2,$%
\begin{eqnarray*}
&&d\ _{2}\widehat{s}^{2} =g_{a_{2}}(x^{i_{1}},y^{3})(\mathbf{e}^{a_{2}})^{2}=
\\
&&-\{\frac{4[(|\ ^{\shortparallel }\zeta _{4}\ \ ^{\shortparallel }\mathring{%
g}_{4}|^{1/2})^{\diamond }]^{2}}{\ ^{\shortparallel }\mathring{g}_{3}|\int
dy^{3}\{(~_{2}^{\shortparallel }\mathcal{K})(\ ^{\shortparallel }\zeta _{4}\
^{\shortparallel }\mathring{g}_{4})^{\diamond }\}|}-\kappa \lbrack \frac{(\
^{\shortparallel }\chi _{4}|\ ^{\shortparallel }\zeta _{4}\ ^{\shortparallel
}\mathring{g}_{4}|^{1/2})^{\diamond }}{4(|\ ^{\shortparallel }\zeta _{4}\
^{\shortparallel }\mathring{g}_{4}|^{1/2})^{\diamond }}-\frac{\int
dy^{3}\{(~_{2}^{\shortparallel }\mathcal{K})[(~^{\shortparallel }\zeta _{4}\
^{\shortparallel }\mathring{g}_{4})\ ^{\shortparallel }\chi _{4}]^{\diamond
}\}}{\int dy^{3}\{(~_{2}^{\shortparallel }\mathcal{K})(\ ^{\shortparallel
}\zeta _{4}\ ^{\shortparallel }\mathring{g}_{4})^{\diamond }\}}]\}\
^{\shortparallel }\mathring{g}_{3} \\
&&+\{dy^{3}+[\frac{\partial _{i_{1}}\ \int dy^{3}(~_{2}^{\shortparallel }%
\mathcal{K})\ (\ ^{\shortparallel }\zeta _{4})^{\diamond }}{(\
^{\shortparallel }\mathring{N}_{i_{1}}^{3})(~_{2}^{\shortparallel }\mathcal{K%
})(\ ^{\shortparallel }\zeta _{4})^{\diamond }}+\kappa (\frac{\partial
_{i_{1}}[\int dy^{3}(~_{2}^{\shortparallel }\mathcal{K})(\ ^{\shortparallel
}\zeta _{4}\ ^{\shortparallel }\chi _{4})^{\diamond }]}{\partial _{i_{1}}\
[\int dy^{3}(~_{2}^{\shortparallel }\mathcal{K})(\ ^{\shortparallel }\zeta
_{4})^{\diamond }]}-\frac{(\ ^{\shortparallel }\zeta _{4}\ ^{\shortparallel
}\chi _{4})^{\diamond }}{(\ ^{\shortparallel }\zeta _{4})^{\diamond }})](\
^{\shortparallel }\mathring{N}_{i_{1}}^{3})dx^{i_{1}}\}^{2} \\
&&+\ ^{\shortparallel }\zeta _{4}(1+\kappa \ ^{\shortparallel }\chi _{4})\
^{\shortparallel }\mathring{g}_{4}\{dt+[(\ ^{\shortparallel }\mathring{N}%
_{k_{1}}^{4})^{-1}[\ _{1}^{\shortparallel }n_{k_{1}}+16\
_{2}^{\shortparallel }n_{k_{1}}[\int dy^{3}\{\frac{\left( [(\
^{\shortparallel }\zeta _{4}\ ^{\shortparallel }\mathring{g}%
_{4})^{-1/4}]^{\diamond }\right) ^{2}}{|\int dy^{3}[(~_{2}^{\shortparallel }%
\mathcal{K})(\ ^{\shortparallel }\zeta _{4}\ \ ^{\shortparallel }\mathring{g}%
_{4})]^{\diamond }|}] \\
&&-\kappa \frac{16\ _{2}^{\shortparallel }n_{k_{1}}\int dy^{3}\frac{\left(
[(\ ^{\shortparallel }\ \zeta _{4}\ \ ^{\shortparallel }\mathring{g}%
_{4})^{-1/4}]^{\diamond }\right) ^{2}}{|\int dy^{3}[(~_{2}^{\shortparallel }%
\mathcal{K})(\ ^{\shortparallel }\zeta _{4}\ ^{\shortparallel }\mathring{g}%
_{4})]^{\diamond }|}(\frac{[(\ ^{\shortparallel }\ \zeta _{4}\ \
^{\shortparallel }\mathring{g}_{4})^{-1/4}\chi _{4})]^{\diamond }}{2[(\
^{\shortparallel }\zeta _{4}\ ^{\shortparallel }\mathring{g}%
_{4})^{-1/4}]^{\diamond }}+\frac{\int dy^{3}[(~_{2}^{\shortparallel }%
\mathcal{K})(\ ^{\shortparallel }\zeta _{4}\ ^{\shortparallel }\chi _{4}\ \
^{\shortparallel }\mathring{g}_{4})]^{\diamond }}{\int
dy^{3}[(~_{2}^{\shortparallel }\mathcal{K})(\ ^{\shortparallel }\zeta _{4}\
\ ^{\shortparallel }\mathring{g}_{4})]^{\diamond }})}{\ _{1}^{\shortparallel
}n_{k_{1}}+16\ _{2}^{\shortparallel }n_{k_{1}}[\int dy^{3}\frac{\left( [(\
^{\shortparallel }\zeta _{4}\ \ ^{\shortparallel }\mathring{g}%
_{4})^{-1/4}]^{\diamond }\right) ^{2}}{|\int dy^{3}[(~_{2}^{\shortparallel }%
\mathcal{K})(\ ^{\shortparallel }\zeta _{4}\ ^{\shortparallel }\mathring{g}%
_{4})]^{\diamond }|}]}](\ ^{\shortparallel }\mathring{N}_{k_{1}}^{4})dx^{%
\acute{k}_{1}}\}
\end{eqnarray*}%
is a solution of the system :
\begin{eqnarray*}
&&(\ _{2}\varpi )^{\diamond }g_{4}^{\diamond }=2g_{3}g_{4}\ \
_{2}^{\shortparallel }\mathcal{K(}x^{i_{1}},y^{3}\mathcal{)}, \\
&&\ _{2}\beta \ w_{j_{1}}-\alpha _{j_{1}}=0,\ n_{k_{1}}^{\diamond \diamond
}+\ _{2}\gamma \ n_{k_{1}}^{\diamond }=0,\mbox{ where } \\
&&\left\{
\begin{array}{c}
\alpha _{i_{1}}=g_{4}^{\diamond }\partial _{i_{1}}(\ _{2}\varpi ),\
_{2}\beta =g_{4}^{\diamond }(\ _{2}\varpi )^{\diamond },\ _{2}\gamma =(\ln
\frac{|g_{4}|^{3/2}}{|g_{3}|})^{\diamond } \\
\mbox{ for generating function }\ _{2}\Psi =\exp (\ _{2}\varpi ),\
_{2}\varpi =\ln \left\vert g_{4}^{\diamond }/\sqrt{|g_{3}g_{4}}|\right\vert%
\end{array}%
\right. .
\end{eqnarray*}%
On co-fiber space, for shell $s=3,$
\begin{eqnarray*}
&&d\ _{3}\widehat{s}^{2}=\ ^{\shortparallel
}g^{a_{3}}(x^{i_{1}},y^{a_{2}},~^{\shortparallel }p_{6})(\ ^{\shortparallel }%
\mathbf{e}_{a_{3}})^{2}= \\
&&\ ^{\shortparallel }\zeta ^{5}(1+\kappa\ ^{\shortparallel}\chi ^{5})\
^{\shortparallel }\mathring{g}^{5}\{d\ ^{\shortparallel }p_{5}+ [(\
^{\shortparallel }\mathring{N}_{i_{2}5})^{-1}[\ _{1}^{\shortparallel
}n_{i_{2}}+16\ _{2}^{\shortparallel }n_{i_{2}}[\int d~^{\shortparallel }p_{6}%
\frac{\left( ~^{\shortparallel }\partial ^{6}[(~^{\shortparallel }\zeta
^{5}\ ~^{\shortparallel }\mathring{g}^{5})^{-1/4}]\right) ^{2}}{|\int
d~^{\shortparallel }p_{6}\ ^{\shortparallel }\partial
^{6}[(~_{3}^{\shortparallel }\mathcal{K})(~^{\shortparallel }\zeta
^{5}~^{\shortparallel }\mathring{g}^{5})]|}]+ \kappa \times \\
&&\frac{16\ _{2}^{\shortparallel }n_{i_{2}}\int d~^{\shortparallel }p_{6}%
\frac{\left( ~^{\shortparallel }\partial ^{6}[(~^{\shortparallel }\zeta
^{5}\ ~^{\shortparallel }\mathring{g}^{5})^{-1/4}]\right) ^{2}}{|\int d\
^{\shortparallel }p_{6}(~_{3}^{\shortparallel }\mathcal{K})\ \
^{\shortparallel }\partial ^{6}[(\ ^{\shortparallel }\zeta ^{5}\
^{\shortparallel} \mathring{g}^{5})]|}(\frac{\ ^{\shortparallel }\partial
^{6}[(\ ^{\shortparallel }\zeta ^{5}\ ^{\shortparallel }\mathring{g}%
^{5})^{-1/4}\ ^{\shortparallel}\chi ^{5})]} {2\ ^{\shortparallel }\partial
^{6}[(\ ^{\shortparallel }\zeta ^{5} \ ^{\shortparallel }\mathring{g}%
^{5})^{-1/4}]}+\frac{\int d\ ^{\shortparallel }p_{6}\ (\
_{3}^{\shortparallel }\mathcal{K})~^{\shortparallel }\partial
^{6}[(~^{\shortparallel }\zeta ^{5}\ ~^{\shortparallel }\mathring{g}%
^{5})~^{\shortparallel }\chi ^{5}]}{\int d~^{\shortparallel
}p_{6}(~_{3}^{\shortparallel }\mathcal{K})\ ~^{\shortparallel }\partial
^{6}[(~^{\shortparallel }\zeta ^{5}\ ~^{\shortparallel }\mathring{g}^{5})]})%
}{\ _{1}^{\shortparallel }n_{i_{2}}+16\ _{2}^{\shortparallel }n_{i_{2}}[\int
d~^{\shortparallel }p_{6}\frac{\left(\ ^{\shortparallel }\partial ^{6}[(\
^{\shortparallel }\zeta ^{5}\ ~^{\shortparallel }\mathring{g}%
^{5})^{-1/4}]\right) ^{2}}{|\int d\ ^{\shortparallel}p_{6} (\
_{3}^{\shortparallel}\mathcal{K})\ ^{\shortparallel}\partial ^{6}[(\
^{\shortparallel}\zeta ^{5}\ ^{\shortparallel}\mathring{g}^{5})]|}]}] \
^{\shortparallel}\mathring{N}_{i_{2}5}dx^{i_{2}}\}
\end{eqnarray*}%
\begin{eqnarray*}
&&-\{\frac{4[~^{\shortparallel }\partial ^{6}(|~^{\shortparallel }\zeta
^{5}~^{\shortparallel }\mathring{g}^{5}|^{1/2})]^{2}}{~^{\shortparallel }%
\mathring{g}^{6}|\int d~^{\shortparallel }p_{6}\{(~_{3}^{\shortparallel }%
\mathcal{K})~^{\shortparallel }\partial ^{6}[(~^{\shortparallel }\zeta
^{5}~^{\shortparallel }\mathring{g}^{5})]\}|}-\kappa \lbrack \frac{\partial
_{i_{2}}[\int d~^{\shortparallel }p_{6}(~_{3}^{\shortparallel }\mathcal{K}%
)~^{\shortparallel }\partial ^{6}(~^{\shortparallel }\zeta
^{5}~^{\shortparallel }\mathring{g}^{5})]}{\partial _{i_{2}}\ [\int
d~^{\shortparallel }p_{6}(~_{3}^{\shortparallel }\mathcal{K}%
)~^{\shortparallel }\partial ^{6}(~^{\shortparallel }\zeta ^{5})]}-\frac{%
^{\shortparallel }\partial ^{6}(~^{\shortparallel }\zeta
^{5}~^{\shortparallel }\mathring{g}^{5})}{~^{\shortparallel }\partial
^{6}(~^{\shortparallel }\zeta ^{5})}]\} \\
&&\ ^{\shortparallel }\mathring{g}^{6}\{d~^{\shortparallel }p_{6}+[\frac{%
\partial _{i_{2}}\ \int d~^{\shortparallel }p_{6}(~_{3}^{\shortparallel }%
\mathcal{K})~^{\shortparallel }\partial ^{6}(~^{\shortparallel }\zeta ^{5})}{%
(\ ~^{\shortparallel }\mathring{N}_{i_{2}6})(~_{3}^{\shortparallel }\mathcal{%
K})~^{\shortparallel }\partial ^{6}(~^{\shortparallel }\zeta ^{5})}+\kappa (%
\frac{\partial _{i_{2}}[\int d~^{\shortparallel }p_{6}(~_{3}^{\shortparallel
}\mathcal{K})~^{\shortparallel }\partial ^{6}(~^{\shortparallel }\zeta
^{5}~^{\shortparallel }\mathring{g}^{5})]}{\partial _{i_{2}}\ [\int
d~^{\shortparallel }p_{6}(~_{3}^{\shortparallel }\mathcal{K}%
)~^{\shortparallel }\partial ^{6}(~^{\shortparallel }\zeta ^{5})]}-\frac{%
~^{\shortparallel }\partial ^{6}(~^{\shortparallel }\zeta
^{5}~^{\shortparallel }\mathring{g}^{5})}{~^{\shortparallel }\partial
^{6}(~^{\shortparallel }\zeta ^{5})})](\ ^{\shortparallel }\mathring{N}%
_{i_{2}6})dx^{i_{2}}\}
\end{eqnarray*}%
is a solution of this system:
\begin{eqnarray*}
&&\ ^{\shortparallel }\partial ^{6}(\ _{3}^{\shortparallel }\varpi )\
^{\shortparallel }\partial ^{6}\ ^{\shortparallel }g^{5}=2\ ^{\shortparallel
}g^{5}\ ^{\shortparallel }g^{6}\ \ _{3}^{\shortparallel }\mathcal{K}(\
^{\shortparallel }x^{i_{2}},~^{\shortparallel }p_{6}), \\
&&\ ^{\shortparallel }\partial ^{66}(\ \ ^{\shortparallel }n_{k_{2}})+\
_{3}^{\shortparallel }\gamma \ ^{\shortparallel }\partial ^{6}(\
^{\shortparallel }n_{k_{2}})=0,\ _{3}^{\shortparallel }\beta \ \
^{\shortparallel }w_{j_{2}}-\ ^{\shortparallel }\alpha _{j_{2}}=0,%
\mbox{
where } \\
&&\left\{
\begin{array}{c}
\ ^{\shortparallel }\alpha _{i_{2}}=(\ ^{\shortparallel }\partial ^{6}\
^{\shortparallel }g^{5})\partial _{i_{2}}(\ _{3}^{\shortparallel }\varpi ),\
_{3}^{\shortparallel }\beta =(\ ^{\shortparallel }\partial ^{6}\
^{\shortparallel }g^{5})\ ^{\shortparallel }\partial ^{6}(\
_{3}^{\shortparallel }\varpi ),\ _{3}^{\shortparallel }\gamma =\
^{\shortparallel }\partial ^{6}(\ln \frac{|\ ^{\shortparallel }g^{5}|^{3/2}}{%
|\ ^{\shortparallel }g^{6}|}) \\
\mbox{ for generating function  }\ _{3}^{\shortparallel }\Psi =\exp (\
_{3}^{\shortparallel }\varpi ),\ _{3}^{\shortparallel }\varpi =\ln
\left\vert \ ^{\shortparallel }\partial ^{6}\ ^{\shortparallel }g^{5}/\sqrt{%
|\ ^{\shortparallel }g^{5}\ ^{\shortparallel }g^{6}}|\right\vert
,(i_{2},j_{2},k_{2}=1,2,3,4)%
\end{array}%
\right. .
\end{eqnarray*}%
The solutions for shell $s=4$ can either have a fixed energy-like coordinate or be dependent of the energy-like coordinate.  We have
\begin{eqnarray}
&&d\ _{4}\widehat{s}^{2}[~^{\shortparallel }E]=\ ^{\shortparallel
}g^{a_{4}}(x^{i_{1}},y^{a_{2}},~^{\shortparallel
}p_{a_{3}},~^{\shortparallel }E)(\ ^{\shortparallel }\mathbf{e}_{a_{4}})^{2}=
\label{aux61a} \\
&&\ ^{\shortparallel }\zeta ^{7}(1+\kappa ~^{\shortparallel }\chi
^{7})~^{\shortparallel }\mathring{g}^{7}\{d~^{\shortparallel
}p_{7}+[(~^{\shortparallel }\mathring{N}_{i_{3}7})^{-1}[\
_{1}^{\shortparallel }n_{i_{3}}+16\ _{2}^{\shortparallel }n_{i_{3}}[\int
d~^{\shortparallel }E\{\frac{\left( [(~^{\shortparallel }\zeta
^{7}~^{\shortparallel }\mathring{g}^{7})^{-1/4}]^{\ast }\right) ^{2}}{|\int
d~^{\shortparallel }E(~_{4}^{\shortparallel }\mathcal{K})[(~^{\shortparallel
}\zeta ^{7}~^{\shortparallel }\mathring{g}^{7})]^{\ast }|}]-  \notag \\
&&\kappa \frac{16\ _{2}^{\shortparallel }n_{i_{3}}\int d~^{\shortparallel }E%
\frac{\left( [(~^{\shortparallel }\zeta ^{7}\ ~^{\shortparallel }\mathring{g}%
^{7})^{-1/4}]^{\ast }\right) ^{2}}{|\int d~^{\shortparallel }E\
(~_{4}^{\shortparallel }\mathcal{K})[(~^{\shortparallel }\zeta
^{7}~^{\shortparallel }\mathring{g}^{7})]^{\ast }|}(\frac{%
[(~^{\shortparallel }\zeta ^{7}~^{\shortparallel }\mathring{g}%
^{7})^{-1/4}~^{\shortparallel }\chi ^{7})]^{\ast }}{2\ [(~^{\shortparallel
}\zeta ^{7}~^{\shortparallel }\mathring{g}^{7})^{-1/4}]^{\ast }}+\frac{\int
d\ ^{\shortparallel }E\ (~_{4}^{\shortparallel }\mathcal{K}%
)[(~^{\shortparallel }\zeta ^{7}\ ~^{\shortparallel }\mathring{g}%
^{7})~^{\shortparallel }\chi ^{7}]^{\ast }}{\int d~^{\shortparallel }E\
(~_{4}^{\shortparallel }\mathcal{K})[(~^{\shortparallel }\zeta ^{7}\
~^{\shortparallel }\mathring{g}^{7})]^{\ast }})}{\ _{1}^{\shortparallel
}n_{i_{3}}+16\ _{2}^{\shortparallel }n_{i_{3}}[\int d~^{\shortparallel }E%
\frac{\left( \ [(~^{\shortparallel }\zeta ^{7}~^{\shortparallel }\mathring{g}%
^{7})^{-1/4}]^{\ast }\right) ^{2}}{|\int d~^{\shortparallel }E\
(~_{4}^{\shortparallel }\mathcal{K})[(~^{\shortparallel }\zeta ^{7}\
~^{\shortparallel }\mathring{g}^{7})]|^{\ast }}]}]\ (\ ^{\shortparallel }%
\mathring{N}_{i_{3}7})dx^{i_{3}}\}  \notag
\end{eqnarray}%
\begin{eqnarray*}
&&-\{4\frac{4[(|~^{\shortparallel }\zeta ^{7}~^{\shortparallel }\mathring{g}%
^{7}|^{1/2})^{\ast }]^{2}}{~^{\shortparallel }\mathring{g}^{8}|\int
d~^{\shortparallel }E\{(~_{4}^{\shortparallel }\mathcal{K}%
)[(~^{\shortparallel }\zeta ^{7}~^{\shortparallel }\mathring{g}^{7})]^{\ast
}\}|}-\kappa \lbrack \frac{(~^{\shortparallel }\chi ^{7}|~^{\shortparallel
}\zeta ^{7}~^{\shortparallel }\mathring{g}^{7}|^{1/2})^{\ast }}{%
4(|~^{\shortparallel }\zeta ^{7}\ ^{\shortparallel }\mathring{g}%
^{7}|^{1/2})^{\ast }}-\frac{\int d~^{\shortparallel
}E\{(~_{4}^{\shortparallel }\mathcal{K})[(~^{\shortparallel }\zeta
^{7}~^{\shortparallel }\mathring{g}^{7})~^{\shortparallel }\chi ^{7}]^{\ast
}\}}{\int d~^{\shortparallel }E\{(~_{4}^{\shortparallel }\mathcal{K}%
)[(~^{\shortparallel }\zeta ^{7}~^{\shortparallel }\mathring{g}^{7})]^{\ast
}\}}]\}~^{\shortparallel }\mathring{g}^{8} \\
&&\{d\ ^{\shortparallel }E+[\frac{~^{\shortparallel }\partial _{i_{3}}\ \int
d\ ^{\shortparallel }E(~_{4}^{\shortparallel }\mathcal{K})\ (\
^{\shortparallel }\zeta ^{7})^{\ast }}{(~^{\shortparallel }\mathring{N}%
_{i_{3}8})(~_{4}^{\shortparallel }\mathcal{K})(~^{\shortparallel }\zeta
^{7})^{\ast }}+\kappa \lbrack \frac{~^{\shortparallel }\partial
_{i_{3}}[\int d~^{\shortparallel }E(~_{4}^{\shortparallel }\mathcal{K}%
)(~^{\shortparallel }\zeta ^{7}~^{\shortparallel }\mathring{g}^{7})^{\ast }]%
}{~^{\shortparallel }\partial _{i_{3}}\ [\int d~^{\shortparallel
}E(~_{4}^{\shortparallel }\mathcal{K})(~^{\shortparallel }\zeta ^{7})^{\ast
}]}-\frac{(~^{\shortparallel }\zeta ^{7}~^{\shortparallel }\mathring{g}%
^{7})^{\ast }}{(~^{\shortparallel }\zeta ^{7})^{\ast }}](~^{\shortparallel }%
\mathring{N}_{i_{3}8})d~^{\shortparallel }x^{i_{3}}\}
\end{eqnarray*}%
is a solution of the system:
\begin{eqnarray}
&&\ (\ _{4}^{\shortparallel }\varpi )^{\ast }(\ ^{\shortparallel
}g^{7})^{\ast }=2\ ^{\shortparallel }g^{7}\ ^{\shortparallel }g^{8}\
_{4}^{\shortparallel }\widehat{\Upsilon }(\ ^{\shortparallel
}x^{i_{3}},~^{\shortparallel }E),  \label{aux62} \\
&&(\ ^{\shortparallel }n_{k_{3}})^{\ast \ast }+\ _{4}^{\shortparallel
}\gamma (\ ^{\shortparallel }n_{k_{3}})^{\ast }=0,\ _{4}^{\shortparallel
}\beta \ \ ^{\shortparallel }w_{j_{3}}-\ ^{\shortparallel }\alpha _{j_{3}}=0,%
\mbox{ where }  \notag \\
&&\left\{
\begin{array}{c}
\ ^{\shortparallel }\alpha _{i_{3}}=(\ ^{\shortparallel }g^{7})^{\ast }\
^{\shortparallel }\partial _{i_{3}}(\ _{3}^{\shortparallel }\varpi ),\
_{4}^{\shortparallel }\beta =(\ ^{\shortparallel }g^{7})^{\ast }(\
_{4}^{\shortparallel }\varpi )^{\ast },\ _{4}^{\shortparallel }\gamma =(\ln
\frac{|\ ^{\shortparallel }g^{7}|^{3/2}}{|\ ^{\shortparallel }g^{8}|})^{\ast
} \\
\mbox{ for }\ _{4}^{\shortparallel }\Psi =\exp (\ _{4}^{\shortparallel
}\varpi ),\ _{4}^{\shortparallel }\varpi =\ln \left\vert (\ ^{\shortparallel
}g^{7})^{\ast }/\sqrt{|\ ^{\shortparallel }g^{7}\ ^{\shortparallel }g^{8}}%
|\right\vert ,(i_{3},j_{3},k_{3}=1,2,...,6)%
\end{array}%
\right. .  \notag
\end{eqnarray}%
For $\ ^{\shortparallel }E=const$ and variable $~^{\shortparallel }p_{7},$
we obtain formulas which are similar to (\ref{aux61a}) and (\ref{aux62}) but
with dependencies on $~^{\shortparallel }p_{7}$ instead of $\
^{\shortparallel }E$.

Any nonlinear quadratic element (\ref{ansatz1}) having on $s=4$ a term $d\
_{4}\widehat{s}^{2}[~^{\shortparallel }E_{0}]$ determines a class of
quasi-stationary phase space solutions of nonassociative vacuum
gravitational equations (\ref{ansatzsourchv}) with Killing symmetry on the
energy type cofiber coordinate. If we consider (\ref{ansatz1}) with $d\ _{4}%
\widehat{s}^{2}[~^{\shortparallel }E]$ (\ref{aux61a}), we obtain s-metrics
derived in general form by formulas (89) in \cite{partner02}.

In the above formulas, the $\eta $-polarization functions may depend on all
phase space coordinates. The values of the indices and
coordinates are: $%
i_{1},j_{1},k_{1},...=1,2;i_{2},j_{2},k_{2},...=1,2,3,4;i_{3},j_{3},k_{3},...=1,2,...6;y^{3}=\varphi ,y^{4}=t,\ ^{\shortparallel }p_{8}=\ ^{\shortparallel }E;
$ and the respective
\begin{eqnarray}
&&\mbox{generating functions: }\psi (\hbar ,\kappa ;x^{k_{1}});\ _{2}\Psi
(\hbar ,\kappa ;x^{k_{1}},y^{3});\ \ _{3}^{\shortparallel }\Psi (\hbar
,\kappa ;x^{k_{2}},\ ^{\shortparallel }p_{6});  \notag \\
&&\ \ _{4}^{\shortparallel }\Psi (\hbar ,\kappa ;\ ^{\shortparallel
}x^{k_{3}},\ ^{\shortparallel }E)\mbox{ or }\ \ _{4}^{\shortparallel }\Psi
(\hbar ,\kappa ;\ ^{\shortparallel }x^{k_{3}},\ \ ^{\shortparallel }p_{7});
\label{genfunctsourc} \\
&&\mbox{generating sources:}\ ~_{1}^{\shortparallel }\mathcal{K}(\hbar
,\kappa ;x^{k_{1}});\ ~_{2}^{\shortparallel }\mathcal{K}(\hbar ,\kappa
;x^{k_{1}},y^{3});\ ~_{3}^{\shortparallel }\mathcal{K}(\hbar ,\kappa
;x^{k_{2}},\ ^{\shortparallel }p_{6});  \notag \\
&&\ ~_{4}^{\shortparallel }\mathcal{K}(\hbar ,\kappa ;\ ^{\shortparallel
}x^{k_{3}},\ ^{\shortparallel }E)\mbox{ or }~_{4}^{\shortparallel }\mathcal{K%
}(\hbar ,\kappa ;\ ^{\shortparallel }x^{k_{3}},\ \ ^{\shortparallel }p_{7});
\notag \\
&&\mbox{integration  functions: }g_{4}^{[0]}(\hbar ,\kappa ;x^{k_{1}}),\
_{1}n_{k_{1}}(\hbar ,\kappa ;x^{j_{1}}),\ _{2}n_{k_{1}}(\hbar ,\kappa
;x^{j_{1}});\ ^{\shortparallel }g_{5}^{[0]}(\hbar ,\kappa ;x^{k_{2}}),\
\notag \\
&&\ _{1}n_{k_{2}}(\hbar ,\kappa ;x^{j_{2}}),\ _{2}n_{k_{2}}(\hbar ,\kappa
;x^{j_{2}});\ ^{\shortparallel }g_{7}^{[0]}(\hbar ,\kappa ;x^{j_{3}}),%
\mbox{
or }\ ^{\shortparallel }g_{8}^{[0]}(\hbar ,\kappa ;x^{j_{3}}),\
_{1}^{\shortparallel }n_{k_{3}}(\hbar ,\kappa ;x^{j_{3}}),\
_{2}^{\shortparallel }n_{k_{3}}(\hbar ,\kappa ;x^{j_{3}}).  \notag
\end{eqnarray}%
We note that any generic, off-diagonal, target solution $\
_{s}^{\shortparallel }\widehat{\mathbf{g}}=\ ^{\shortparallel }\widehat{%
\mathbf{g}}_{\alpha _{s}\beta _{s}}(~_{s}x,\ _{s}^{\shortparallel }p)\
^{\shortparallel }\widehat{\mathbf{e}}^{\alpha _{s}}\mathbf{\otimes \
^{\shortparallel }}\widehat{\mathbf{e}}^{\beta _{s}}$ of type (\ref{ansatz1}%
) determined by the above quasi-stationary conditions depends on $\hbar ,\kappa $ for any R-flux nonassociative data from $~_{s}^{\shortparallel }\mathcal{K}$.

A nonassociative quasi-stationary phase space solution (\ref{ansatz1}) is
characterized by generating sources involving $\eta $-polarization functions (%
\ref{dmpolariz}) and $\kappa $-coefficients (\ref{epstargsm}) with redefined
parametric generating functions (\ref{epsilongenfdecomp}). \ Using such
nonlinear transforms, we can re-define the generating functions and
generating sources into equivalent data with effective cosmological
constants, when some s-metric coefficients or 
polarization functions can be used as generating functions,
\begin{eqnarray}
(\ _{s}\Psi ,\ ~_{s}^{\shortparallel }\mathcal{K}) &\leftrightarrow &(\ _{s}%
\widehat{\mathbf{g}},\ ~_{s}^{\shortparallel }\mathcal{K})\leftrightarrow
(~_{s}^{\shortparallel }\eta \ \ \ \ _{s}^{\shortparallel }\mathring{g}%
_{\alpha _{s}}\sim \ _{s}^{\shortparallel }\zeta (1+\kappa \
_{s}^{\shortparallel }\chi _{\alpha _{s}})\ \ _{s}^{\shortparallel }%
\mathring{g}_{\alpha _{s}},\ ~_{s}^{\shortparallel }\mathcal{K}%
)\leftrightarrow  \label{nonlinsym} \\
(\ _{s}\Phi ,\ _{s}\Lambda _{0}) &\leftrightarrow &(\ _{s}\widehat{\mathbf{g}%
},\ \ _{s}\Lambda _{0})\leftrightarrow (~_{s}^{\shortparallel }\eta \ \ \ \
_{s}^{\shortparallel }\mathring{g}_{\alpha _{s}}\sim \ _{s}^{\shortparallel
}\zeta (1+\kappa \ _{s}^{\shortparallel }\chi _{\alpha _{s}})\ \
_{s}^{\shortparallel }\mathring{g}_{\alpha _{s}},\ \ _{s}\Lambda _{0}).
\notag
\end{eqnarray}%
We provde respective formulas and examples in appendix \ref{apbssns}.

\section{Nonassociative star R-flux distortions of 8-d and (4+4)-d BHs}

\label{sec3} Constructing BH configurations is important in theories of gravity. BH solutions are
exact solutions of vacuum gravitational equations,
which have a horizon that acts as a boundary of a spatial region whose causal
future is trapped. In GR, general BH solutions are determined by three parameters: charge,
mass, and angular momentum. Such solutions are known as the Kerr-Newmann family of
stationary vacuum metrics. There are also generalizations of BH solutions
with non-trivial cosmological constants and/or non-trivial electro-magnetic/
gauge fields,\cite{misner,hawking73,wald82,kramer03}. 
Many of these stationary, vacuum solutions are characterized by a curvature
singularity inside the horizon. Also many of these simple solutions have purely diagonal metrics with spherical/ cylindrical
symmetry. In this way the vacuum Einstein equations are transformed into a
system of nonlinear ordinary differential equations, ODEs. 

Here we investigate nonassociative phase space generalizations
of BH solutions encoding $\star$-products and R-flux contributions.
The corresponding geometric constructions is carried out in 8-d $\star$-product
deformed (co) tangent Lorentz bundles. This can be motivated by
string and M-theory and various nonassociative and noncommutative geometric
generalizations. This approach was carried out in N-adapted and s-adapted forms \cite{partner01,partner02}
allowing for the generation of exact solutions of nonlinear systems
of partial differential equations, PDEs.
For nonassociative higher dimension models, there are several substantial
differences involving certain physical features which are
known for nonassociative, nonholonomic 4-d configurations \cite{partner03}.
There are also similar features for 8-d, 10-d, 11-d theories of associative (super)
string gravity, supergravity and (super) geometric flows \cite%
{bubuianu19,rajpoot17}:

\begin{itemize}
\item In the class of nonassociative vacuum gravity theories of type \cite%
{blumenhagen16,aschieri17,partner01,partner02,partner03}, we have nontrivial,
real, effective sources $~_{s}^{\shortparallel }\mathcal{\check{K}}$ which
via nonlinear symmetries (\ref{nonltransf}) are related to respective shell
cosmological constants $\ _{s}\Lambda _{0}.$ 

\item Nonassociative s-metrics $\ _{\star }^{\shortparallel }\mathbf{q}%
_{\alpha _{s}\beta _{s}}$ are characterized by symmetric $\ _{\star
}^{\shortparallel }\mathbf{\check{q}}_{\alpha _{s}\beta _{s}}$ (\ref{aux40b}%
) and nonsymmetric components $\ _{\star }^{\shortparallel }\mathbf{a}%
_{\alpha _{s}\beta _{s}}$ (\ref{aux40aa}), which are generic off-diagonal
with respect to coordinate frames.

\item Adapting the
geometric constructions to nonholonomic distributions, we can define an
infinite number of (non) linear connections which, in general, are not
metric compatible and with nonzero torsion. For certain subclasses of such
metric compatible connections, we can consider physically important
invariant conditions and generalized conservation laws, elaborate on
nonholonomic Clifford structures with nonassociative modified
Einstein-Dirac-Yang-Mills-Higgs systems. 

\item The nonassociative gravitational field equations (\ref%
{cannonsymparamc2}) consist a nonlinear system of PDEs for the coefficients
of nontrivial N-connection structure and corresponding off-diagonal metrics
depending (in general) on all spacetime and phase space coordinates. Such
PDEs can not be transformed in general form into ODEs using certain
"simplified" diagonal ansatz for metrics with high symmetry and dependence,
for instance, on a radial/ cylindric coordinates. 

\item The geometric and physical properties of certain classes
of nonsymmetric nonholonomic solutions is investigated, We begin with
some prime metrics for BH spacetimes and phase space commutative
configurations and elaborate on some small parametric deformations to
nonassociative models. By prime metric we mean a well known solution to ordinary GR.
\end{itemize}

Here, we focus our research on quasi-stationary s-metrics with
star R-flux deformations of certain BH metrics embedded into $\mathbb{M%
}_{4}\times \ ^{\shortparallel }\mathbb{M}_{4};$ with distortions of 6-d
Tangherlini solutions to 8-d, or to double 4-d BH metrics, when the
target s-metrics define solutions of nonassociative vacuum
gravitational equations (\ref{cannonsymparamc2}).

\subsection{Prime metrics for higher dimensional and phase space BHs}

Here we analyze two classes of prime, phase space BH metrics taken as diagonal
associative solutions of nonassociative vacuum gravitational equations. The
first class of solutions we consider is of the Tangherlini type \cite{tangherlini63,pappas16} but having a
different signature and with phase space dimensions. The
second class of solutions is for configurations with double BHs -- with the base being spacetime coordinates and with the cofiber
having momentum coordinates. Such prime metrics were used for generating associative and
commutative stationary BH solutions in relativistic
Finsler-Lagrange-Hamilton gravity \cite{bubuianu19}. In the next subsections, we
shall nonholonomically deform both classes of BH solutions to
off-diagonal configurations generating exact and parametric solutions of
nonassociative vacuum gravitational equations. Via effective sources, such
symmetric and nonsymmetric metrics encode R-flux contributions and spacetime
- cofiber correlations.

\subsubsection{Tangerlini 6-d BHs embedded and distorted in curved 8-d phase
space}

\label{tangph}For a 5-d phase space with local coordinates $(x^{\grave{\imath}%
},p_{5},p_{6})$ [when $\grave{\imath}=1,2,3;x^{4}=t$] and metrics of
signature $(+...+),$ we introduce a corresponding radial coordinate $ ^{\shortmid }r$, 
\begin{eqnarray*}
\ ^{\shortmid }r &=&\sqrt{%
(x^{1})^{2}+(x^{2})^{2}+(y^{3})^{2}+(p_{5})^{2}+(p_{6})^{2}}
\end{eqnarray*}
or we can introduce a different radial coordinate with complex momentum variables
\begin{eqnarray*}
\ ^{\shortparallel }r &=&\sqrt{(x^{1})^{2}+(x^{2})^{2}+(y^{3})^{2}-\hbar
^{2}[(\ ^{\shortparallel }p_{5})^{2}+(\ ^{\shortparallel }p_{6})^{2}]}.
\end{eqnarray*}%
In the above we are working in units where the dimensions of position and momentum are equivalent -- see also conventions on coordinates and indices (\ref%
{coordshell}). The 6-d spherical coordinates are parameterized as $x^1 = \ ^{\shortmid}r,x^{2}=\theta ,y^{3}=\varphi ,x^{4}=t;p_{5},p_{6}$ and the shell $s=4$ is parameterized by 
 momentum coordinates $p_{7},p_{8}=E$. For the cofiber space we use a trivial
embedding into a 8-d phase space. For the Tangerlini solution we find it necessary to define the radial, metric function
\begin{equation}
\ h(\ ^{\shortmid }r)=1-\frac{\ ^{\shortmid }\mu }{(\ ^{\shortmid }r)^{3}}-%
\frac{\ ^{\shortmid }\kappa _{6}^{2}\ ^{\shortmid }\Lambda }{10}(\
^{\shortmid }r)^{2}.  \label{radialf}
\end{equation}%
Here $\ ^{\shortmid }\kappa _{6}$ is a constant determined by the effective
gravitational constant in $6$-d and the constant $\ ^{\shortmid
}\mu =(\ ^{\shortmid }\kappa _{6}^{2}\Gamma \lbrack 5/2]/4\pi ^{5/2})\
^{\shortmid }M$ is for an effective mass $\ ^{\shortmid }M$ and with $\Gamma \lbrack 5/2],$ being the gamma function -- more details can be found in \cite%
{tangherlini63,pappas16,partner02}. Such an $\ h(\ ^{\shortmid }r)$ is used
for constructing 6-d Schwarzschild - de Sitter solutions with an effective
cosmological constant $\ ^{\shortmid }\Lambda .$

Static 6-d phase space s-metrics of type $\ _{s}^{\shortparallel }\mathbf{%
\mathring{g}}$ (\ref{primedm}) can be generated by quadratic linear elements
\begin{eqnarray}
d\ \ _{z}^{\shortparallel }\mathring{s}^{2} &=&\ _{z}^{\shortmid }\mathring{g%
}_{\alpha _{s}\beta _{s}}(x^{i_{s}},p_{a_{s}})d\ ~^{\shortmid }u^{\alpha
_{s}}d\ ~^{\shortmid }u^{\beta _{s}}=\ h^{-1}(\ ^{\shortmid }r)(d\
^{\shortmid }r)^{2}-\ h(\ ^{\shortmid }r)dt^{2}+(\ ^{\shortmid
}r)^{2}d\Omega _{4}^{2}+(dp_{7})^{2}-dE^{2}  \label{pmtang} \\
&=&\ _{z}^{\shortparallel }\mathring{g}_{\alpha _{s}\beta
_{s}}(x^{i_{s}},p_{a_{s}})d\ ~^{\shortparallel }u^{\alpha _{s}}d\
~^{\shortparallel }u^{\beta _{s}}=\ h^{-1}(\ ^{\shortparallel }r)(d\
^{\shortparallel }r)^{2}-\ h(\ ^{\shortparallel }r)dt^{2}+(\
^{\shortparallel }r)^{2}d\ ^{\shortparallel }\Omega _{4}^{2}-\hbar ^{2}[(d\
^{\shortparallel }p_{7})^{2}-d\ ^{\shortparallel }E^{2}],  \notag
\end{eqnarray}%
where the area of the 5-dimensional unit sphere is parameterized by
coordinates on shells $s=1,2,3,$%
\begin{equation*}
d\Omega _{4}^{2}=d\theta _{3}^{2}+\sin ^{2}\theta _{3}[d\theta _{2}^{2}+\sin
^{2}\theta _{2}(d\theta _{1}^{2}+\sin ^{2}\theta _{1}d\varphi ^{2})],
\end{equation*}%
for $\theta _{1}=\theta$ is the standard polar angle. Such a diagonal metric is a trivial extension on
coordinates $u^{7}$ and $u^{8},$. Equation (\ref{pmtang}) defines an exact vacuum
solution of nonassociative gravitational equations (\ref{cannonsymparamc2}),
with $\ _{1}\Lambda _{0}=\ _{2}\Lambda _{0}=\ _{3}\Lambda _{0}=\
^{\shortmid}\Lambda $ and $\ _{4}\Lambda _{0}=0$ on shells $s=1,2,3.$ In
standard form, such a solution is associative and commutative. However, on phase
space the solution may involve star R-flux information if it is constructed as a
diagonal limit of a subclass of solutions determined by generating functions
related to effective sources and shell cosmological constants via nonlinear
symmetries (\ref{nonlinsym}). In general, for 8-d extensions and nontrivial
4-shell sources, such phase space modifications of the 6-d Tangherlini metrics
\cite{tangherlini63} for higher dimension generalization of the
Schwarzschild BH, are not exact solutions of the nonassociative vacuum
equations. Nevertheless, it is possible to construct off-diagonal
deformations (\ref{offdiagdef}) of any prime metric (\ref{pmtang}) to
certain target quasi-stationary, nonassociative, vacuum phase spaces as we
shall show in next subsections. 

\subsubsection{Spacetime and co-fiber phase space double BHs configurations}

\label{prim2bh} Turning now to phase space BHs, we show that it is possible to construct double 4-d BH
configurationson shells $s=1,2$ on the base spacetime
and shells $s=3,4$ co-fiber spaces. We will use the spherical coordinate parametrization (%
\ref{pmtang}) with local coordinates $x^{1}=r,x^{2}=\theta ,y^{3}=\varphi
,y^{4}=t$ for the spacetime coordinates and $p_{5}=\ ^{p}r,p_{6}=\ ^{p}\theta ,p_{7}=\ ^{p}\varphi ,p_{8}=E$ for the momentum-energy phase space coordinates. 
More details about the conventions on coordinates and indices can be found in Appendix (\ref{coordshell}). There are two conventional radial coordinates: one in coordinate space and one in momentum space define as%
\begin{eqnarray}
r &=&\sqrt{(x^{1^{\prime }})^{2}+(x^{2^{\prime }})^{2}+(y^{3^{\prime }})^{2}}%
\mbox{ and }  \label{doublesph} \\
\ ^{p}r &=&\sqrt{(p_{5^{\prime }})^{2}+(p_{6^{\prime }})^{2}+(p_{7^{\prime
}})^{2}},\mbox{ or }\ _{\shortparallel }^{p}r=\sqrt{(~^{\shortparallel
}p_{5^{\prime }})^{2}+(~^{\shortparallel }p_{6^{\prime
}})^{2}+(~^{\shortparallel }p_{7^{\prime }})^{2}}=\ ^{p}r/i\hbar .  \notag
\end{eqnarray}%
In these formulas, prime indices indicate Cartesian coordinates {\it e.g.} $x^{1'} = r \sin \theta \cos \varphi$ {\it etc.}. The
label ``p" on the left side indicates spherical type coordinates on
co-fibers. The prime phase space metric is taken as in section 2.4.1 of \cite%
{bubuianu19},
\begin{eqnarray}
d\ \ \mathring{s}^{2} &=&\ ^{\shortmid }\mathring{g}_{\alpha _{s}\beta
_{s}}(r,\ ^{p}r)d\ ~^{\shortmid }u^{\alpha _{s}}d\ ~^{\shortmid }u^{\beta
_{s}}=\ ^{\shortparallel }\mathring{g}_{\alpha _{s}\beta _{s}}(r,\ ^{p}r)d\
~^{\shortparallel }u^{\alpha _{s}}d\ ~^{\shortparallel }u^{\beta _{s}}
\label{doublepm} \\
&=&f^{-1}(r)(d\ r)^{2}+r^{2}d\Omega ^{2}-f(r)dt^{2}+\ ^{p}f^{-1}(\ ^{p}r)(d\
\ ^{p}r)^{2}+\ ^{p}r^{2}d\ \ ^{p}\Omega ^{2}-\ ^{p}f(\ ^{p}r)dE^{2}  \notag
\\
&=&f^{-1}(r)(d\ r)^{2}+r^{2}d\Omega ^{2}-f(r)dt^{2}+\ \ _{\shortparallel
}^{p}f^{-1}(\ \ _{\shortparallel }^{p}r)(d\ \ \ _{\shortparallel
}^{p}r)^{2}+\ \ _{\shortparallel }^{p}r^{2}d\ \ \ _{\shortparallel
}^{p}\Omega ^{2}-\ \ _{\shortparallel }^{p}f(\ \ _{\shortparallel }^{p}r)d\
_{\shortparallel }^{p}E^{2},  \notag
\end{eqnarray}%
with radial functions
\begin{eqnarray}
f(r) &=&1-\frac{\mu }{r}-\frac{\kappa _{4}^{2}\Lambda r^{2}}{3}\mbox{ and }
\label{doublef} \\
\ ^{p}f(\ ^{p}r) &=&1-\frac{\ ^{p}\mu }{\ ^{p}r}-\frac{(\ ^{p}\kappa )^{2}(\
^{p}\Lambda )(\ ^{p}r)^{2}}{3},\mbox{ or }\ _{\shortparallel }^{p}f(\
_{\shortparallel }^{p}r)=1-\frac{\ \ _{\shortparallel }^{p}\mu }{\
_{\shortparallel }^{p}r}-\frac{(\ ^{p}\kappa )^{2}(\ _{\shortparallel
}^{p}\Lambda )(\ \ _{\shortparallel }^{p}r)^{2}}{3},  \notag
\end{eqnarray}%
and the 2-d solid angle given by 
\begin{eqnarray*}
d\Omega ^{2} &=&d\theta ^{2}+\sin ^{2}\theta d\varphi ^{2}\mbox{ and } \\
d\ ^{p}\Omega ^{2} &=&d\ ^{p}\theta ^{2}+\sin ^{2}(\ ^{p}\theta )d(\
^{p}\varphi )^{2},\mbox{ or }d\ \ _{\shortparallel }^{p}\Omega ^{2}=d\ \
_{\shortparallel }^{p}\theta ^{2}+\sin ^{2}(\ \ _{\shortparallel }^{p}\theta
)d(\ \ _{\shortparallel }^{p}\varphi )^{2}.
\end{eqnarray*}%
In the above formulas $8\pi G_{4}=\kappa _{4}^{2}\Lambda ,$ where $G_{4}$ is the
Newtonian gravitational constant, $\mu =\kappa _{4}^{2}M\Gamma
(3/2)/2\pi ^{3/2}$ with $M$ being the BH mass and $
\Gamma (3/2),$ is the Gamma function. Further details can be found in \cite{pappas16}. In the co-fiber phase space one can write down similar relationship among constants {\it e.g.} $8\pi \ ^{p}G=\
^{p}\kappa ^{2}\ ^{p}\Lambda ,$. The superscript $p$ indicates these are relationships in the co-fiber space.  We will chose nonholonomic distributions to be of the form so that formulas labeled by "$^{\shortmid }$" and "$\
^{\shortparallel }$" are equivalent.

A prime s-metric may define an exact vacuum solution of nonassociative
gravitational equations (\ref{cannonsymparamc2}), if the s-adapted
nonholonomic distributions are subjected to nonlinear symmetries (\ref{nonlinsym}) when $\
_{1}\Lambda _{0}=\ _{2}\Lambda _{0}=\Lambda $ and $\ _{3}\Lambda _{0}=\
_{4}\Lambda _{0}=\ ^{p}\Lambda .$  In this work we omit explicit
computations of s-adapted components of nonsymmetric metrics and complex
valued Ricci tensors for complex phase space double BH solutions. Such phase
space and quasi-stationary solutions have different nonholonomic
structures compared to similar solutions in Finsler-Lagrange-Hamilton gravity
\cite{bubuianu19} where the effective sources/cosmological constants are not
related to R-flux contributions. On the base Lorentz manifold, the radial
function $f(r)$ from (\ref{doublepm}) defines a Schwarzschild - de Sitter
configuration as in GR. A similar configuration is found for the co-fiber
space, but is determined by $\ ^{p}f(\ ^{p}r).$

\subsection{Nonassociative quasi-stationary deformations of BH solutions
with fixed energy parameter}

\label{ssqstfe} Here we will construct explicit 
off-diagonal solutions of the type (\ref%
{ansatz1}) for a fixed $E_{0}$ describing quasi stationary nonholonomic
deformations of prime BH metrics (\ref{pmtang}) or (\ref{doublepm}). For
simplicity, we consider only linear $\kappa $-dependent nonassociative
contributions with the other generating functions and generating sources
subject to nonlinear symmetries as in appendix \ref{apbssns}, see
formulas (\ref{nonltransf}). Such solutions have nonhlonomic induced
torsion but we can always extract LC-configurations. Prescribing generating sources $\ _{s}^{\shortparallel }\mathcal{K}$ (%
\ref{cannonsymparamc2a}) we impose certain nonholonomic constraints on
distributions of nonassociative sources which allow us to find generic
off-diagonal solutions in explicit form. 

\subsubsection{Off-diagonal parametric R-flux deformations of Tangerlini BHs}

In this subsection we consider the prime diagonal metric (\ref{pmtang}) but under a  non-singular coordinate transformation $(\ ^{\shortmid }r,\theta ,\varphi
,\theta _{2},\theta _{3},p_{7},E)$ $\rightarrow ~^{\shortmid }u^{\beta
_{s}}, $$\ ~^{\shortparallel }%
\mathring{g}_{\alpha _{s}}(~^{\shortmid }u^{\beta _{s}})=\
~_{h}^{\shortparallel }\mathring{g}_{\alpha _{s}}[~^{\shortmid }u^{\beta
_{s}}(\ ^{\shortmid }r,\theta ,\varphi ,\theta _{2},\theta _{3},p_{7})]$, \ \
and with non-zero N-coefficients$~^{\shortparallel }\mathring{N}%
_{i_{s-1}}^{a_{s}}(~^{\shortmid }u^{\beta _{s}})$. In this way  the non-zero s-adapted coefficients of metric \eqref{pmtang} become
\begin{eqnarray}
\ ~_{h}^{\shortparallel }\mathring{g}_{1} &=&\ h^{-1}(\ ^{\shortmid }r),\
~_{h}^{\shortparallel }\mathring{g}_{2}=(\ ^{\shortmid }r)^{2}\sin
^{2}\theta _{2}\cdot \sin ^{2}\theta _{3},\ ~_{h}^{\shortparallel }\mathring{%
g}_{3}=(\ ^{\shortmid }r)^{2}\sin ^{2}\theta \cdot \sin ^{2}\theta _{2}\cdot
\sin ^{2}\theta _{3},\ ~_{h}^{\shortparallel }\mathring{g}_{4}=-\ h(\
^{\shortmid }r),  \notag \\
~_{h}^{\shortparallel }\mathring{g}_{5} &=&(\ ^{\shortmid }r)^{2}\sin
^{2}\theta _{3},~_{h}^{\shortparallel }\mathring{g}_{6}=(\ ^{\shortmid
}r)^{2},~_{h}^{\shortparallel }\mathring{g}_{7}=1,~_{h}^{\shortparallel }%
\mathring{g}_{7}=-1;  \label{primetangd} \\
&&\ _{h}^{\shortparallel }\mathring{N}_{i_{s-1}}^{a_{s}}\neq 0,\mbox{ but }%
~_{h}^{\shortparallel }\mathring{N}_{i_{s-1}}^{a_{s}}(~^{\shortmid }u^{\beta
_{s}})\rightarrow ~^{\shortparallel }\mathring{N}_{i_{s-1}}^{a_{s}}(\
^{\shortmid }r,\theta ,\varphi ,\theta _{2},\theta _{3},p_{7},E)=0.  \notag
\end{eqnarray}%
In these formulas, we use a left label $h$ in order to emphasize that such
diagonalizable s-metrics are determined by a radial function $\ h(\
^{\shortmid }r)$ (\ref{radialf}). We have used similar labels for the N-connection
coefficients $\ _{h}^{\shortparallel }\mathring{N}_{i_{s-1}}^{a_{s}}$. 
For this class of prime metrics, denoted with over-circles, the form of the metric components and N-coefficients are given by
coordinate transforms and encode physical parameters of the prime BHs.

We now apply the parametric deformations of type (\ref{ansatz1}), with a fixed energy
parameter $E_{0}=const$, to the metric and N-coefficients of \eqref{primetangd}. The result is:
\begin{eqnarray}
d\ ^{\shortparallel }\widehat{s}^{2} &=&\ ^{\shortparallel }\widehat{g}%
_{\alpha _{s}\beta _{s}}(\ ^{\shortmid }r,\theta ,\varphi ,\theta
_{2},\theta _{3},p_{7},E_{0};\ h(\ ^{\shortmid }r),\ ~_{s}^{\shortparallel }%
\mathcal{K},\ \ _{s}^{\shortparallel }\Lambda _{0},\kappa )d\
^{\shortparallel }u^{\alpha _{s}}d\ ^{\shortparallel }u^{\beta _{s}}=
\label{offtangfix} \\
&=&d\ _{1}\widehat{s}^{2}(\ ^{\shortmid }r,\varphi ;\ h(\ ^{\shortmid
}r),~_{1}^{\shortparallel }\mathcal{K},\ \ _{1}^{\shortparallel }\Lambda
_{0},\kappa )+d\ _{2}\widehat{s}^{2}(\ ^{\shortmid }r,\varphi ,\theta ;\ h(\
^{\shortmid }r),~~_{2}^{\shortparallel }\mathcal{K},\ \ _{2}^{\shortparallel
}\Lambda _{0},\kappa )+  \notag \\
&&d\ _{3}^{\shortparallel }\widehat{s}^{2}(\ ^{\shortmid }r,\theta ,\varphi
,\theta _{2},\theta _{3};\ h(\ ^{\shortmid }r),~_{3}^{\shortparallel }%
\mathcal{K},\ _{3}^{\shortparallel }\Lambda _{0},\kappa )+d\
_{4}^{\shortparallel }\widehat{s}^{2}(\ ^{\shortmid }r,\theta ,\varphi
,\theta _{2},\theta _{3},p_{7};\ h(\ ^{\shortmid }r),\ ~_{4}^{\shortparallel
}\mathcal{K},E_{0},\ \ _{4}^{\shortparallel }\Lambda _{0},\kappa ).  \notag
\end{eqnarray}%
On shell 1 the components are:
\begin{eqnarray}
d\ _{1}\widehat{s}^{2} &=&\widehat{g}%
_{i_{1}}(x^{k_{1}})[(dx^{i_{1}})^{2}]=e^{\psi _{0}(x^{k_{1}})}[1+\kappa \
^{\psi }\ ^{\shortparallel }\chi (x^{k_{1}})][(dx^{1})^{2}+(dx^{2})^{2}]
\label{offtangfixs1} \\
&=&~^{\shortparallel }\eta _{i_{i}}\ ~^{\shortparallel }\mathring{g}%
_{i_{1}}[(dx^{i_{1}})^{2}]=~^{\shortparallel }\zeta _{i_{1}}(1+\kappa
~^{\shortparallel }\chi _{i_{1}})\ ~^{\shortparallel }\mathring{g}%
_{i_{1}}[(dx^{i_{1}})^{2}]  \notag \\
&=&[\ ~^{\shortparallel }\eta _{1}(r)\ h^{-1}(\ ^{\shortmid }r)(d\ \
^{\shortmid }r)^{2}+~^{\shortparallel }\eta _{2}\ (\ ^{\shortmid }r,\theta
,\varphi ,\theta _{2},\theta _{3})(\ ^{\shortmid }r)^{2}\sin ^{2}\theta
_{3}\ \sin ^{2}\theta _{2}\ \sin ^{2}\theta \ d\varphi ],  \notag
\end{eqnarray}%
where the polarization functions are $\ ^{\shortparallel }\eta _{i_{i}}$ and $%
\ ^{\shortparallel }\zeta _{i_{1}}(1+\kappa \ ^{\shortparallel }\chi
_{i_{1}})$ -- see formulas (\ref{aux02}). The coordinate transformations on
shells $s=1,2,3,4$ are prescribed so that $\psi (x^{k_{1}}(\
^{\shortmid }r,\theta ))=\psi _{0}(\ ^{\shortmid }r,\theta )+\kappa \ ^{\psi
}\ ^{\shortparallel }\chi (\ ^{\shortmid }r,\theta )$ is a solution of the 2-d
Poisson equation (\ref{aux01}) with R-flux source $\ _{1}^{\shortparallel }%
\mathcal{K}.$

On shell $s=2,$ the s-metric (\ref{offtangfix}) is determined by
\begin{eqnarray}
d\ _{2}\widehat{s}^{2} &=&g_{a_{2}}(^{\shortmid }r,\varphi ,\theta )[\mathbf{%
e}^{a_{2}}(^{\shortmid }r,\varphi ,\theta )]^{2}=  \label{offtangfixs2} \\
&&-\{\frac{4[(|\ ^{\shortparallel }\zeta _{4} \ _{h}^{\shortparallel }%
\mathring{g}_{4}|^{1/2})^{\diamond }]^{2}}{\ _{h}^{\shortparallel }\mathring{%
g}_{3}|\int d\varphi \{(\ _{2}^{\shortparallel }\mathcal{K})(\
^{\shortparallel }\zeta _{4}\ ^{\shortparallel }\mathring{g}_{4})^{\diamond
}\}|}-\kappa \lbrack \frac{(\ ^{\shortparallel }\chi _{4}|\ ^{\shortparallel
}\zeta _{4}\ _{h}^{\shortparallel }\mathring{g}_{4}|^{1/2})^{\diamond }}{%
4(|\ ^{\shortparallel }\zeta _{4}\ _{h}^{\shortparallel }\mathring{g}%
_{4}|^{1/2})^{\diamond }}-\frac{\int d\varphi \{(~_{2}^{\shortparallel }%
\mathcal{K})[(~^{\shortparallel }\zeta _{4}\ _{h}^{\shortparallel }\mathring{%
g}_{4})\ ^{\shortparallel }\chi _{4}]^{\diamond }\}}{\int d\varphi
\{(~_{2}^{\shortparallel }\mathcal{K})(\ ^{\shortparallel }\zeta _{4}\
_{h}^{\shortparallel }\mathring{g}_{4})^{\diamond }\}}]\}\
_{h}^{\shortparallel }\mathring{g}_{3}+  \notag \\
&&\{d\varphi +[\frac{\partial _{i_{1}}\ \int d\varphi (~_{2}^{\shortparallel
}\mathcal{K})\ (\ ^{\shortparallel }\zeta _{4})^{\diamond }}{(\
^{\shortparallel }\mathring{N}_{i_{1}}^{3})(~_{2}^{\shortparallel }\mathcal{K%
})(\ ^{\shortparallel }\zeta _{4})^{\diamond }}+\kappa (\frac{\partial
_{i_{1}}[\int d\varphi (~_{2}^{\shortparallel }\mathcal{K})(\
^{\shortparallel }\zeta _{4}\ ^{\shortparallel }\chi _{4})^{\diamond }]}{%
\partial _{i_{1}}\ [\int d\varphi (~_{2}^{\shortparallel }\mathcal{K})(\
^{\shortparallel }\zeta _{4})^{\diamond }]}-\frac{(\ ^{\shortparallel }\zeta
_{4}\ ^{\shortparallel }\chi _{4})^{\diamond }}{(\ ^{\shortparallel }\zeta
_{4})^{\diamond }})](\ ^{\shortparallel }\mathring{N}_{i_{1}}^{3})dx^{i_{1}}%
\}^{2}+  \notag
\end{eqnarray}
\begin{eqnarray}
&&\ ^{\shortparallel }\zeta _{4}(1+\kappa \ ^{\shortparallel }\chi _{4})\
_{h}^{\shortparallel }\mathring{g}_{4}\{dt+[(\ ^{\shortparallel }\mathring{N}%
_{k_{1}}^{4})^{-1}[\ _{1}^{\shortparallel }n_{k_{1}}+16\
_{2}^{\shortparallel }n_{k_{1}}[\int d\varphi \{\frac{\left( [(\
^{\shortparallel }\zeta _{4}\ _{h}^{\shortparallel }\mathring{g}%
_{4})^{-1/4}]^{\diamond }\right) ^{2}}{|\int d\varphi \lbrack
(~_{2}^{\shortparallel }\mathcal{K})(\ ^{\shortparallel }\zeta _{4}\ \
_{h}^{\shortparallel }\mathring{g}_{4})]^{\diamond }|}]-  \notag \\
&&\kappa \frac{16\ _{2}^{\shortparallel }n_{k_{1}}\int d\varphi \frac{\left(
\lbrack (\ ^{\shortparallel }\ \zeta _{4}\ \ _{h}^{\shortparallel }\mathring{%
g}_{4})^{-1/4}]^{\diamond }\right) ^{2}}{|\int d\varphi \lbrack
(~_{2}^{\shortparallel }\mathcal{K})(\ ^{\shortparallel }\zeta _{4}\
_{h}^{\shortparallel }\mathring{g}_{4})]^{\diamond }|}(\frac{[(\
^{\shortparallel }\ \zeta _{4}\ \ _{h}^{\shortparallel }\mathring{g}%
_{4})^{-1/4}\chi _{4})]^{\diamond }}{2[(\ ^{\shortparallel }\zeta _{4}\
_{h}^{\shortparallel }\mathring{g}_{4})^{-1/4}]^{\diamond }}+\frac{\int
d\varphi \lbrack (~_{2}^{\shortparallel }\mathcal{K})(\ ^{\shortparallel
}\zeta _{4}\ ^{\shortparallel }\chi _{4}\ \ _{h}^{\shortparallel }\mathring{g%
}_{4})]^{\diamond }}{\int d\varphi \lbrack (~_{2}^{\shortparallel }\mathcal{K%
})(\ ^{\shortparallel }\zeta _{4}\ \ _{h}^{\shortparallel }\mathring{g}%
_{4})]^{\diamond }})}{\ _{1}^{\shortparallel }n_{k_{1}}+16\
_{2}^{\shortparallel }n_{k_{1}}[\int d\varphi \frac{\left( \lbrack (\
^{\shortparallel }\zeta _{4}\ \ _{h}^{\shortparallel }\mathring{g}%
_{4})^{-1/4}]^{\diamond }\right) ^{2}}{|\int d\varphi \lbrack
(~_{2}^{\shortparallel }\mathcal{K})(\ ^{\shortparallel }\zeta _{4}\
_{h}^{\shortparallel }\mathring{g}_{4})]^{\diamond }|}]}](\ ^{\shortparallel
}\mathring{N}_{k_{1}}^{4})dx^{\acute{k}_{1}}\},  \notag
\end{eqnarray}%
with generating functions $\ ^{\shortparallel }\zeta _{4}(\
^{\shortmid}r,\theta ,\varphi )$ and $\ ^{\shortparallel }\chi _{4}(\
^{\shortmid }r,\theta ,\varphi );$ R-flux effective source $\
_{2}^{\shortparallel }\mathcal{K}(\ ^{\shortmid }r,\theta ,\varphi )$;
integration functions $\ _{1}^{\shortparallel }n_{k_{1}}(\ ^{\shortmid
}r,\theta )$ and $\ _{2}^{\shortparallel }n_{k_{2}}(\ ^{\shortmid }r,\theta
);$ and components of the prime metric $\ _{h}^{\shortparallel }\mathring{g}_{3}(\ ^{\shortmid
}r,\theta )$ and $\ _{h}^{\shortparallel }\mathring{g}_{4}(\ ^{\shortmid
}r,\theta )$ from (\ref{primetangd}). The non-zero prime N-connection coefficients, $\
^{\shortparallel }\mathring{N}_{k_{1}}^{a_{2}}$, are introduced via coordinate transforms $\ ^{\shortparallel }\mathring{N}%
_{k_{1}}^{a_{2}}[\ ^{\shortmid }u^{\beta _{2}}(\ ^{\shortmid }r,\theta
)]\rightarrow \ ^{\shortparallel }\mathring{N}_{k_{1}}^{a_{2}}[~^{\shortmid
}u^{\beta _{2}}(\ ^{\shortmid }r,\theta ,\varphi )]$, and, in general, give rise to generic off-diagonal metrics.

For shell $s=3$ on the co-fiber space, the nonlinear quadratic component of (\ref%
{offtangfix}), with Killing symmetry on $\theta _{2}$ and
$(\ ^{\shortparallel }p_{5},~^{\shortparallel }p_{6})\rightarrow (\theta
_{2},\theta _{3})$, is
\begin{eqnarray}
&&d\ _{3}\widehat{s}^{2}=\ ^{\shortparallel }g^{a_{3}}(\ ^{\shortmid
}u^{\beta _{3}}(\ ^{\shortmid }r,\theta ,\varphi ,\theta _{2},\theta _{3})[\
^{\shortparallel }\mathbf{e}_{a_{3}}(\ ^{\shortmid }u^{\beta _{3}}(\
^{\shortmid }r,\theta ,\varphi ,\theta _{2},\theta _{3})]^{2}=
\label{offtangfixs3} \\
&&\ ^{\shortparallel }\zeta ^{5}(1+\kappa \ ^{\shortparallel }\chi ^{5})\
_{h}^{\shortparallel }\mathring{g}^{5}\{d\theta _{2}+[(\ ^{\shortparallel }%
\mathring{N}_{i_{2}5})^{-1}[\ _{1}^{\shortparallel }n_{i_{2}}+16\
_{2}^{\shortparallel }n_{i_{2}}[\int d\theta _{3}\frac{\left(
~^{\shortparallel }\partial _{\theta _{3}}[(~^{\shortparallel }\zeta ^{5}\
~_{h}^{\shortparallel }\mathring{g}^{5})^{-1/4}]\right) ^{2}}{|\int d\theta
_{3}~^{\shortparallel }\partial _{\theta _{3}}[(~_{3}^{\shortparallel }%
\mathcal{K})(~^{\shortparallel }\zeta ^{5}~_{h}^{\shortparallel }\mathring{g}%
^{5})]|}]+  \notag \\
&&\kappa \frac{16\ _{2}^{\shortparallel }n_{i_{2}}\int d\theta _{3}\frac{%
\left( ~^{\shortparallel }\partial _{\theta _{3}}[(~^{\shortparallel }\zeta
^{5}\ ~_{h}^{\shortparallel }\mathring{g}^{5})^{-1/4}]\right) ^{2}}{|\int
d\theta _{3}(~_{3}^{\shortparallel }\mathcal{K})\ ~^{\shortparallel
}\partial ^{6}[(~^{\shortparallel }\zeta ^{5}\ ~^{\shortparallel }\mathring{g%
}^{5})]|}(\frac{\ ~^{\shortparallel }\partial _{\theta _{3}}[(\
~^{\shortparallel }\zeta ^{5}\ ~_{h}^{\shortparallel }\mathring{g}%
^{5})^{-1/4}~^{\shortparallel }\chi ^{5})]}{2\ ~^{\shortparallel }\partial
_{\theta _{3}}[(~^{\shortparallel }\zeta ^{5}~_{h}^{\shortparallel }%
\mathring{g}^{5})^{-1/4}]}+\frac{\int d\theta _{3}\ (~_{3}^{\shortparallel }%
\mathcal{K})~^{\shortparallel }\partial _{\theta _{3}}[(~^{\shortparallel
}\zeta ^{5}\ ~_{h}^{\shortparallel }\mathring{g}^{5})~^{\shortparallel }\chi
^{5}]}{\int d\theta _{3}(~_{3}^{\shortparallel }\mathcal{K})\
~^{\shortparallel }\partial _{\theta _{3}}[(~^{\shortparallel }\zeta ^{5}\
~_{h}^{\shortparallel }\mathring{g}^{5})]})}{\ _{1}^{\shortparallel
}n_{i_{2}}+16\ _{2}^{\shortparallel }n_{i_{2}}[\int d\theta _{3}\frac{\left(
\ ~^{\shortparallel }\partial _{\theta _{3}}[(\ ~^{\shortparallel }\zeta
^{5}\ ~_{h}^{\shortparallel }\mathring{g}^{5})^{-1/4}]\right) ^{2}}{|\int
d\theta _{3}\ (~_{3}^{\shortparallel }\mathcal{K})\ ~^{\shortparallel
}\partial _{\theta _{3}}[(~^{\shortparallel }\zeta ^{5}\
~_{h}^{\shortparallel }\mathring{g}^{5})]|}]}]  \notag
\end{eqnarray}%
\begin{eqnarray*}
&&(\ ^{\shortparallel }\mathring{N}_{i_{2}5})dx^{i_{2}}\}-\{\frac{%
4[~^{\shortparallel }\partial _{\theta _{3}}(|~^{\shortparallel }\zeta
^{5}~_{h}^{\shortparallel }\mathring{g}^{5}|^{1/2})]^{2}}{%
~_{h}^{\shortparallel }\mathring{g}^{6}|\int d\theta
_{3}\{(~_{3}^{\shortparallel }\mathcal{K})~^{\shortparallel }\partial
_{\theta _{3}}[(~^{\shortparallel }\zeta ^{5}~_{h}^{\shortparallel }%
\mathring{g}^{5})]\}|}-\kappa \lbrack \frac{\partial _{i_{2}}[\int d\theta
_{3}(~_{3}^{\shortparallel }\mathcal{K})~^{\shortparallel }\partial _{\theta
_{3}}(~^{\shortparallel }\zeta ^{5}~_{h}^{\shortparallel }\mathring{g}^{5})]%
}{\partial _{i_{2}}\ [\int d\theta _{3}(~_{3}^{\shortparallel }\mathcal{K}%
)~^{\shortparallel }\partial _{\theta _{3}}(~^{\shortparallel }\zeta ^{5})]}-%
\frac{^{\shortparallel }\partial _{\theta _{3}}(~^{\shortparallel }\zeta
^{5}~_{h}^{\shortparallel }\mathring{g}^{5})}{~^{\shortparallel }\partial
_{\theta _{3}}(~^{\shortparallel }\zeta ^{5})}]\} \\
&&\ _{h}^{\shortparallel }\mathring{g}^{6}\{d\theta _{3}+[\frac{\partial
_{i_{2}}\ \int d\theta _{3}(~_{3}^{\shortparallel }\mathcal{K}%
)~^{\shortparallel }\partial _{\theta _{3}}(~^{\shortparallel }\zeta ^{5})}{%
(\ ~^{\shortparallel }\mathring{N}_{i_{2}6})(~_{3}^{\shortparallel }\mathcal{%
K})~^{\shortparallel }\partial _{\theta _{3}}(~^{\shortparallel }\zeta ^{5})}%
+\kappa (\frac{\partial _{i_{2}}[\int d\theta _{3}(~_{3}^{\shortparallel }%
\mathcal{K})~^{\shortparallel }\partial _{\theta _{3}}(~^{\shortparallel
}\zeta ^{5}~_{h}^{\shortparallel }\mathring{g}^{5})]}{\partial _{i_{2}}\
[\int d\theta _{3}(~_{3}^{\shortparallel }\mathcal{K})~^{\shortparallel
}\partial _{\theta _{3}}(~^{\shortparallel }\zeta ^{5})]}-\frac{%
~^{\shortparallel }\partial _{\theta _{3}}(~^{\shortparallel }\zeta
^{5}~_{h}^{\shortparallel }\mathring{g}^{5})}{~^{\shortparallel }\partial
_{\theta _{3}}(~^{\shortparallel }\zeta ^{5})})](\ ^{\shortparallel }%
\mathring{N}_{i_{2}6})dx^{i_{2}}\},
\end{eqnarray*}%
with generating functions $~^{\shortparallel }\zeta ^{5}(\ ^{\shortmid
}r,\theta ,\varphi ,\theta _{3})$ and $\ ^{\shortparallel }\chi ^{5}(\
^{\shortmid }r,\theta ,\varphi ,\theta _{3});$ R-flux effective source $%
~_{3}^{\shortparallel }\mathcal{K}(\ ^{\shortmid }r,\theta ,\varphi ,\theta
_{3})$; \newline
integration functions $\ _{1}^{\shortparallel }n_{k_{2}}(\ ^{\shortmid
}r,\theta ,\varphi )$ and $\ _{2}^{\shortparallel }n_{k_{2}}(\ ^{\shortmid
}r,\theta ,\varphi );$ and prime metric components $\ _{h}^{\shortparallel }\mathring{g}%
^{5}(\ ^{\shortmid }r,\theta ,\varphi )$ and $\ _{h}^{\shortparallel }%
\mathring{g}^{6}(\ ^{\shortmid }r,\theta ,\varphi )$ from (\ref{primetangd}%
). The non-zero, prime N-connection coefficients, $\ ^{\shortparallel }\mathring{N}%
_{k_{2}}^{a_{3}}$, arise via coordinate transformations
$\ ^{\shortparallel }\mathring{N}_{k_{2}}^{a_{3}}[~^{\shortmid }u^{\beta
_{3}}(\ ^{\shortmid }r,\theta ,\varphi )]\rightarrow \ ^{\shortparallel }%
\mathring{N}_{k_{2}}^{a_{3}}[~^{\shortmid }u^{\beta _{3}}(\ ^{\shortmid
}r,\theta ,\varphi ,\theta _{3})]$ $\ $ and lead to generic off-diagonal metrics.

Finally for shell 4, we consider solutions (\ref{offtangfix}) with $\
^{\shortparallel }E=\ ^{\shortparallel }E_{0}=const,$ {\it i.e.} with Killing
symmetry on $\ ^{\shortparallel }p^{8},$ and variable $\
^{\shortparallel }p_{7},$ on the co-fiber space. This yields the metric%
\begin{eqnarray}
&&d\ _{4}\widehat{s}^{2}[\ ^{\shortparallel }E_{0}]=\ ^{\shortparallel
}g^{a_{4}}(\ ^{\shortmid }r,\theta ,\varphi ,\theta _{2},\theta
_{3},~^{\shortparallel }p_{7},\ ^{\shortparallel }E_{0})[\ ^{\shortparallel }%
\mathbf{e}_{a_{4}}(\ ^{\shortmid }r,\theta ,\varphi ,\theta _{2},\theta
_{3},~^{\shortparallel }p_{7},\ ^{\shortparallel }E_{0})]^{2}=  \notag \\
&&-\{4\frac{4[~^{\shortparallel }\partial ^{7}(|~^{\shortparallel }\zeta
^{8}~_{h}^{\shortparallel }\mathring{g}^{8}|^{1/2})]^{2}}{%
~_{h}^{\shortparallel }\mathring{g}^{7}|\int d~^{\shortparallel
}p_{7}\{(~_{4}^{\shortparallel }\mathcal{K})~^{\shortparallel }\partial
^{7}[(~^{\shortparallel }\zeta ^{8}~_{h}^{\shortparallel }\mathring{g}%
^{8})]\}|}-\kappa \lbrack \frac{~^{\shortparallel }\partial ^{7}(\
^{\shortparallel }\chi ^{8}|\ ^{\shortparallel }\zeta
^{8}~_{h}^{\shortparallel }\mathring{g}^{8}|^{1/2})}{4~^{\shortparallel
}\partial ^{7}(|~^{\shortparallel }\zeta ^{8}\ _{h}^{\shortparallel }%
\mathring{g}^{8}|^{1/2})}-  \label{offtangfixs4} \\
&&\frac{\int d~^{\shortparallel }p_{7}\{(~_{4}^{\shortparallel }\mathcal{K}%
)~^{\shortparallel }\partial ^{7}[(~^{\shortparallel }\zeta
^{8}~_{h}^{\shortparallel }\mathring{g}^{8})~^{\shortparallel }\chi ^{8}]\}}{%
\int d~^{\shortparallel }p_{7}\{(~_{4}^{\shortparallel }\mathcal{K}%
)~^{\shortparallel }\partial ^{7}[(~^{\shortparallel }\zeta
^{8}~_{h}^{\shortparallel }\mathring{g}^{8})]\}}]\}~_{h}^{\shortparallel }%
\mathring{g}^{7}\{d\ ^{\shortparallel }p_{7}+[\frac{~^{\shortparallel
}\partial _{i_{3}}\ \int d\ ^{\shortparallel }p_{7}(~_{4}^{\shortparallel }%
\mathcal{K})\ ~^{\shortparallel }\partial ^{7}(\ ^{\shortparallel }\zeta
^{8})}{(~^{\shortparallel }\mathring{N}_{i_{3}7})(~_{4}^{\shortparallel }%
\mathcal{K})~^{\shortparallel }\partial ^{7}(~^{\shortparallel }\zeta ^{8})}
\notag \\
&&+\kappa \lbrack \frac{~^{\shortparallel }\partial _{i_{3}}[\int
d~^{\shortparallel }p_{7}(~_{4}^{\shortparallel }\mathcal{K}%
)~^{\shortparallel }\partial ^{7}(~^{\shortparallel }\zeta
^{8}~_{h}^{\shortparallel }\mathring{g}^{8})]}{~^{\shortparallel }\partial
_{i_{3}}\ [\int d~^{\shortparallel }p_{7}(~_{4}^{\shortparallel }\mathcal{K}%
)~^{\shortparallel }\partial ^{7}(~^{\shortparallel }\zeta ^{8})]}-\frac{%
~^{\shortparallel }\partial ^{7}(~^{\shortparallel }\zeta
^{8}~_{h}^{\shortparallel }\mathring{g}^{8})}{~^{\shortparallel }\partial
^{7}(~^{\shortparallel }\zeta ^{8})}]](~^{\shortparallel }\mathring{N}%
_{i_{3}7})d~^{\shortparallel }x^{i_{3}}\}+  \notag
\end{eqnarray}%
\begin{eqnarray*}
&&\ ^{\shortparallel }\zeta ^{8}(1+\kappa \ ^{\shortparallel }\chi ^{8})\
_{h}^{\shortparallel }\mathring{g}^{8}\{d\ ^{\shortparallel }E+[(\
^{\shortparallel }\mathring{N}_{i_{3}8})^{-1}[\ _{1}^{\shortparallel
}n_{i_{3}}+16\ _{2}^{\shortparallel }n_{i_{3}}[\int d\ ^{\shortparallel
}p_{7}\{\frac{\left( \ ^{\shortparallel }\partial ^{7}[(\ ^{\shortparallel
}\zeta ^{8}~_{h}^{\shortparallel }\mathring{g}^{8})^{-1/4}]\right) ^{2}}{%
|\int d\ ^{\shortparallel }p_{7}(~_{4}^{\shortparallel }\mathcal{K})\
^{\shortparallel }\partial ^{7}[(\ ^{\shortparallel }\zeta ^{8}\
_{h}^{\shortparallel }\mathring{g}^{8})]|}]-\kappa \times \\
&&\frac{16\ _{2}^{\shortparallel }n_{i_{3}}\int d\ ^{\shortparallel }p_{7}%
\frac{\left( \ ^{\shortparallel }\partial ^{7}[(\ ^{\shortparallel }\zeta
^{8}\ _{h}^{\shortparallel }\mathring{g}^{8})^{-1/4}]\right) ^{2}}{|\int d\
^{\shortparallel }p_{7}\ (\ _{4}^{\shortparallel }\mathcal{K})\
^{\shortparallel }\partial ^{7}[(\ ^{\shortparallel }\zeta ^{8}\
_{h}^{\shortparallel }\mathring{g}^{8})]|}(\frac{\ ^{\shortparallel
}\partial ^{7}[(\ ^{\shortparallel }\zeta ^{8}\ ^{\shortparallel }\mathring{g%
}^{8})^{-1/4}\ ^{\shortparallel }\chi ^{8})]}{2\ ^{\shortparallel }\partial
^{7}[(\ ^{\shortparallel }\zeta ^{8}\ ^{\shortparallel }\mathring{g}%
^{8})^{-1/4}]}+\frac{\int d\ ^{\shortparallel }p_{7}(\ _{4}^{\shortparallel }%
\mathcal{K})\ ^{\shortparallel }\partial ^{7}[(\ ^{\shortparallel }\zeta
^{8}\ _{h}^{\shortparallel }\mathring{g}^{8})\ ^{\shortparallel }\chi ^{8}]}{%
\int d\ ^{\shortparallel }p_{7}(\ _{4}^{\shortparallel }\mathcal{K})\
^{\shortparallel }\partial ^{7}[(\ ^{\shortparallel }\zeta ^{8}\
_{h}^{\shortparallel }\mathring{g}^{8})]})}{\ _{1}^{\shortparallel
}n_{i_{3}}+16\ _{2}^{\shortparallel }n_{i_{3}}[\int d\ ^{\shortparallel
}p_{7}\frac{\left( \ ^{\shortparallel }\partial ^{7}[(\ ^{\shortparallel
}\zeta ^{8}\ _{h}^{\shortparallel }\mathring{g}^{8})^{-1/4}]\right) ^{2}}{\
^{\shortparallel }\partial ^{7}|\int d\ ^{\shortparallel }p_{7}\ (\
_{4}^{\shortparallel }\mathcal{K})[(\ ^{\shortparallel }\zeta ^{8}\ \
_{h}^{\shortparallel }\mathring{g}^{8})]|}]}](\ ^{\shortparallel }\mathring{N%
}_{i_{3}8})dx^{i_{3}}\}
\end{eqnarray*}%
with generating functions $~^{\shortparallel }\zeta ^{8}(\ ^{\shortmid
}r,\theta ,\varphi ,\theta _{2},\theta _{3},~^{\shortparallel }p_{7})$ and $%
\ ^{\shortparallel }\chi ^{8}(\ ^{\shortmid }r,\theta ,\varphi ,\theta
_{2},\theta _{3},~^{\shortparallel }p_{7});$ R-flux effective source \newline
$~_{4}^{\shortparallel }\mathcal{K}(\ ^{\shortmid }r,\theta ,\varphi ,\theta
_{2},\theta _{3},~^{\shortparallel }p_{7})$; integration functions $\
_{1}^{\shortparallel }n_{k_{3}}(\ ^{\shortmid }r,\theta ,\varphi ,\theta
_{2},\theta _{3})$ and $\ _{2}^{\shortparallel }n_{k_{3}}(\ ^{\shortmid
}r,\theta ,\varphi ,\theta _{2},\theta _{3});$ and components of the prime metric $\
_{h}^{\shortparallel }\mathring{g}^{7}$ and $\ _{h}^{\shortparallel }%
\mathring{g}^{8}$ from (\ref{primetangd}). The non-zero, prime N-connection
coefficients $\ ^{\shortparallel }\mathring{N}_{k_{3}}^{a_{4}}$ arise via the coordinate transformations $\ ^{\shortparallel }%
\mathring{N}_{k_{3}}^{a_{4}}[~^{\shortmid }u^{\beta _{4}}(\ ^{\shortmid
}r,\theta ,\varphi ,\theta _{2},\theta _{3})]\rightarrow \ ^{\shortparallel }%
\mathring{N}_{k_{3}}^{a_{4}}[~^{\shortmid }u^{\beta _{4}}(\ ^{\shortmid
}r,\theta ,\varphi ,\theta _{2},\theta _{3},~^{\shortparallel }p_{7})]$, and
result in generic off-diagonal metrics.

We can deform the solutions of (\ref{offtangfix}) into ellipsoidal
form by choosing the generating functions in (\ref{offtangfixs2}%
) as:
\begin{equation}
\ ^{\shortparallel }\chi _{4}=\ \chi _{4}=\ ^{e}\chi _{4}(\ ^{\shortmid
}r,\theta ,\varphi )=2\underline{\chi }(\ ^{\shortmid }r,\theta )\sin
(\omega _{0}\varphi +\varphi _{0}),  \label{ellipsconf}
\end{equation}%
The function $\underline{\chi }(\ ^{\shortmid }r,\theta )$ can be a smooth
function, or a constant, and $\omega _{0}$ and $\varphi _{0}$ are some
constants. On the spacetime manifold, this s-metric has an ellipsoidal horizon
with eccentricity $\kappa $ given by
\begin{equation*}
(1+\kappa \ ^{\shortparallel }\chi _{4})\ _{h}^{\shortparallel }\mathring{g}%
_{4}=1-\frac{\ ^{\shortmid }\mu }{(\ ^{\shortmid }r)^{3}}-\frac{\
^{\shortmid }\kappa _{6}^{2}\ ^{\shortmid }\Lambda }{10}(\ ^{\shortmid
}r)^{2}+\kappa \ ^{\shortparallel }\chi _{4}=0
\end{equation*}%
Considering configurations with $\frac{\ ^{\shortmid }\kappa _{6}^{2}\
^{\shortmid }\Lambda }{10}(\ ^{\shortmid }r)^{2}\approx 0,$ we can
approximate $\ ^{\shortmid }r\approx \ ^{\shortmid }\mu ^{1/3}/(1-\frac{%
\kappa \ }{3}^{\shortparallel }\chi _{4})$ which defines an ellipsoidal horizon.

Next the Tangerlini BH metric in (\ref{offtangfix}) can be deformed into a family of nonassociative, quasi-stationary BHs,
with a fixed energy parameter. The deformed metric splits in a symmetric, $\ _{\star
}^{\shortparallel }\mathbf{\check{q}}_{\alpha _{s}\beta _{s}},$ and
antisymmetric, $\ _{\star }^{\shortparallel }\mathbf{a}_{\alpha _{s}\beta
_{s}},$ s-metrics. The s-adapted coefficients of $\ _{\star
}^{\shortparallel }\mathbf{\check{q}}_{\alpha _{s}\beta _{s}},$ and
 $\ _{\star }^{\shortparallel }\mathbf{a}_{\alpha _{s}\beta
_{s}},$ can be computed
in explicit form for any prescribed R-flux, $\overline{\mathcal{R}}%
_{\quad \beta _{s}}^{\tau _{s}\xi _{s}}$, by introducing
the s-metric in (\ref{aux40b}) and the s-connection in (\ref{aux40aa}) yielding
\begin{eqnarray*}
\ _{\star }^{\shortparallel }\mathbf{\check{q}}_{\alpha _{s}\beta _{s}}&:=&
\ _{\star }^{\shortparallel }\mathbf{g}_{\alpha _{s}\beta _{s}}-\frac{%
i\kappa }{2}\left( \overline{\mathcal{R}}_{\quad \beta _{s}}^{\tau _{s}\xi
_{s}}\ \mathbf{^{\shortparallel }e}_{\xi _{s}}\ _{\star }^{\shortparallel }%
\mathbf{g}_{\tau _{s}\alpha _{s}}+\overline{\mathcal{R}}_{\quad \alpha
_{s}}^{\tau _{s}\xi _{s}}\ ^{\shortparallel }\mathbf{e}_{\xi _{s}}\ _{\star
}^{\shortparallel }\mathbf{g}_{\beta _{s}\tau _{s}}\right) ,\mbox{ and } \\
\ _{\star }^{\shortparallel }\mathbf{a}_{\alpha _{s}\beta _{s}}&:= &\frac{%
i\kappa }{2}\left( \overline{\mathcal{R}}_{\quad \beta _{s}}^{\tau _{s}\xi
_{s}}\ \mathbf{^{\shortparallel }e}_{\xi _{s}}\ _{\star }^{\shortparallel }%
\mathbf{g}_{\tau _{s}\alpha _{s}}-\overline{\mathcal{R}}_{\quad \alpha
_{s}}^{\tau _{s}\xi _{s}}\ \mathbf{^{\shortparallel }e}_{\xi _{s}}\ _{\star
}^{\shortparallel }\mathbf{g}_{\beta _{s}\tau _{s}}\right)
\end{eqnarray*}%
We omit the technical details for computing $\
_{\star }^{\shortparallel }\mathbf{q}_{\alpha _{s}\beta _{s}}=\ _{\star
}^{\shortparallel }\mathbf{\check{q}}_{\alpha _{s}\beta _{s}}+\ _{\star
}^{\shortparallel }\mathbf{a}_{\alpha _{s}\beta _{s}}$ given in (\ref{aux40b}).  For quasi-stationary R-flux
deformations considered in this section, such nonassociative s-metrics are
induced and completely determined by $(\
^{\shortparallel }\mathbf{g}_{\alpha _{s}\beta _{s}},^{\shortparallel
}N_{i_{s-1}}^{a_{s}}),$ with coefficients computed above or in other forms
with different generating functions. Ellipsoidal base spacetime configurations (%
\ref{ellipsconf}) are modelled for  functions $\ ^{\shortparallel }\chi _{4}=\ ^{e}\chi _{4}(\ ^{\shortmid
}r,\theta ,\varphi ).$

\subsubsection{Spacetime and co-fiber phase space double BHs and BEs with
fixed energy parameter}

Starting with the prime metric (\ref{doublepm}) we can generate nonassociative configurations with metric components of the form $\
~^{\shortparallel }\mathring{g}_{\alpha _{s}}(~^{\shortmid }u^{\beta
_{s}})=\ ~^{\shortparallel }\mathring{g}_{\alpha _{s}}[~^{\shortparallel
}u^{\beta _{s}}(r,\theta ,\varphi ;\ _{\shortparallel }^{p}r,\
_{\shortparallel }^{p}\theta ,\ _{\shortparallel }^{p}\varphi )]$ \ \ and
nontrivial N-coefficients$~\ ~^{\shortparallel }\mathring{N}%
_{i_{s-1}}^{a_{s}}(~^{\shortparallel }u^{\beta _{s}})$ determined by some
non-singular coordinate transforms $(r,\theta ,\varphi ,t,\ ^{p}r,\
^{p}\theta ,\ ^{p}\varphi ,E)$ $\rightarrow ~^{\shortmid }u^{\beta _{s}},$
or \newline
$(r,\theta ,\varphi ,t,\ \ _{\shortparallel }^{p}r,\ \ _{\shortparallel
}^{p}\theta ,\ \ _{\shortparallel }^{p}\varphi ,~^{\shortparallel }E)$ $%
\rightarrow ~^{\shortparallel }u^{\beta _{s}}$. The metric components and non-zero N-coefficients are
\begin{eqnarray}
\ ^{\shortparallel }\mathring{g}_{1} &=&\ f^{-1}(r),\ ~^{\shortparallel }%
\mathring{g}_{2}=r^{2},\ ~^{\shortparallel }\mathring{g}_{3}=r^{2}\sin
^{2}\theta ,\ ~~^{\shortparallel }\mathring{g}_{4}=-\ f(r),
\label{pr2bhdata} \\
~^{\shortparallel }\mathring{g}_{5} &=&\ _{\shortparallel }^{p}f^{-1}(\ \
_{\shortparallel }^{p}r),\ ~^{\shortparallel }\mathring{g}_{6}=(\ \
_{\shortparallel }^{p}r)^{2},\ ~^{\shortparallel }\mathring{g}_{7}=(\ \
_{\shortparallel }^{p}r)^{2}\sin ^{2}\ _{\shortparallel }^{p}\theta ,\
~~^{\shortparallel }\mathring{g}_{8}=-\ _{\shortparallel }^{p}f^{-1}(\ \
_{\shortparallel }^{p}r);  \notag \\
&&\ ~^{\shortparallel }\mathring{N}_{i_{s-1}}^{a_{s}}\neq 0,\mbox{ but }%
~~^{\shortparallel }\mathring{N}_{i_{s-1}}^{a_{s}}(~^{\shortparallel
}u^{\beta _{s}})\rightarrow ~^{\shortparallel }\mathring{N}%
_{i_{s-1}}^{a_{s}}(r,\theta ,\varphi ,t,\ \ _{\shortparallel }^{p}r,\ \
_{\shortparallel }^{p}\theta ,\ \ _{\shortparallel }^{p}\varphi
,~^{\shortparallel }E)=0.  \notag
\end{eqnarray}%
In these formulas, a left label $f$ emphasizes that such diagonalizable
s-metrics are determined respectively by two radial functions $\ \
f^{-1}(r)$ and $\ _{\shortparallel }^{p}f^{-1}(\ \ _{\shortparallel }^{p}r).$
The nontrivial N-connection coefficients $\ ~~^{\shortparallel }\mathring{N}%
_{i_{s-1}}^{a_{s}}(~^{\shortparallel }u^{\beta _{s}})$ can be generated by
coordinate transformations and encode physical parameters of the prime BH metric. Substituting, shell by shell, the prime metric components, $%
\ ~^{\shortparallel }\mathring{g}_{\alpha _{s}}(~^{\shortmid }u^{\beta
_{s}}) $, from (\ref{pr2bhdata}) into
formulas (\ref{offtangfixs1}), (\ref{offtangfixs2}), (\ref{offtangfixs3}),
and (\ref{offtangfixs4}), we can construct a nonlinear quadratic element of
type (\ref{offtangfix}). These parametric, generic off-diagonal solutions
describe nonassociative, R-flux deformations of double BH configurations, where
the first prime metric is on the base spacetime manifold and the second prime
metric in the co-fiber subspace. In a more compact form, this class of
solutions can be written in terms of  $\ ~^{\shortparallel }\mathring{g}_{\alpha _{s}}$ and $\
~^{\shortparallel }\mathring{N}_{i_{s-1}}^{a_{s}}$ as 
\begin{eqnarray}
\ _{\star }^{\shortparallel }\mathbf{g}_{\tau _{s}\alpha _{s}}
&=&diag[g_{1}=\ \eta _{1}(r)\ \ f^{-1}(r),g_{2}=\eta _{2}\ (r,\theta
,\varphi )r^{2},  \label{2bhscoef} \\
g_{3} &=&-\{\frac{4[\partial _{\varphi }(|\zeta _{4}\ (r,\theta ,\varphi )\
\ f(r)|^{1/2})]^{2}}{\ r^{2}\sin ^{2}\theta |\int d\varphi \{[\ _{2}\mathcal{%
K}\ (r,\theta ,\varphi )]\partial _{\varphi }[\ \zeta _{4}\ (r,\theta
,\varphi )\ \ f(r)]\}|}-  \notag \\
&&\kappa \lbrack \frac{\partial _{\varphi }[\chi _{4}\ (r,\theta ,\varphi
)|\zeta _{4}\ (r,\theta ,\varphi )\ \ f(r)|^{1/2}]}{4\partial _{\varphi
}[|\zeta _{4}\ (r,\theta ,\varphi )\ \ f(r)|^{1/2}]}  \notag \\
&&-\frac{\int d\varphi \{[~_{2}\mathcal{K}\ (r,\theta ,\varphi )]\partial
_{\varphi }[(\zeta _{4}\ (r,\theta ,\varphi )\ \ f(r))\ \chi _{4}\ (r,\theta
,\varphi )]\}}{\int d\varphi \{[~_{2}\mathcal{K}\ (r,\theta ,\varphi
)]\partial _{\varphi }[\zeta _{4}\ (r,\theta ,\varphi )\ \ f(r)]\}}]\}\
r^{2}\sin ^{2}\theta ,  \notag \\
g_{4} &=&-\zeta _{4}(r,\theta ,\varphi )[1+\kappa \ \chi _{4}(r,\theta
,\varphi )]\ \ f(r),  \notag
\end{eqnarray}%
\begin{eqnarray*}
~^{\shortparallel }g^{5} &=&\ ^{\shortparallel }\zeta ^{5}(\
_{\shortparallel }^{p}r,\ \ _{\shortparallel }^{p}\theta ,\ \
_{\shortparallel }^{p}\varphi )[1+\kappa \ ^{\shortparallel }\chi ^{5}(\
_{\shortparallel }^{p}r,\ \ _{\shortparallel }^{p}\theta ,\ \
_{\shortparallel }^{p}\varphi )]\ \ _{\shortparallel }^{p}f^{-1}(\ \
_{\shortparallel }^{p}r), \\
~^{\shortparallel }g^{6} &=&-\{\frac{4[~^{\shortparallel }\partial _{\
_{\shortparallel }^{p}\theta }(|~^{\shortparallel }\zeta ^{5}(\
_{\shortparallel }^{p}r,\ \ _{\shortparallel }^{p}\theta ,\ \
_{\shortparallel }^{p}\varphi )~\ _{\shortparallel }^{p}f^{-1}(\ \
_{\shortparallel }^{p}r)|^{1/2})]^{2}}{(\ \ _{\shortparallel
}^{p}r)^{2}|\int d\ \ \ _{\shortparallel }^{p}\theta
\{[~_{3}^{\shortparallel }\mathcal{K}(\ _{\shortparallel }^{p}r,\ \
_{\shortparallel }^{p}\theta ,\ \ _{\shortparallel }^{p}\varphi
)]~^{\shortparallel }\partial _{\theta _{3}}[(~^{\shortparallel }\zeta
^{5}(\ _{\shortparallel }^{p}r,\ \ _{\shortparallel }^{p}\theta ,\ \
_{\shortparallel }^{p}\varphi )~\ _{\shortparallel }^{p}f^{-1}(\ \
_{\shortparallel }^{p}r))]\}|}- \\
&&\kappa \lbrack \frac{\partial _{i_{2}}[\int d\ \ _{\shortparallel
}^{p}\theta \lbrack ~_{3}^{\shortparallel }\mathcal{K}(\ _{\shortparallel
}^{p}r,\ \ _{\shortparallel }^{p}\theta ,\ \ _{\shortparallel }^{p}\varphi
)]~^{\shortparallel }\partial _{\ _{\shortparallel }^{p}\theta
}(~^{\shortparallel }\zeta ^{5}~(\ _{\shortparallel }^{p}r,\ \
_{\shortparallel }^{p}\theta ,\ \ _{\shortparallel }^{p}\varphi )\ \
_{\shortparallel }^{p}f^{-1}(\ \ _{\shortparallel }^{p}r))]}{\partial
_{i_{2}}\ [\int d\ \ _{\shortparallel }^{p}\theta \lbrack
~_{3}^{\shortparallel }\mathcal{K}(\ _{\shortparallel }^{p}r,\ \
_{\shortparallel }^{p}\theta ,\ \ _{\shortparallel }^{p}\varphi
)]~^{\shortparallel }\partial _{\ _{\shortparallel }^{p}\theta
}(~^{\shortparallel }\zeta ^{5}(\ _{\shortparallel }^{p}r,\ \
_{\shortparallel }^{p}\theta ,\ \ _{\shortparallel }^{p}\varphi ))]} \\
&&-\frac{^{\shortparallel }\partial _{\ _{\shortparallel }^{p}\theta
}(~^{\shortparallel }\zeta ^{5}(\ _{\shortparallel }^{p}r,\ \
_{\shortparallel }^{p}\theta ,\ \ _{\shortparallel }^{p}\varphi )~\
_{\shortparallel }^{p}f^{-1}(\ \ _{\shortparallel }^{p}r))}{%
~^{\shortparallel }\partial _{\ \ _{\shortparallel }^{p}\theta
}(~^{\shortparallel }\zeta ^{5}(\ _{\shortparallel }^{p}r,\ \
_{\shortparallel }^{p}\theta ,\ \ _{\shortparallel }^{p}\varphi ))}]\}~(\ \
_{\shortparallel }^{p}r)^{2},
\end{eqnarray*}%
\begin{eqnarray*}
~^{\shortparallel }g^{7} &=&-\{4\frac{4[~^{\shortparallel }\partial _{\
_{\shortparallel }^{p}\varphi }(|~^{\shortparallel }\zeta ^{8}(\
_{\shortparallel }^{p}r,\ \ _{\shortparallel }^{p}\theta ,\ \
_{\shortparallel }^{p}\varphi )~\ _{\shortparallel }^{p}f^{-1}(\ \
_{\shortparallel }^{p}r)|^{1/2})]^{2}}{(\ \ _{\shortparallel }^{p}r)^{2}\sin
^{2}\ _{\shortparallel }^{p}\theta |\int d~\ _{\shortparallel }^{p}\varphi
\{(~_{4}^{\shortparallel }\mathcal{K}(\ _{\shortparallel }^{p}r,\ \
_{\shortparallel }^{p}\theta ,\ \ _{\shortparallel }^{p}\varphi
))~^{\shortparallel }\partial _{\ _{\shortparallel }^{p}\varphi
}[(~^{\shortparallel }\zeta ^{8}(\ _{\shortparallel }^{p}r,\ \
_{\shortparallel }^{p}\theta ,\ \ _{\shortparallel }^{p}\varphi )~\
_{\shortparallel }^{p}f^{-1}(\ \ _{\shortparallel }^{p}r))]\}|}- \\
&&\kappa \lbrack \frac{~^{\shortparallel }\partial _{\ _{\shortparallel
}^{p}\varphi }(\ ^{\shortparallel }\chi ^{8}(\ _{\shortparallel }^{p}r,\ \
_{\shortparallel }^{p}\theta ,\ \ _{\shortparallel }^{p}\varphi )|\
^{\shortparallel }\zeta ^{8}(\ _{\shortparallel }^{p}r,\ \ _{\shortparallel
}^{p}\theta ,\ \ _{\shortparallel }^{p}\varphi )~\ _{\shortparallel
}^{p}f^{-1}(\ \ _{\shortparallel }^{p}r)|^{1/2})}{4~^{\shortparallel
}\partial _{\ _{\shortparallel }^{p}\varphi }(|~^{\shortparallel }\zeta
^{8}(\ _{\shortparallel }^{p}r,\ \ _{\shortparallel }^{p}\theta ,\ \
_{\shortparallel }^{p}\varphi )\ \ _{\shortparallel }^{p}f^{-1}(\ \
_{\shortparallel }^{p}r)|^{1/2})} \\
&&+\frac{\int d~\ _{\shortparallel }^{p}\varphi \{(~_{4}^{\shortparallel }%
\mathcal{K}(\ _{\shortparallel }^{p}r,\ \ _{\shortparallel }^{p}\theta ,\ \
_{\shortparallel }^{p}\varphi ))~^{\shortparallel }\partial _{\
_{\shortparallel }^{p}\varphi }[(~^{\shortparallel }\zeta ^{8}(\
_{\shortparallel }^{p}r,\ \ _{\shortparallel }^{p}\theta ,\ \
_{\shortparallel }^{p}\varphi )~\ _{\shortparallel }^{p}f^{-1}(\ \
_{\shortparallel }^{p}r))~^{\shortparallel }\chi ^{8}]\}}{\int d~\
_{\shortparallel }^{p}\varphi \{(~_{4}^{\shortparallel }\mathcal{K}(\
_{\shortparallel }^{p}r,\ \ _{\shortparallel }^{p}\theta ,\ \
_{\shortparallel }^{p}\varphi ))~^{\shortparallel }\partial _{\
_{\shortparallel }^{p}\varphi }[(~^{\shortparallel }\zeta ^{8}(\
_{\shortparallel }^{p}r,\ \ _{\shortparallel }^{p}\theta ,\ \
_{\shortparallel }^{p}\varphi )~\ _{\shortparallel }^{p}f^{-1}(\ \
_{\shortparallel }^{p}r))]\}}]\}~(\ \ _{\shortparallel }^{p}r)^{2}\sin ^{2}\
_{\shortparallel }^{p}\theta , \\
~^{\shortparallel }g^{8} &=&-\ ^{\shortparallel }\zeta ^{8}(\
_{\shortparallel }^{p}r,\ \ _{\shortparallel }^{p}\theta ,\ \
_{\shortparallel }^{p}\varphi )[1+\kappa \ ^{\shortparallel }\chi ^{8}(\
_{\shortparallel }^{p}r,\ \ _{\shortparallel }^{p}\theta ,\ \
_{\shortparallel }^{p}\varphi )]\ \ _{\shortparallel }^{p}f^{-1}(\ \
_{\shortparallel }^{p}r)],
\end{eqnarray*}%
and for the N-connection%
\begin{eqnarray}
\ ^{\shortparallel }\ N_{i_{1}}^{3} &=&[\frac{\partial _{i_{1}}\ \int
d\varphi (~_{2}\mathcal{K}\ (r,\theta ,\varphi )\ )\ \partial _{\varphi
}(\zeta _{4}\ (r,\theta ,\varphi )\ )}{(~^{\shortparallel }\mathring{N}%
_{i_{1}}^{3})(~_{2}\mathcal{K}\ (r,\theta ,\varphi )\ )\partial _{\varphi
}(\zeta _{4}\ (r,\theta ,\varphi )\ )}+\kappa (\frac{\partial _{i_{1}}[\int
d\varphi (~_{2}\mathcal{K}\ (r,\theta ,\varphi )\ )\partial _{\varphi
}(\zeta _{4}\ (r,\theta ,\varphi )\ \chi _{4}\ (r,\theta ,\varphi )\ )]}{%
\partial _{i_{1}}\ [\int d\varphi (~_{2}\mathcal{K}\ (r,\theta ,\varphi )\
)\partial _{\varphi }(\zeta _{4}\ (r,\theta ,\varphi )\ )]}  \notag \\
&&-\frac{\partial _{\varphi }(\zeta _{4}\ (r,\theta ,\varphi )\ \ \chi _{4}\
(r,\theta ,\varphi )\ )}{\partial _{\varphi }(\zeta _{4}\ (r,\theta ,\varphi
)\ )})](~^{\shortparallel }\mathring{N}_{i_{1}}^{3}),  \notag
\end{eqnarray}%
\begin{eqnarray}
\ ^{\shortparallel }\ N_{k_{1}}^{4} &=&[(~^{\shortparallel }\mathring{N}%
_{k_{1}}^{4})^{-1}[\ _{1}n_{k_{1}}+16\ _{2}n_{k_{1}}[\int d\varphi \{\frac{%
\left( \partial _{\varphi }[(\zeta _{4}\ \ (r,\theta ,\varphi )\
f(r))^{-1/4}]\right) ^{2}}{|\int d\varphi \partial _{\varphi }[(~_{2}%
\mathcal{K}\ (r,\theta ,\varphi )\ )(\zeta _{4}\ (r,\theta ,\varphi )\ \ \
f(r)]|}]-  \label{2bhnscoef} \\
&&\kappa \frac{16\ _{2}n_{k_{1}}\int d\varphi \frac{\left( \partial
_{\varphi }[(\ \zeta _{4}\ (r,\theta ,\varphi )\ \ \ f(r))^{-1/4}]\right)
^{2}}{|\int d\varphi \partial _{\varphi }[(~_{2}\mathcal{K}\ (r,\theta
,\varphi )\ )(\ \zeta _{4}\ (r,\theta ,\varphi )\ \ \ f(r))]|}(\frac{%
\partial _{\varphi }[(\zeta _{4}\ (r,\theta ,\varphi )\ \ \ f(r))^{-1/4}\chi
_{4})]}{2\partial _{\varphi }[(\ \zeta _{4}\ (r,\theta ,\varphi )\ \ \
f(r))^{-1/4}]}}{\ _{1}n_{k_{1}}\ (r,\theta )\ +16\ _{2}n_{k_{1}}\ (r,\theta
)\ [\int d\varphi \frac{\left( \partial _{\varphi }[(\zeta _{4}\ (r,\theta
,\varphi )\ \ \ f(r))^{-1/4}]\right) ^{2}}{|\int d\varphi \partial _{\varphi
}[(~_{2}\mathcal{K}\ (r,\theta ,\varphi )\ )(\zeta _{4}\ (r,\theta ,\varphi
)\ \ f(r))]|}]}  \notag \\
&&+\frac{\frac{\int d\varphi \partial _{\varphi }[(~_{2}\mathcal{K}\
(r,\theta ,\varphi )\ )(\zeta _{4}\ (r,\theta ,\varphi )\ \ \chi _{4}\
(r,\theta ,\varphi )\ \ \ f(r))]}{\int d\varphi \partial _{\varphi }[(~_{2}%
\mathcal{K}\ (r,\theta ,\varphi )\ )(\zeta _{4}\ (r,\theta ,\varphi )\ \ \
f(r))]})}{\ _{1}n_{k_{1}}\ (r,\theta )\ +16\ _{2}n_{k_{1}}\ (r,\theta )\
[\int d\varphi \frac{\left( \partial _{\varphi }[(\zeta _{4}\ (r,\theta
,\varphi )\ \ \ f(r))^{-1/4}]\right) ^{2}}{|\int d\varphi \partial _{\varphi
}[(~_{2}\mathcal{K}\ (r,\theta ,\varphi )\ )(\zeta _{4}\ (r,\theta ,\varphi
)\ \ f(r))]|}]})](~^{\shortparallel }\mathring{N}_{k_{1}}^{4});  \notag
\end{eqnarray}%
\begin{eqnarray*}
\ ^{\shortparallel }N_{i_{2}5} &=&[(~^{\shortparallel }\mathring{N}%
_{i_{2}5})^{-1}[\ _{1}^{\shortparallel }n_{i_{2}}(\ \ _{\shortparallel
}^{p}r)+16\ _{2}^{\shortparallel }n_{i_{2}}(\ \ _{\shortparallel
}^{p}r)[\int d\ _{\shortparallel }^{p}\theta \frac{\left( ~^{\shortparallel
}\partial _{\ _{\shortparallel }^{p}\theta }[(~^{\shortparallel }\zeta
^{5}(\ ^{p}r,\ ^{p}\theta )\ \ \ _{\shortparallel }^{p}f^{-1}(\ \
_{\shortparallel }^{p}r))^{-1/4}]\right) ^{2}}{|\int d\ _{\shortparallel
}^{p}\theta ~^{\shortparallel }\partial _{\ _{\shortparallel }^{p}\theta
}[(~_{3}^{\shortparallel }\mathcal{K})(~^{\shortparallel }\zeta ^{5}(\
^{p}r,\ ^{p}\theta )\ ~\ _{\shortparallel }^{p}f^{-1}(\ \ _{\shortparallel
}^{p}r))]|}] \\
&&+\kappa \frac{16\ _{2}^{\shortparallel }n_{i_{2}}(\ \ _{\shortparallel
}^{p}r)}{\ _{1}^{\shortparallel }n_{i_{2}}(\ \ _{\shortparallel }^{p}r)+16\
_{2}^{\shortparallel }n_{i_{2}}(\ \ _{\shortparallel }^{p}r)[\int d\
_{\shortparallel }^{p}\theta \frac{\left( \ ~^{\shortparallel }\partial _{\
_{\shortparallel }^{p}\theta }[(\ ~^{\shortparallel }\zeta ^{5}(\ ^{p}r,\
^{p}\theta )\ \ \ _{\shortparallel }^{p}f^{-1}(\ \ _{\shortparallel
}^{p}r))^{-1/4}]\right) ^{2}}{|\int d\ _{\shortparallel }^{p}\theta \
(~_{3}^{\shortparallel }\mathcal{K}(\ ^{p}r,\ ^{p}\theta )\ )\
~^{\shortparallel }\partial _{\ _{\shortparallel }^{p}\theta
}[(~^{\shortparallel }\zeta ^{5}(\ ^{p}r,\ ^{p}\theta )\ \ \
_{\shortparallel }^{p}f^{-1}(\ \ _{\shortparallel }^{p}r))]|}]}\times \\
&&\int d\ _{\shortparallel }^{p}\theta \frac{\left( ~^{\shortparallel
}\partial _{\ _{\shortparallel }^{p}\theta }[(~^{\shortparallel }\zeta
^{5}(\ ^{p}r,\ ^{p}\theta )\ \ \ _{\shortparallel }^{p}f^{-1}(\ \
_{\shortparallel }^{p}r))^{-1/4}]\right) ^{2}}{|\int d\ _{\shortparallel
}^{p}\theta (~_{3}^{\shortparallel }\mathcal{K})\ ~^{\shortparallel
}\partial _{_{\shortparallel }^{p}\theta }[(~^{\shortparallel }\zeta ^{5}(\
^{p}r,\ ^{p}\theta )\ \ \ _{\shortparallel }^{p}f^{-1}(\ \ _{\shortparallel
}^{p}r))]|} \\
&&(\frac{\ ~^{\shortparallel }\partial _{\ _{\shortparallel }^{p}\theta }[(\
~^{\shortparallel }\zeta ^{5}(\ ^{p}r,\ ^{p}\theta )\ \ \ _{\shortparallel
}^{p}f^{-1}(\ \ _{\shortparallel }^{p}r))^{-1/4}~^{\shortparallel }\chi
^{5}(\ ^{p}r,\ ^{p}\theta )\ )]}{2\ ~^{\shortparallel }\partial _{\
_{\shortparallel }^{p}\theta }[(~^{\shortparallel }\zeta ^{5}(\ ^{p}r,\
^{p}\theta )\ ~\ _{\shortparallel }^{p}f^{-1}(\ \ _{\shortparallel
}^{p}r))^{-1/4}]}+ \\
&&\frac{\int d\ _{\shortparallel }^{p}\theta \ (~_{3}^{\shortparallel }%
\mathcal{K}(\ ^{p}r,\ ^{p}\theta )\ )~^{\shortparallel }\partial _{\
_{\shortparallel }^{p}\theta }[(~^{\shortparallel }\zeta ^{5}(\ ^{p}r,\
^{p}\theta )\ \ \ _{\shortparallel }^{p}f^{-1}(\ \ _{\shortparallel
}^{p}r))~^{\shortparallel }\chi ^{5}(\ ^{p}r,\ ^{p}\theta )\ ]}{\int d\
_{\shortparallel }^{p}\theta (~_{3}^{\shortparallel }\mathcal{K}(\ ^{p}r,\
^{p}\theta )\ )\ ~^{\shortparallel }\partial _{\ _{\shortparallel
}^{p}\theta }[(~^{\shortparallel }\zeta ^{5}(\ ^{p}r,\ ^{p}\theta )\ \ \
_{\shortparallel }^{p}f^{-1}(\ \ _{\shortparallel }^{p}r))]}%
)](~^{\shortparallel }\mathring{N}_{i_{2}5}), \\
\ ^{\shortparallel }N_{i_{2}6} &=&[\frac{\partial _{i_{2}}\ \int d\
_{\shortparallel }^{p}\theta (~_{3}^{\shortparallel }\mathcal{K}(\ ^{p}r,\
^{p}\theta )\ )~^{\shortparallel }\partial _{\ _{\shortparallel }^{p}\theta
}(~^{\shortparallel }\zeta ^{5}(\ ^{p}r,\ ^{p}\theta )\ )}{%
(~^{\shortparallel }\mathring{N}_{i_{2}6})(~_{3}^{\shortparallel }\mathcal{K}%
(\ ^{p}r,\ ^{p}\theta )\ )~^{\shortparallel }\partial _{\ _{\shortparallel
}^{p}\theta }(~^{\shortparallel }\zeta ^{5}(\ ^{p}r,\ ^{p}\theta )\ )} \\
&&+\kappa (\frac{\partial _{i_{2}}[\int d\ _{\shortparallel }^{p}\theta
(~_{3}^{\shortparallel }\mathcal{K}(\ ^{p}r,\ ^{p}\theta )\
)~^{\shortparallel }\partial _{\ _{\shortparallel }^{p}\theta
}(~^{\shortparallel }\zeta ^{5}(\ ^{p}r,\ ^{p}\theta )\ ~\ _{\shortparallel
}^{p}f^{-1}(\ \ _{\shortparallel }^{p}r))]}{\partial _{i_{2}}\ [\int d\
_{\shortparallel }^{p}\theta (~_{3}^{\shortparallel }\mathcal{K}(\ ^{p}r,\
^{p}\theta )\ )~^{\shortparallel }\partial _{\ _{\shortparallel }^{p}\theta
}(~^{\shortparallel }\zeta ^{5}(\ ^{p}r,\ ^{p}\theta )\ )]} \\
&&-\frac{~^{\shortparallel }\partial _{\ _{\shortparallel }^{p}\theta
}(~^{\shortparallel }\zeta ^{5}(\ ^{p}r,\ ^{p}\theta )\ ~\ _{\shortparallel
}^{p}f^{-1}(\ \ _{\shortparallel }^{p}r))}{~^{\shortparallel }\partial _{\
_{\shortparallel }^{p}\theta }(~^{\shortparallel }\zeta ^{5}(\ ^{p}r,\
^{p}\theta )\ )})](~_{\circ \circ }^{\shortparallel }N_{i_{2}6});
\end{eqnarray*}%
\begin{eqnarray*}
\ ^{\shortparallel }N_{i_{3}7} &=&[\frac{~^{\shortparallel }\partial
_{i_{3}}\ \int d\ \ _{\shortparallel }^{p}\varphi (~_{4}^{\shortparallel }%
\mathcal{K}(\ ^{p}r,\ ^{p}\theta ,\ ^{p}\varphi ))\ ~^{\shortparallel
}\partial _{\ \ _{\shortparallel }^{p}\varphi }(\ ^{\shortparallel }\zeta
^{8}(\ ^{p}r,\ ^{p}\theta ,\ ^{p}\varphi ))}{(~_{\circ \circ
}^{\shortparallel }N_{i_{3}7})(~_{4}^{\shortparallel }\mathcal{K}(\ ^{p}r,\
^{p}\theta ,\ ^{p}\varphi ))~^{\shortparallel }\partial _{\ \
_{\shortparallel }^{p}\varphi }(~^{\shortparallel }\zeta ^{8}(\ ^{p}r,\
^{p}\theta ,\ ^{p}\varphi ))}+ \\
&&\kappa \lbrack \frac{~^{\shortparallel }\partial _{i_{3}}[\int d\
_{\shortparallel }^{p}\varphi (~_{4}^{\shortparallel }\mathcal{K}(\ ^{p}r,\
^{p}\theta ,\ ^{p}\varphi ))~^{\shortparallel }\partial _{\ _{\shortparallel
}^{p}\varphi }(~^{\shortparallel }\zeta ^{8}(\ ^{p}r,\ ^{p}\theta ,\
^{p}\varphi )~~\ _{\shortparallel }^{p}f(\ \ _{\shortparallel }^{p}r))]}{%
~^{\shortparallel }\partial _{i_{3}}\ [\int d\ _{\shortparallel }^{p}\varphi
(~_{4}^{\shortparallel }\mathcal{K}(\ ^{p}r,\ ^{p}\theta ,\ ^{p}\varphi
))~^{\shortparallel }\partial _{\ _{\shortparallel }^{p}\varphi
}(~^{\shortparallel }\zeta ^{8}(\ ^{p}r,\ ^{p}\theta ,\ ^{p}\varphi ))]}- \\
&&\frac{~^{\shortparallel }\partial _{\ _{\shortparallel }^{p}\varphi
}(~^{\shortparallel }\zeta ^{8}~(\ ^{p}r,\ ^{p}\theta ,\ ^{p}\varphi )\
_{\shortparallel }^{p}f(\ \ _{\shortparallel }^{p}r))}{~^{\shortparallel
}\partial _{\ _{\shortparallel }^{p}\varphi }(~^{\shortparallel }\zeta
^{8}(\ ^{p}r,\ ^{p}\theta ,\ ^{p}\varphi ))}]](~^{\shortparallel }\mathring{N%
}_{i_{3}7}),
\end{eqnarray*}%
\begin{eqnarray*}
\ ^{\shortparallel }N_{i_{3}8} &=&[(~^{\shortparallel }\mathring{N}%
_{i_{3}8})^{-1}[\ _{1}^{\shortparallel }n_{i_{3}}(\ ^{p}r,\ ^{p}\theta )+16\
_{2}^{\shortparallel }n_{i_{3}}(\ ^{p}r,\ ^{p}\theta )\times \\
&&[\int d\ \ _{\shortparallel }^{p}\varphi \{\frac{\left( \ ^{\shortparallel
}\partial _{\ ^{p}\varphi }[(\ ^{\shortparallel }\zeta ^{8}(\ ^{p}r,\
^{p}\theta ,\ ^{p}\varphi )~~\ _{\shortparallel }^{p}f^{1}(\ \
_{\shortparallel }^{p}r))^{-1/4}]\right) ^{2}}{|\int d\ ^{p}\varphi
(~_{4}^{\shortparallel }\mathcal{K}(\ ^{p}r,\ ^{p}\theta ,\ ^{p}\varphi ))\
^{\shortparallel }\partial _{\ ^{p}\varphi }[(\ ^{\shortparallel }\zeta
^{8}(\ ^{p}r,\ ^{p}\theta ,\ ^{p}\varphi )\ ~\ _{\shortparallel }^{p}f(\ \
_{\shortparallel }^{p}r))]|}]- \\
&&\frac{\kappa }{\ _{1}^{\shortparallel }n_{i_{3}}(\ ^{p}r,\ ^{p}\theta
)+16\ _{2}^{\shortparallel }n_{i_{3}}(\ ^{p}r,\ ^{p}\theta )[\int d\
_{\shortparallel }^{p}\varphi \frac{\left( \ ^{\shortparallel }\partial _{\
\ _{\shortparallel }^{p}\varphi }[(\ ^{\shortparallel }\zeta ^{8}\ (\
^{p}r,\ ^{p}\theta ,\ ^{p}\varphi )~\ _{\shortparallel }^{p}f(\ \
_{\shortparallel }^{p}r))^{-1/4}]\right) ^{2}}{\ ^{\shortparallel }\partial
\ ^{p}\varphi |\int d\ \ _{\shortparallel }^{p}\varphi \ (\
_{4}^{\shortparallel }\mathcal{K}(\ ^{p}r,\ ^{p}\theta ,\ ^{p}\varphi ))[(\
^{\shortparallel }\zeta ^{8}\ (\ ^{p}r,\ ^{p}\theta ,\ ^{p}\varphi )\ ~\
_{\shortparallel }^{p}f(\ \ _{\shortparallel }^{p}r))]|}]}\times \\
&&16\ _{2}^{\shortparallel }n_{i_{3}}(\ ^{p}r,\ ^{p}\theta )\int d\
_{\shortparallel }^{p}\varphi \frac{\left( \ ^{\shortparallel }\partial _{\
^{p}\varphi }[(\ ^{\shortparallel }\zeta ^{8}(\ ^{p}r,\ ^{p}\theta ,\
^{p}\varphi )\ ~\ _{\shortparallel }^{p}f(\ \ _{\shortparallel
}^{p}r))^{-1/4}]\right) ^{2}}{|\int d\ \ _{\shortparallel }^{p}\varphi \ (\
_{4}^{\shortparallel }\mathcal{K}(\ ^{p}r,\ ^{p}\theta ,\ ^{p}\varphi ))\
^{\shortparallel }\partial _{\ _{\shortparallel }^{p}\varphi }[(\
^{\shortparallel }\zeta ^{8}\ (\ ^{p}r,\ ^{p}\theta ,\ ^{p}\varphi )~\
_{\shortparallel }^{p}f(\ \ _{\shortparallel }^{p}r))]|} \\
&&(\frac{\ ^{\shortparallel }\partial _{\ _{\shortparallel }^{p}\varphi }[(\
^{\shortparallel }\zeta ^{8}(\ ^{p}r,\ ^{p}\theta ,\ ^{p}\varphi )\
^{\shortparallel }~\ _{\shortparallel }^{p}f(\ \ _{\shortparallel
}^{p}r))^{-1/4}\ ^{\shortparallel }\chi ^{8}(\ ^{p}r,\ ^{p}\theta ,\
^{p}\varphi ))]}{2\ ^{\shortparallel }\partial _{\ _{\shortparallel
}^{p}\varphi }[(\ ^{\shortparallel }\zeta ^{8}\ (\ ^{p}r,\ ^{p}\theta ,\
^{p}\varphi )~\ _{\shortparallel }^{p}f(\ \ _{\shortparallel }^{p}r))^{-1/4}]%
}+ \\
&&\frac{\int d\ _{\shortparallel }^{p}\varphi (\ _{4}^{\shortparallel }%
\mathcal{K}(\ ^{p}r,\ ^{p}\theta ,\ ^{p}\varphi ))\ ^{\shortparallel
}\partial _{\ _{\shortparallel }^{p}\varphi }[(\ ^{\shortparallel }\zeta
^{8}(\ ^{p}r,\ ^{p}\theta ,\ ^{p}\varphi )\ ~\ _{\shortparallel }^{p}f(\ \
_{\shortparallel }^{p}r))\ ^{\shortparallel }\chi ^{8}(\ ^{p}r,\ ^{p}\theta
,\ ^{p}\varphi )]}{\int d\ _{\shortparallel }^{p}\varphi (\
_{4}^{\shortparallel }\mathcal{K}(\ ^{p}r,\ ^{p}\theta ,\ ^{p}\varphi ))\
^{\shortparallel }\partial _{\ _{\shortparallel }^{p}\varphi }[(\
^{\shortparallel }\zeta ^{8}(\ ^{p}r,\ ^{p}\theta ,\ ^{p}\varphi )\ ~\
_{\shortparallel }^{p}f(\ \ _{\shortparallel }^{p}r(\ ^{p}r,\ ^{p}\theta ,\
^{p}\varphi )))]})](~^{\shortparallel }\mathring{N}_{i_{3}8}).
\end{eqnarray*}

The metric components from (\ref{2bhscoef}) and the N-connections from (\ref{2bhnscoef}) generate
ellipsoidal configurations if we chose generating functions of the form:
\begin{eqnarray}
\ ^{\shortparallel }\chi _{4} &=&\ \chi _{4}=\ ^{e}\chi _{4}(r,\theta
,\varphi )=2\underline{\chi }(r,\theta )\sin (\omega _{0}\varphi +\varphi
_{0}),  \label{delipgenf} \\
\ ^{\shortparallel }\chi ^{8} &=&\ \ ^{e}\chi (\ _{\shortparallel }^{p}r,\ \
_{\shortparallel }^{p}\theta ,\ \ _{\shortparallel }^{p}\varphi )=2\overline{%
\chi }(\ \ _{\shortparallel }^{p}r,\ \ _{\shortparallel }^{p}\theta )\sin (\
\ _{\shortparallel }^{p}\omega _{0}\ _{\shortparallel }^{p}\varphi +\
_{\shortparallel }^{p}\varphi _{0}).  \notag
\end{eqnarray}%
Here $\underline{\chi }(r,\theta )$ and $\overline{\chi }(\
_{\shortparallel }^{p}r,\ _{\shortparallel }^{p}\theta )$ are smooth
functions (or constants), and the pairs $(\omega _{0},$ $\varphi _{0})$ and $(\
_{\shortparallel }^{p}\omega _{0},$ $\ _{\shortparallel }^{p}\varphi _{0})$
are constants. Both on the spacetime manifold, and on the typical fiber
space, such an s-metric possesses two distinct ellipsoidal horizons with
eccentricities, $\kappa _{4}$\ and $\ ^{p}\kappa $, given by
\begin{eqnarray*}
(1+\kappa \ \chi _{4})\ ~^{\shortparallel }\mathring{g}_{4} &=&1-\frac{\ \mu
}{r^{3}}-\frac{\ \kappa _{4}^{2}\ \Lambda }{6}r^{2}+\kappa \ \chi _{4}=0%
\mbox{ and } \\
(1+\ ^{p}\kappa \ \ ^{\shortparallel }\chi ^{8})\ ~^{\shortparallel }%
\mathring{g}^{8} &=&1-\frac{\ ^{p}\mu }{\ ^{p}r}-\frac{(\ ^{p}\kappa )^{2}(\
^{p}\Lambda )(\ ^{p}r)^{2}}{2}+\ ^{p}\kappa \ \ ^{\shortparallel }\chi ^{8}.
\end{eqnarray*}%
Here $\ ^{\shortparallel }\zeta _{4}\neq 0$ and $\ ^{\shortparallel }\zeta
^{8}\neq 0$. The physical meaning of these constants is explained just after
formulas (\ref{doublef}). For small parametric deformations with $\kappa _{4}^{2}\Lambda r^{2}/6\approx 0$ and $(\
^{p}\kappa )^{2}(\ ^{p}\Lambda )(\ ^{p}r)^{2}/6\approx 0,$ we can make the 
approximations
\begin{equation*}
r\approx \ \mu ^{1/3}/(1-\frac{\kappa \ }{3}\chi _{4})\mbox{ and }\
^{p}r\approx (\ ^{p}\mu )^{1/3}/(1-\frac{\ ^{p}\kappa \ }{3}\ \
^{\shortparallel }\chi ^{8}).
\end{equation*}%
These are parametric formulas for ellipsoidal horizons are defined by
small gravitational R-flux polarizations. In the limit when the
eccentricities go to zero, these double BE configurations transform into standard prime double BH metric given in \eqref{pr2bhdata}.

We conclude this section by noting the following two points:

\begin{itemize}
\item We could also used AFCDM to construct such parametric, off-diagonal solutions describing nonassociative R-flux
deformations of Tangherlini or double BH and BE metrics but with a Killing
symmetry on $\ ^{\shortmid }p ^{7}$ and with explicit dependence on $p_{8}=E.$ Such quasi-stationary solutions were found in
\cite{bubuianu19} in MGTs with MDRs (modified dispersion relations and
generalized Finsler-Lagrange-Hamilton gravity). 

\item The nonlinear quadratic elements, with off-diagonal metric components given by  (\ref{2bhscoef}), (%
\ref{2bhnscoef})) are characterized by nontrivial, nonholonomically induced
s-torsion structures (\ref{twoconsstar}). Nevertheless, we can always
extract zero torsion, LC-configurations for more special classes of generating functions and effective sources.
\end{itemize}

\section{Generalized Bekenstein-Hawking and Perelman thermodynamics of
non\-associative BHs} 

\label{sec4} In this section we investigate the
thermodynamics for the quasi-stationary, parametric
solutions outlined in the previous section. We find results
similar to those in Section 5 of \cite{bubuianu19}. Here the
effective sources and generating functions are chosen so as to encode nonassociative, R-flux deformations. We find that the entropy
is completely determined by the 8-d phase space quantities 
$(\ ^{\shortparallel }%
\mathbf{g}_{\alpha _{s}\beta _{s}},^{\shortparallel}N_{i_{s-1}}^{a_{s}})$ . 
Also only for very special classes of small
parametric deformations of the original, prime metrics 
is it possible to formulate
thermodynamics using Bekenstein-Hawking entropy \cite%
{bek1,bek2,haw1,haw2}. In general, nonassociative, R-flux deformations of metrics will result in nonholonomic, off-diagonal configurations without conventional horizons. It is this lack of a horizon that makes it impossible to define a Bekenstein-Hawking entropy for these spacetimes. 
In contrast we find that for all of these quasi-stationary solutions one can define an entropy by using the generalized Perelman entropy functionals for nonholonomic Ricci solitons \cite{perelman1,rajpoot17,ibubuianu20,ibubuianu21,bubuianu19}.
These nonholonomic, Ricci solitons are defined as self-similar geometric 
flow configurations for a fixed flow parameter.

\subsection{Phase space generalizations of the Bekenstein-Hawking entropy
for prime and nonassociative target s-metrics}

For parametric solutions with conventional spherical/ ellipsoidal horizons in
(non) associative/ commutative phase space gravity, we can apply standard Bekenstein-Hawking BH thermodynamics. In this subsection, we study two such examples:

\subsubsection{The entropy and temperature of Tangerlini BHs and BEs}

The prime Tangherlini-like s-metric $\ _{s}^{\shortparallel }%
\mathbf{\mathring{g}}$ given in (\ref{pmtang}) is a trivial embedding of a 6-d BH
into a 8-d phase space. For this spacetime, we can define the Hawking
temperature as that of a Schwarzschild BH with dimension $n^{\prime }=4+m^{\prime }=6$ - see \cite{pourhassan,bubuianu19}. This Hawking temperature is similar to that of a Schwarzschild BH in $n^{\prime}\geq 4$
spacetime  dimensions,
\begin{equation}
T=\frac{m^{\prime }+1}{4\pi }\left( \frac{\Omega _{m^{\prime }+2}}{4}\right)
^{1/(m^{\prime }+2)}S_{0}^{-1/(m^{\prime }+2)}.  \label{hawtemp}
\end{equation}%
Here the Bekenstein-Hawking entropy is given by 
\begin{equation}
S_{0}=\frac{\Omega _{m^{\prime }+2}\times (\hat{r}_{0})^{m^{\prime }+2}}{4}.
\label{bekhawentr}
\end{equation}%
Here the volume of the unit sphere is, $\Omega _{m^{\prime }+2}=\frac{\pi
^{4+m^{\prime }/2}}{((4+m^{\prime })/2)!}$, and the horizon radius, $\hat{r}%
_{0}=(\frac{8\pi \widehat{M}}{(m^{\prime }+2)\Omega _{m^{\prime }+2}}%
)^{1/(m^{\prime }+1)}$, is determined by the BH mass $\widehat{M}$. The
time coordinate in this case is given by $y^4=t$ and the radial coordinate is given by $\hat{r}=\sqrt{%
(x^{1})^{2}+(x^{2})^{2}+(y^{3})^{2}+(y^{5})^{2}+...+(y^{m^{\prime }})^{2}}$. The results in \eqref{hawtemp} and \eqref{bekhawentr} differ from the standard temperature and entropy, by the number of dimensions and by the dependence of ${\hat r}$ on both coordinate and momentum variables.  

We now deform the metric in (\ref{offtangfix}) by the function $\chi _{4}=2\underline{\chi }\sin (\omega _{0}\varphi
+\varphi _{0}),$ with $\underline{\chi }=const$  and $\hat{r}_{0}=(\frac{2\pi
\widehat{M}}{\Omega _{4}})^{1/3}=\ ^{\shortmid }\mu ^{1/3}$ as in (\ref%
{ellipsconf}). This deformed metric has an ellipsoidal horizon with eccentricity $\varepsilon =2%
\underline{\chi }\kappa /3$, and is described by the parametric formula $\
^{\shortmid }r\approx \ ^{\shortmid }\mu ^{1/3}/(1-\varepsilon \sin (\omega
_{0}\varphi +\varphi _{0})).$ For small $\varepsilon $ and $\hat{r}_{0}\rightarrow \hat{r}_{0}(\varphi )$, we obtain the following
anisotropic modifications of the Hawking temperature and Bekenstein-Hawking
entropy,
\begin{equation}
T(\varphi )=\frac{3}{4\pi }\left( \frac{\Omega _{4}}{4}\right)
^{1/4}S_{0}^{-1/4}(\varphi )\mbox{~~ with ~~}S_{0}(\varphi )=\frac{\Omega
_{4}\times (\ ^{\shortmid }\mu )^{4/3}}{4}(1+4\underline{\chi }\kappa /3\sin
(\omega _{0}\varphi +\varphi _{0})).  \label{1beconf}
\end{equation}%
This example is a very special class of R-flux deformations
described by parametric solutions with ellipsoidal horizons and integration
constant $\underline{\chi }.$ The results in \eqref{1beconf} become the temperature and entropy in (\ref{hawtemp}) and (\ref{bekhawentr}) if the nonassociative
deformations are taken to zero, $\underline{\chi }\kappa \rightarrow 0$.  

\subsubsection{Phase space thermodynamics for double 4-d Schwarzschild BHs
and BEs}

\label{sss412} In this subsection we show that one can extend the concept of the Bekenstein-Hawking entropy to the double 4-d BH and BE configurations of  (\ref{2bhscoef}) and (\ref{2bhnscoef}).

For the two BEs from (\ref{1beconf}) the Bekenstein-Hawkign temperature and entropy are
\begin{eqnarray*}
T(\varphi ) &=&\frac{1}{4\pi }\left( \frac{\Omega _{2}}{4}\right) ^{1/2}%
\left[ S_{0}^{-1/2}(\varphi )+\ _{\shortmid }S_{0}^{-1/2}(\ _{\shortmid
}^{p}\varphi )\right] ,\mbox{ with } \\
S_{0}(\varphi ) &=&\frac{\Omega _{2}\times (\mu )^{4/3}}{4}(1+4\underline{%
\chi }\kappa /3\sin (\omega _{0}\varphi +\varphi _{0}))\mbox{ and }\
_{\shortmid }S_{0}(\ _{\shortmid }^{p}\varphi )=\frac{\Omega _{2}\times (\
^{\shortmid }\mu )^{4/3}}{4}(1+4\overline{\chi }\kappa /3(\ _{\shortmid
}^{p}\omega _{0}\ _{\shortmid }^{p}\varphi +\ _{\shortmid }^{p}\varphi
_{0}))~,
\end{eqnarray*}%
where $\Omega _{2}=\frac{%
\pi ^{2}}{(2)!},$  is the volume of the unit-sphere in 4-d. 
The horizon in the 4-d spacetime and the horizon in the
4-d co-fiber are computed using conventional mass parameters using (\ref{doublesph}) and (\ref{doublef}). If $%
\underline{\chi }\kappa \rightarrow 0$ the BEs transform into BH configurations, with
the standard temperatures.

For the spacetimes considered in this subsection and the previous subsection, it was possible to define Bekenstein-Hawking thermodynamic variables. In the next section  we examine generalizes spacetimes, with both regular coordinates and momentum coordinates, for which it is not to define Bekenstein-Hawking thermodynamics variables, but for which it is possible to define thermodynamic variables using Perelman's entropy functional \cite{perelman1}. 

\subsection{Perelman's thermodynamic variables for nonassociative
quasi-stationary vacuum configurations}

\label{pthvarqs}A Hawking temperature and a Bekenstein-Hawking
entropy can often not not be defined for solutions of modified GR which
do not possess horizons. In reference \cite%
{rajpoot17,ibubuianu20,ibubuianu21,bubuianu19} we studied statistical and geometric thermodynamics models which were derived from Perelman's
W-entropy functional \cite{perelman1} for relativistic flows of phase
spaces. Perelman's W-entropy can be applied to 
general spacetimes from various gravity theories \cite%
{partner01,partner02} for which Bekenstein-Hawking thermodynamics do not apply.  Despite the difficulty of
formulating a theory of nonassociative geometric flows we can apply Perelman's thermodynamic variables to
parametric, quasi-stationary solutions in nonassociative MGTs. For 
certain classes of generating functions and sources, which encode R-flux deformations, we
can formulate statistical thermodynamic models which are 
determined by $(\ ^{\shortparallel }\mathbf{g}_{\alpha
_{s}\beta _{s}},\ ^{\shortparallel }N_{i_{s-1}}^{a_{s}})$. 

\subsubsection{A statistical analogy for nonholonomic phase space Ricci flows%
}

Section 5 of \cite{perelman1} proposes a geometric,
thermodynamic model for geometric flows of Riemannian metrics on a closed
manifold $V^{\prime },\dim V^{\prime }=n^{\prime }$. These geometric flows describe the evolution of a metric $g_{\alpha ^{\prime }\beta ^{\prime }}(\tau )\simeq g_{\alpha
^{\prime }\beta ^{\prime }}(\tau ,u^{\gamma ^{\prime }}),$ as a function of the positive definite, temperature-like parameter $\tau$. A similar
statistical analogy can be formulated on a phase space endowed with $\tau $%
-parametric flows of the quantities $(\ ^{\shortmid }\mathbf{g}_{\alpha
_{s}\beta _{s}}(\tau ),\ ^{\shortmid }N_{i_{s-1}}^{a_{s}}(\tau ))$, and
for the flows of the canonical s-connection $\ _{s}^{\shortmid }\widehat{%
\mathbf{D}}(\tau )$, Ricci s-tensor, $\ ^{\shortmid }\widehat{\mathbf{R}}%
_{\beta _{s}\gamma _{s}}(\tau )$, and Ricci s-scalar, $\ _{s}^{\shortmid }%
\widehat{\mathbf{R}}sc(\tau )$, see formulas (\ref{paramsricci}).

The Perelman F-functional and W-functional can be defined for the canonical quantities $(\
_{s}^{\shortmid }\mathbf{g}(\tau ),\ _{s}^{\shortmid }\widehat{\mathbf{D}}%
(\tau ))$ as \cite%
{rajpoot17,ibubuianu20,ibubuianu21,bubuianu19}:
\begin{eqnarray}
\ _{s}^{\shortmid }\widehat{\mathcal{F}}(\tau ) &=&\int_{\ _{s}^{\shortmid }%
\widehat{\Xi }}e^{-\ \ _{s}^{\shortmid }\widehat{f}}\sqrt{\ _{s}^{\shortmid }%
\mathbf{g}_{\alpha \beta }|}\ _{s}^{\shortmid }\delta \ _{s}^{\shortmid }u\
(\ _{s}^{\shortmid }\widehat{\mathbf{R}}sc+|\ _{s}^{\shortmid }\widehat{%
\mathbf{D}}\ _{s}^{\shortmid }\widehat{f}|^{2}),\mbox{ and }  \notag \\
\ _{s}^{\shortmid }\widehat{\mathcal{W}}(\tau ) &=&\int_{\ _{s}^{\shortmid }%
\widehat{\Xi }}\left( 4\pi \tau \right) ^{-4}e^{-\ _{s}^{\shortmid }\widehat{%
f}}\sqrt{|\ _{s}^{\shortmid }\mathbf{g}_{\alpha _{s}\beta _{s}}|}\
_{s}^{\shortmid }\delta \ _{s}^{\shortmid }u\ [\tau (\ _{s}^{\shortmid }%
\widehat{\mathbf{R}}sc+\sum\nolimits_{s}|\ _{s}^{\shortmid }\widehat{\mathbf{%
D}}\ _{s}^{\shortmid }\widehat{f}|)^{2}+\ _{s}^{\shortmid }\widehat{f}-8].
\label{wfperelmctl}
\end{eqnarray}%
The W-functional from (\ref{wfperelmctl}) can be treated as a ``minus entropy"
for projections on space-like hypersurfaces. 

In above formulas, the integrals and normalizing functions $\
_{s}^{\shortmid }\widehat{f}(\tau ,\ _{s}^{\shortmid }u)$ 
satisfy the condition
\begin{equation*}
\int_{\ _{s}^{\shortmid }\widehat{\Xi }}\ _{s}^{\shortmid }\widehat{\nu }%
\sqrt{|\ _{s}^{\shortmid }\mathbf{g}_{\alpha _{s}\beta _{s}}|}\
_{s}^{\shortmid }\delta \ _{s}^{\shortmid }u:=\int_{t_{1}}^{t_{2}}\int_{%
\widehat{\Xi }_{t}}\ \int_{\ ^{\shortmid }\widehat{\Xi }_{E}}\
_{s}^{\shortmid }\widehat{\nu }\sqrt{|\ _{s}^{\shortmid }\mathbf{g}_{\alpha
_{s}\beta _{s}}|}\ _{s}^{\shortmid }\delta \ _{s}^{\shortmid }u=1,
\end{equation*}%
for integration measures $\ _{s}^{\shortmid }\widehat{\nu }=\left( 4\pi \tau \right) ^{-4}e^{-\ _{s}^{\shortmid }\widehat{f}}$.  The integration measures are
chosen as in \cite{perelman1} but for $8-d$ with shell splitting
and adapting normalizing functions in order to simplify
formulas for s-adapted coefficients. Defining a class of normalizing
shell functions, we define relativistic flow models of geometric
objects. The W-functional (\ref{wfperelmctl}) can be treated as a ``minus entropy" only
for projections on space like hypersurfaces. We can formulate
relativistic thermodynamics models for a solution of (\ref{cannonsymparamc2})  if $\ _{s}^{\shortparallel }\mathbf{g\simeq \{\ ^{\shortmid }\mathbf{%
g}}_{\alpha _{s}\beta _{s}}\}$ (\ref{sdm}) has a causal
(3+1)+(3+1) splitting. We write
\begin{eqnarray*}
&&\ _{s}^{\shortmid }\mathbf{g}(\tau )=\ ^{\shortmid }\mathbf{g}_{\alpha
^{\prime }\beta ^{\prime }}(\tau ,\ _{s}^{\shortmid }u)d\ ^{\shortmid }%
\mathbf{e}^{\alpha ^{\prime }}\otimes d\ ^{\shortmid }\mathbf{e}^{\beta
^{\prime }}=q_{i}(\tau ,x^{k})dx^{i}\otimes dx^{i}+q_{3}(\tau ,x^{k},y^{3})%
\mathbf{e}^{3}\otimes \mathbf{e}^{3}-\breve{N}^{2}(\tau ,x^{k},y^{3})\mathbf{%
e}^{4}\otimes \mathbf{e}^{4}+ \\
&&\ ^{\shortmid }q^{a_{2}}(\tau ,x^{k},y^{3},p_{b_{2}})\ ^{\shortmid }%
\mathbf{e}_{a_{2}}\otimes \ ^{\shortmid }\mathbf{e}_{a_{2}}+\ ^{\shortmid
}q^{7}(\tau ,x^{k},y^{3},p_{b_{2}},p_{b_{3}})\ ^{\shortmid }\mathbf{e}%
_{7}\otimes \ ^{\shortmid }\mathbf{e}_{7}-\ ^{\shortmid }\breve{N}^{2}(\tau
,x^{k},y^{3},p_{b_{2}},p_{b_{3}})\ ^{\shortmid }\mathbf{e}_{8}\otimes \
^{\shortmid }\mathbf{e}_{8},
\end{eqnarray*}%
where $\ ^{\shortmid }\mathbf{e}^{\alpha _{s}}$ are N-adapted bases in total
phase space. Such an ansatz for 8--d phase space
s-metrics are written as an extension of a couple of 3--d metrics, $%
q_{ij}=diag(q_{\grave{\imath}})=(q_{i},q_{3})$ on a hypersurface $\widehat{%
\Xi }_{t}$ and $\ ^{\shortmid }q^{\grave{a}\grave{b}}=diag(\ ^{\shortmid }q^{%
\grave{a}})=(\ ^{\shortmid }q^{a_{2}},\ ^{\shortmid }q^{7})$ on a
hypersurface $\ ^{\shortmid }\widehat{\Xi }_{E},$ \ if
\begin{equation}
q_{1}=g_{1},q_{2}=g_{2},q_{3}=g_{3},\breve{N}^{2}=-g_{4}\mbox{ and }\
^{\shortmid }q^{5}=\ ^{\shortmid }g^{5},\ ^{\shortmid }q^{6}=\ ^{\shortmid
}g^{6},\ ^{\shortmid }q^{7}=\ ^{\shortmid }g^{7},\ ^{\shortmid }\breve{N}%
^{2}=-\ ^{\shortmid }g^{8}.  \label{shift1}
\end{equation}%
In these formulas, $\breve{N}$ is the lapse function on the base and $\
^{\shortmid }\breve{N}^{2}$ is the lapse function in the co-fiber.

Next we introduce a general statistical partition function
\begin{equation}
\ \ _{s}^{\shortmid }\widehat{Z}(\tau )=\exp [\int_{\ ^{\shortmid }\widehat{%
\Xi }}[-\ \ _{s}^{\shortmid }\widehat{f}+8]~\left( 4\pi \tau \right)
^{-4}e^{-\ _{s}^{\shortmid }\widehat{f}}\ ^{\shortmid }\delta \ ^{\shortmid }%
\mathcal{V}(\tau ),  \label{spf}
\end{equation}
with the volume element
\begin{eqnarray}
\ ^{\shortmid }\delta \ ^{\shortmid }\mathcal{V}(\tau ):= &&\sqrt{%
|q_{1}(\tau )q_{2}(\tau )q_{3}(\tau )\breve{N}^{2}(\tau )\ ^{\shortmid
}q^{5}(\tau )\ ^{\shortmid }q^{6}(\tau )\ ^{\shortmid }q^{7}(\tau )\
^{\shortmid }\breve{N}^{2}(\tau )|}  \notag \\
&&dx^{1}dx^{2}\delta y^{3}\delta y^{4}\ ^{\shortmid }\delta \ ^{\shortmid
}u_{5}(\tau )\ ^{\shortmid }\delta \ ^{\shortmid }u_{6}(\tau )\ ^{\shortmid
}\delta \ ^{\shortmid }u_{7}(\tau )\ ^{\shortmid }\delta \ ^{\shortmid
}u_{8}(\tau ).  \label{volume}
\end{eqnarray}

Using the quantities $\ _{s}^{\shortmid }\widehat{Z}$ from (\ref{spf}) and $\
_{s}^{\shortmid }\widehat{\mathcal{W}}$ from  (\ref{wfperelmctl}), we can define
the Perelman thermodynamic values of energy, entropy and fluctuations:
\begin{eqnarray}
\ _{s}^{\shortmid }\widehat{\mathcal{E}}\ &=&-\tau ^{2}\int_{\
_{s}^{\shortmid }\widehat{\Xi }}\ \left( \ _{s}^{\shortmid }\widehat{\mathbf{%
R}}sc+|\ _{s}^{\shortmid }\widehat{\mathbf{D}}\ \ _{s}^{\shortmid }\tilde{f}%
|^{2}-\frac{4}{\tau }\right) \left( 4\pi \tau \right) ^{-4}e^{-\
_{s}^{\shortmid }\widehat{f}}\ ^{\shortmid }\delta \ ^{\shortmid }\mathcal{V}%
(\tau ),  \label{gthermodvalues} \\
\ \ _{s}^{\shortmid }\widehat{S} &=&-\int_{\ _{s}^{\shortmid }\widehat{\Xi }%
}\left( \tau (\ _{s}^{\shortmid }\widehat{\mathbf{R}}sc+|\ _{s}^{\shortmid }%
\widehat{\mathbf{D}}\ \ _{s}^{\shortmid }\tilde{f}|^{2})+\ _{s}^{\shortmid }%
\tilde{f}-8\right) \left( 4\pi \tau \right) ^{-4}e^{-\ _{s}^{\shortmid }%
\widehat{f}}\ ^{\shortmid }\delta \ ^{\shortmid }\mathcal{V}(\tau ),  \notag
\\
\ _{s}^{\shortmid }\widehat{\sigma } &=&2\ \tau ^{4}\int_{\ _{s}^{\shortmid }%
\widehat{\Xi }}|\ \ ^{\shortmid }\widehat{\mathbf{R}}_{\alpha _{s}\beta
_{s}}+\ ^{\shortmid }\widehat{\mathbf{D}}_{\alpha _{s}}\ ^{\shortmid }%
\widehat{\mathbf{D}}_{\beta _{s}}\ _{s}^{\shortmid }\tilde{f}-\frac{1}{2\tau
}q_{\grave{\imath}\grave{j}}|^{2}\left( 4\pi \tau \right) ^{-4}e^{-\
_{s}^{\shortmid }\widehat{f}}\ ^{\shortmid }\delta \ ^{\shortmid }\mathcal{V}%
(\tau )  \notag
\end{eqnarray}

In the next sections, we show how Perelman's W-entropy (\ref{wfperelmctl})
and related thermodynamic variables (\ref{gthermodvalues}) can be computed
for quasi-stationary, phase space, deformed BE and BH  solutions.

\subsubsection{Phase space integration functions, measures and effective
cosmological constants}

To formulate geometric thermodynamic models, which give quasi-stationary
solutions of the nonholonomic Einstein phase space equations $\ ^{\shortparallel }%
\widehat{\mathbf{R}}_{\beta _{s}\gamma _{s}}~(\tau _{0})= \ ^{\shortparallel
}\mathbf{K}_{_{\beta _{s}\gamma _{s}}} \left\lceil \tau _{0},\hbar ,\kappa
\right\rceil$ (\ref{cannonsymparamc2}), we will chose constant values
for the normalizing functions, $\ _{s}^{\shortmid }\widehat{f}=\
_{s}^{\shortmid }\widehat{f}_{0}=const=0$.  These quasi-stationary solutions 
define a subclass of noncommutative Ricci solitons, which are self-similar
configurations under nonholonomic Ricci flows, when the effective temperature
is fixed as $\tau = \tau _{0}.$ We will
study examples where the constants for integration functions will
substantially simplify the computation of Perelman's thermodynamic variables. Then, using
general frame/coordinate transformation to an arbitrary system of reference, the solutions can be given in general covariant form.

We now compute $\ _{s}^{\shortmid }\widehat{\mathcal{W}}$ (\ref%
{wfperelmctl}), $\ _{s}^{\shortmid }\widehat{Z}$ (\ref{spf}) and $\
_{s}^{\shortmid }\widehat{\mathcal{E}} , \ _{s}^{\shortmid }\widehat{S}$ (%
\ref{gthermodvalues}). An s-metric $\ _{s}^{\shortmid }\mathbf{g}(\tau )$,
a $\tau $-dependent solution of (\ref{cannonsymparamc2}), is
generated by shift and laps s-coefficients, with integration
functions fixed to constant values\footnote{\label{fn12}Here for
simplicity we set $g_{4}^{[0]}=0,\ ^{\shortmid }g_{[0]}^{5}=0,$ $\ ^{\shortmid
}g_{[0]}^{8}=0;$ $\ _{1}n_{k_{1}}=0,\ _{2}n_{k_{1}}=0;$ $\ \ _{1}^{\shortmid
}n_{k_{2}}=0,\ \ _{2}^{\shortmid }n_{k_{2}}=0;\ _{1}^{\shortmid
}n_{k_{3}}=0,\ \ _{2}^{\shortmid }n_{k_{3}}=0$.}:
\begin{eqnarray*}
q_{1}(\tau ,x^{k_{1}}) &=&q_{2}(\tau ,x^{k_{1}})=e^{\ \psi (\tau )},\mathbf{q%
}_{3}(\tau ,x^{k},y^{3})= \frac{4\{\partial _{3}[\ _{2}\Phi (\tau
)]^{2}\}^{2}}{|\int dy^{3}\ _{2}\mathcal{K}(\tau )\{\partial _{3}[\ _{2}\Phi
(\tau)]^{2}\}^{2}|\ }, \\
\lbrack \breve{N}(\tau )]^{2} &=&-g_{4}(\tau )=-g_{4}(\tau,x^{k_{1}},y^{3})=
-\frac{[\ _{2}\Phi (\tau )]^{2}}{4\ _{2}\Lambda _{0}(\tau )}; \\
\ ^{\shortmid }q^{5}(\tau ) &=&\ ^{\shortmid
}q^{5}(\tau,x^{k_{1}},y^{a_{2}},\ ^{\shortmid }p_{6})\ = -\frac{[\
_{3}^{\shortmid }\Phi (\tau )]^{2}}{4\ \ _{3}^{\shortmid }\Lambda _{0}(\tau )%
}, \\
\ ^{\shortmid }q^{6}(\tau ) &=&\ ^{\shortmid
}q^{6}(\tau,x^{k_{1}},y^{a_{2}},\ ^{\shortmid }p_{6})\ = \frac{4\{\
^{\shortmid}\partial ^{6}[\ _{3}^{\shortmid }\Phi (\tau )]^{2}\}^{2}}{|\int
d\ ^{\shortmid }p_{6} \ _{3}^{\shortmid }\mathcal{K}(\tau )\{\
^{\shortmid}\partial ^{6}[\ _{3}^{\shortmid }\Phi (\tau )]^{2}\}^{2}|\ }, \\
\ ^{\shortmid }q^{7}(\tau ) &=&\ ^{\shortmid }q^{7}(\tau,
x^{k_{1}},y^{a_{2}},\ ^{\shortmid }p_{a_{3}},\ ^{\shortmid }p_{7})\ = \frac{%
4\{\ ^{\shortmid }\partial ^{7}[\ _{4}^{\shortmid }\Phi (\tau )]^{2}\}^{2}} {%
|\int d\ ^{\shortmid }p_{7} \ _{4}^{\shortmid }\mathcal{K}(\tau )\{\
^{\shortmid }\partial ^{7}[ \ _{4}^{\shortmid }\Phi (\tau )]^{2}\}^{2}|\ },
\\
\lbrack \ ^{\shortmid }\breve{N}(\tau )]^{2} &=&-\ ^{\shortmid
}g^{8}(\tau)=-\ ^{\shortmid }g^{8}(\tau ,x^{k_{1}},y^{a_{2}},\ ^{\shortmid
}p_{a_{3}},\ ^{\shortmid }p_{7})=-\frac{[\ _{4}^{\shortmid }\Phi (\tau )]^{2}%
}{4\ \ _{4}^{\shortmid }\Lambda _{0}(\tau )}.
\end{eqnarray*}%
Inserting these values into (\ref{wfperelmctl}) we obtain
\begin{equation}
\mathcal{W}(\tau _{0})=\int\nolimits_{\tau ^{\prime }}^{\tau _{0}}\frac{%
d\tau }{\tau ^{4}} \int_{\ _{s}^{\shortmid }\widehat{\Xi }}\sqrt{|\
_{s}^{\shortmid }\mathbf{g}[\ _{s}^{\shortmid }\Phi (\tau )]|} \
_{s}^{\shortmid }\delta ^{8}\ _{s}^{\shortmid }u(\tau ) \left( \tau \lbrack
\sum\nolimits_{s}\ _{s}^{\shortmid }\Lambda (\tau )]^{2}-8\right) ,
\label{wn}
\end{equation}%
where integration over the temperature parameter, $\tau ^{\prime }<\tau _{0},$ is
defined so as to avoid singularity conditions and adapted to the spacetime
shell configurations so as to give well-defined thermodynamic
variables. In (\ref{wn}),
\begin{eqnarray*}
&& \sqrt{ \ _{s}^{\shortmid }\mathbf{g}[ \ _{s}^{\shortmid }\Phi (\tau )]|}
= \sqrt{|q_{1}(\tau )\ q_{2}(\tau )\ \mathbf{q}_{3}|}\ |\ \breve{N}(\tau )|%
\sqrt{|\ ^{\shortmid }q^{5}(\tau )\ ^{\shortmid }q^{6}(\tau )\ ^{\shortmid
}q^{7}(\tau )|}|\ ^{\shortmid }\breve{N}(\tau )|= \\
&&e^{\ \psi (\tau )}\frac{|\ _{2}\Phi (\tau )\partial _{3}[\ _{2}\Phi (\tau
)]^{2}|}{|\ _{2}\Lambda _{0}(\tau )\int dy^{3}\ _{2}\mathcal{K}(\tau
)\{\partial _{3}[\ _{2}\Phi (\tau )]^{2}\}^{2}|^{1/2}\ }\times \\
&&\frac{|\ _{3}^{\shortmid }\Phi (\tau )\ ^{\shortmid }\partial ^{6}[\
_{3}^{\shortmid }\Phi (\tau )]^{2}|}{| \ _{3}^{\shortmid }\Lambda _{0}(\tau
)\int d\ ^{\shortmid }p_{6} \ _{3}^{\shortmid }\mathcal{K}(\tau )\{\
^{\shortmid }\partial ^{6}[ \ _{3}^{\shortmid }\Phi (\tau
)]^{2}\}^{2}|^{1/2}\ }\times \frac{|\ _{4}^{\shortmid }\Phi (\tau )\
^{\shortmid }\partial ^{7}[\ _{4}^{\shortmid }\Phi (\tau )]^{2}|}{|\
_{4}^{\shortmid }\Lambda _{0}(\tau )\int d\ ^{\shortmid }p_{7} \ \
_{4}^{\shortmid }\mathcal{K}(\tau )\{\ ^{\shortmid }\partial ^{7}[ \
_{4}^{\shortmid }\Phi (\tau )]^{2}\}^{2}|^{1/2}\ }.
\end{eqnarray*}%
Using the assumptions as in footnote \ref{fn12}, we find that the $\ _{s}^{\shortmid }\delta ^{8}\ _{s}^{\shortmid }u(\tau )$ from \eqref{wn} is
\begin{eqnarray*}
\ _{s}^{\shortmid }\delta ^{8}\ _{s}^{\shortmid }u(\tau )
&=&dx^{1}dx^{2}\delta u^{3}(\tau )\ \delta u^{4}(\tau )\ ^{\shortmid }\delta
\ ^{\shortmid }u_{5}(\tau )\ ^{\shortmid }\delta \ ^{\shortmid }u_{6}(\tau)\
^{\shortmid }\delta \ ^{\shortmid }u_{7}(\tau )\ ^{\shortmid }\delta \
^{\shortmid }u_{8}(\tau ) \\
&=&dx^{1}dx^{2}[dy^{3}+w_{i_{1}}(\tau )dx^{i_{1}}]dtdp_{5}[dp_{6}+\
^{\shortmid }w_{i_{2}}(\tau )dx^{i_{2}}][dp_{7}+\ ^{\shortmid}w_{i_{3}}(\tau
)d\ ^{\shortmid }x^{i_{3}}]dE.
\end{eqnarray*}%
In these formulas, the shell N-connection coefficients are split into $w_{{i}_{1,2,3}}$ and $n_{{i}_{1,2,3}}$ as:
\begin{eqnarray*}
N_{i_{1}}^{a_{2}}(\tau ) &\to&\left[ w_{i_{1}}(\tau )=\frac{\partial _{i_{1}}\left(
\int dy^{3}\ \ _{2}\mathcal{K}(\tau )\partial _{3}[\ _{2}\Phi (\tau
)]^{2}\right) }{\ _{2}\mathcal{K}(\tau )\partial _{3}[\ _{2}\Phi (\tau )]^{2}%
} ~~~,~~~ n_{i_{1}}(\tau )=0 \right], \\
\ ^{\shortmid }N_{i_{2}a_{3}}(\tau ) &\to& \left[ \ ^{\shortmid }n_{i_{2}}(\tau )=0 ~~~,~~~\
^{\shortmid }w_{i_{2}}(\tau )=\frac{\ ^{\shortmid }\partial _{i_{2}}\left(
\int dp_{6}\ \ _{3}^{\shortmid }\mathcal{K}(\tau )\ ^{\shortmid }\partial
^{6}[\ _{3}^{\shortmid }\Phi (\tau )]^{2}\right) }{\ \ _{3}^{\shortmid }%
\mathcal{K}(\tau )\ ^{\shortmid }\partial ^{6}[\ _{3}^{\shortmid }\Phi (\tau
)]^{2}} \right], \\
\ ^{\shortmid }N_{i_{3}a_{5}}(\tau ) &\to& \left[\ ^{\shortmid }w_{i_{3}}(\tau )=%
\frac{\ ^{\shortmid }\partial _{i_{3}}\left( \int dp_{7}\ \ _{4}^{\shortmid }%
\mathcal{K}(\tau )\ ^{\shortmid }\partial ^{7}[\ _{4}^{\shortmid }\Phi (\tau
)]^{2}\right) }{\ \ _{4}^{\shortmid }\mathcal{K}(\tau )\ ^{\shortmid
}\partial ^{7}[\ _{4}^{\shortmid }\Phi (\tau )]^{2}} ~~~,~~~~\ ^{\shortmid
}n_{i_{4}}(\tau )=0 \right].
\end{eqnarray*}

For the above s-connection and N-connection coefficients and N-elongated
differentials, the effective volume integration functional is
\begin{eqnarray}
\ ^{\shortmid }\delta \ ^{\shortmid }\mathcal{V}(\tau ) &=&e^{\psi (\tau )}%
\frac{|\ _{2}\Phi (\tau )\partial _{3}[\ _{2}\Phi (\tau )]^{2}|}{|\
_{2}\Lambda _{0}(\tau ) \int dy^{3}\ _{2}\mathcal{K}(\tau )\{\partial _{3}[\
_{2}\Phi (\tau )]^{2}\}^{2}|^{1/2}\ }\left[dy^{3}+\frac{\partial _{i_{1}}\left(
\int dy^{3}\ _{2}\mathcal{K}(\tau )\partial _{3}[\ _{2}\Phi
(\tau)]^{2}\right) }{\ _{2}\mathcal{K}(\tau ) \partial _{3}[\ _{2}\Phi (\tau
)]^{2}}dx^{i_{1}} \right]dx^{1}dx^{2}dt  \notag \\
&&\frac{|\ _{3}^{\shortmid }\Phi (\tau )\ ^{\shortmid }\partial ^{6}[\
_{3}^{\shortmid }\Phi (\tau )]^{2}|}{\ _{3}^{\shortmid }\Lambda
_{0}(\tau)\int d\ ^{\shortmid }p_{6} \ _{3}^{\shortmid }\mathcal{K}(\tau )\
^{\shortmid }\partial ^{6}[\ _{3}^{\shortmid }\Phi (\tau )]^{2}|^{1/2}\ }%
\left[dp_{6}+\frac{\ ^{\shortmid }\partial _{i_{2}}\left( \int dp_{6} \
_{3}^{\shortmid }\mathcal{K}(\tau )\ ^{\shortmid }\partial ^{6}[\
_{3}^{\shortmid }\Phi (\tau )]^{2}\right) } {\ _{3}^{\shortmid }\mathcal{K}%
(\tau )\ ^{\shortmid }\partial ^{6}[\ _{3}^{\shortmid }\Phi (\tau )]^{2}}%
dx^{i_{2}}\right]  \label{volumf} \\
&&\frac{|\ _{4}^{\shortmid }\Phi (\tau )\ ^{\shortmid }\partial ^{7}[\
_{4}^{\shortmid }\Phi (\tau )]^{2}|} {|\ _{4}^{\shortmid }\Lambda
_{0}(\tau)\int d\ ^{\shortmid }p_{7} \ \ _{4}^{\shortmid }\mathcal{K}(\tau )
\ ^{\shortmid }\partial ^{7}[\ \ _{4}^{\shortmid }\Phi (\tau )]^{2}|^{1/2}\ }%
\left[dp_{7}+\frac{\ ^{\shortmid }\partial _{i_{3}}\left( \int dp_{7} \
_{4}^{\shortmid }\mathcal{K}(\tau ) \ ^{\shortmid }\partial ^{7}[\
_{4}^{\shortmid }\Phi (\tau )]^{2}\right) }{\ \ _{4}^{\shortmid }\mathcal{K}%
(\tau )\ ^{\shortmid }\partial ^{7}[\ _{4}^{\shortmid }\Phi (\tau )]^{2}}
\right]dp_{5}dE.  \notag
\end{eqnarray}%
This volume integation functional allows us to compute all statistical and geometric
variables by integrating over the $\tau$ parameter and phase space
coordinates. It can be restricted to LC-configurations by imposing
additional nonholonomic constraints. 

\subsubsection{Perelman thermodynamic variables for quasi-stationary
configurations and effective cosmological constants}

The thermodynamic variables for a fixed temperature parameter $\tau _{0}$
and with $\ ^{\shortmid }\delta \ ^{\shortmid }\mathcal{V}(\tau )$ computed for
quasi-stationary solutions are expressed in the form {\small
\begin{eqnarray}
\ _{s}^{\shortmid }\widehat{\mathcal{W}} &=&\int\nolimits_{\tau ^{\prime
}}^{\tau _{0}}\frac{d\tau }{(4\pi \tau )^{4}}\int_{\ ^{\shortmid }\widehat{%
\Xi }}\left( \tau \lbrack \sum\nolimits_{s}\ _{s}^{\shortmid }\Lambda (\tau
)]^{2}-8\right) \ ^{\shortmid }\delta \ ^{\shortmid }\mathcal{V}(\tau
),\quad \ _{s}^{\shortmid }\widehat{\mathcal{Z}}=\exp \left[
\int\nolimits_{\tau ^{\prime }}^{\tau _{0}}\frac{d\tau }{(2\pi \tau )^{4}}%
\int_{\ ^{\shortmid }\widehat{\Xi }}\ ^{\shortmid }\delta \ ^{\shortmid }%
\mathcal{V}(\tau )\right]  \label{thvcann} \\
\ _{s}^{\shortmid }\widehat{\mathcal{E}} &=&-\int\nolimits_{\tau ^{\prime
}}^{\tau _{0}}\frac{d\tau }{(4\pi )^{4}\tau ^{2}}\int_{\ ^{\shortmid }%
\widehat{\Xi }}\left( [\sum\nolimits_{s}\ _{s}^{\shortmid }\Lambda (\tau )]-%
\frac{4}{\tau }\right) \ ^{\shortmid }\delta \ ^{\shortmid }\mathcal{V}(\tau
),\ _{s}^{\shortmid }\widehat{\mathcal{S}}=-\int\nolimits_{\tau ^{\prime
}}^{\tau _{0}}\frac{d\tau }{(4\pi \tau )^{4}}\int_{\ ^{\shortmid }\widehat{%
\Xi }}\left( \tau \lbrack \sum\nolimits_{s}\ _{s}^{\shortmid }\Lambda (\tau
)]-8\right) \ ^{\shortmid }\delta \ ^{\shortmid }\mathcal{V}(\tau ).  \notag
\end{eqnarray}%
} These integrals can be computed for quasi-stationary solutions of the
generalized vacuum equations, with the s-metrics being $\tau$-dependent. 
We consider only well-defined thermodynamic variables on
certain classes of spacetimes. To study possible physical applications, we set $\
_{s}^{\shortmid }\Lambda (\tau )=\ _{s}^{\shortmid }\Lambda _{0}$.

\subsubsection{Effective volume elements for quasi-stationary generating
functions \& sources}

The effective volume element (\ref{volumf}) is a parametric functional $\
^{\shortmid }\delta \ ^{\shortmid }\mathcal{V[}\tau , \
_{s}^{\shortmid}\Lambda (\tau ),\ _{s}^{\shortmid }\mathcal{K}(\tau ); \psi
(\tau ),\ _{s}^{\shortmid }\Phi (\tau )]$ which allows us to compute
thermodynamic variables (\ref{thvcann}) for a corresponding class of
generating functions $\ _{s}^{\shortmid }\Phi (\tau )$. Using the nonlinear
symmetries of (\ref{nonlinsym}) and the transforms of (\ref{nonltransf}), we can
redefine the generating functions to give
deformations of the prime s-metrics into target quasi-stationary metric, $\
_{s}^{\shortmid }\mathbf{\mathring{g}}(\tau )\rightarrow \ _{s}^{\shortmid }%
\mathbf{g}(\tau )$ (\ref{offdiagdef}). For the $\eta $-polarization functions
and with decompositions on the small parameter $\kappa ,$ we find
\begin{eqnarray}
\ _{2}\Phi (\tau ) &=&2\sqrt{|\ _{2}\Lambda (\tau )\ g_{4}(\tau )|}=\ 2\sqrt{%
|\ _{2}\Lambda (\tau )\ \eta _{4}(\tau )\mathring{g}_{4}(\tau )|}\simeq 2%
\sqrt{|\ _{2}\Lambda (\tau )\ \zeta _{4}(\tau )\mathring{g}_{4}(\tau )|}[1-%
\frac{\kappa }{2}\chi _{4}(\tau )],  \label{genf1} \\
\ \ _{3}^{\shortmid }\Phi (\tau ) &=&2\sqrt{|\ _{3}^{\shortmid }\Lambda
(\tau )~^{\shortmid }g^{6}(\tau )|}=\ 2\sqrt{|\ _{3}^{\shortmid }\Lambda
(\tau )~^{\shortmid }\eta ^{6}(\tau )~^{\shortmid }\mathring{g}^{6}(\tau )|}%
\simeq 2\sqrt{|\ _{3}^{\shortmid }\Lambda (\tau )~^{\shortmid }\zeta
^{6}(\tau )~^{\shortmid }\mathring{g}^{6}(\tau )|}[1-\frac{\kappa }{2}%
~^{\shortmid }\chi ^{6}(\tau )],  \notag \\
\ \ _{4}^{\shortmid }\Phi (\tau ) &=&2\sqrt{|\ _{4}^{\shortmid }\Lambda
(\tau )~^{\shortmid }g^{7}(\tau )|}=\ 2\sqrt{|\ _{4}^{\shortmid }\Lambda
(\tau )~^{\shortmid }\eta ^{7}(\tau )~^{\shortmid }\mathring{g}^{7}(\tau )|}%
\simeq 2\sqrt{|\ _{4}^{\shortmid }\Lambda (\tau )~^{\shortmid }\zeta
^{7}(\tau )~^{\shortmid }\mathring{g}^{7}(\tau )|}[1-\frac{\kappa }{2}%
~^{\shortmid }\chi ^{7}(\tau )].  \notag
\end{eqnarray}%
Inserting the generating functions (\ref{genf1}) into (\ref{volumf}), we find representations of the effective volume functionals
\begin{eqnarray}
\ ^{\shortmid }\delta \ ^{\shortmid }\mathcal{V} &\mathcal{=}&\ ^{\shortmid
}\delta \ ^{\shortmid }\mathcal{V[}\tau ,\ _{s}^{\shortmid }\Lambda (\tau
),\ \ _{s}^{\shortmid }\mathcal{K}(\tau );\psi (\tau ),\ g_{4}(\tau
),~^{\shortmid }g^{6}(\tau ),~^{\shortmid }g^{7}(\tau )]  \label{volumfd} \\
&=&\ ^{\shortmid }\delta \ ^{\shortmid }\mathcal{V[}\tau ,\ _{s}^{\shortmid
}\Lambda (\tau ),\ \ _{s}^{\shortmid }\mathcal{K}(\tau );\psi (\tau ),\
\mathring{g}_{4}(\tau ),~^{\shortmid }\mathring{g}^{6}(\tau ),~^{\shortmid }%
\mathring{g}^{7}(\tau );\eta _{4}(\tau ),~^{\shortmid }\eta ^{6}(\tau
),~^{\shortmid }\eta ^{8}(\tau )]  \notag \\
&=&\ ^{\shortmid }\delta \ ^{\shortmid }\mathcal{V}_{0}\mathcal{[}\tau ,\
_{s}^{\shortmid }\Lambda (\tau ),\ \ _{s}^{\shortmid }\mathcal{K}(\tau
);\psi (\tau ),\ \mathring{g}_{4}(\tau ),~^{\shortmid }\mathring{g}^{6}(\tau
),~^{\shortmid }\mathring{g}^{7}(\tau );\zeta _{4}(\tau ),~^{\shortmid
}\zeta ^{6}(\tau ),~^{\shortmid }\zeta ^{8}(\tau )]+  \notag \\
&&\kappa \ ^{\shortmid }\delta \ ^{\shortmid }\mathcal{V}_{1}\mathcal{[}\tau
,\ _{s}^{\shortmid }\Lambda (\tau ),\ \ _{s}^{\shortmid }\mathcal{K}(\tau
);\psi (\tau ),\ \mathring{g}_{4}(\tau ),~^{\shortmid }\mathring{g}^{6}(\tau
),~^{\shortmid }\mathring{g}^{7}(\tau );\zeta _{4}(\tau ),~^{\shortmid
}\zeta ^{6}(\tau ),~^{\shortmid }\zeta ^{8}(\tau ),\chi _{4}(\tau
),~^{\shortmid }\chi ^{6}(\tau ),~^{\shortmid }\chi ^{8}(\tau )].  \notag
\end{eqnarray}%
Under the above linear $\kappa $--decomposition, with $\
^{\shortmid }\delta \ ^{\shortmid }\mathcal{V}=\ ^{\shortmid }\delta \
^{\shortmid }\mathcal{V}_{0}+ \kappa \ ^{\shortmid }\delta \ ^{\shortmid }%
\mathcal{V}_{1}$, the first term describes a re-definition of the prime
thermodynamic vacuum via $\zeta $-terms and the second term describes
contributions of small $\chi $-terms. This linear $\kappa $--decomposition 
of the effective volume functionals also gives a linear $%
\kappa $--decomposition of the thermodynamic variables from (%
\ref{thvcann}),%
\begin{equation}
\ _{s}^{\shortmid }\widehat{\mathcal{W}}=\ _{s}^{\shortmid }\widehat{%
\mathcal{W}}_{0}+\kappa \ _{s}^{\shortmid }\widehat{\mathcal{W}}_{1}~~,~~\
_{s}^{\shortmid }\widehat{\mathcal{Z}}=\ _{s}^{\shortmid }\widehat{\mathcal{Z%
}}_{0}\ _{s}^{\shortmid }\widehat{\mathcal{Z}}_{1} ~~,~~\ _{s}^{\shortmid }%
\widehat{\mathcal{E}}=\ _{s}^{\shortmid }\widehat{\mathcal{E}}_{0}+\kappa \
_{s}^{\shortmid }\widehat{\mathcal{E}}_{1} ~~,~~\ _{s}^{\shortmid }\widehat{%
\mathcal{S}}=\ _{s}^{\shortmid }\widehat{\mathcal{S}}_{0}+\kappa \
_{s}^{\shortmid }\widehat{\mathcal{S}}_{1}.  \label{klinthvar}
\end{equation}%
These values may be not well-defined in the framework of a relativistic
thermodynamic model for some solutions. We consider such
solutions as unphysical. 

\subsection{Examples of nonassociative BH deformations and Perelman's
thermodynamics}

\label{extherm} In this section, the generating and integration functions
and effective sources are given in a form so that the statistical
geometric thermodynamic variables are determined by the corresponding effective
volume functionals and fixed effective cosmological constants. We show how
these values are computed in parametric form for nonassociative R-flux,
phase space Tangerlini BHs and double 4-d Schwarzschild BHs, AdS BHs and dS BHs.

\subsubsection{The thermodynamic variables for quasi-stationary R-flux
deformed Tangerlini BHs}

Expressing the effective volume (\ref{volumfd}) in terms of $\eta $%
-polarizations and prime metrics \newline
$\ _{h}^{\shortmid }\mathring{g}_{\alpha _{s}}[\ ^{\shortmid }u^{\beta
_{s}}(\ ^{\shortmid }r,\theta ,\varphi ,\theta _{2},\theta _{3},p_{7})]$ (%
\ref{primetangd}) for generating functions (\ref{genf1}), we get the
volume functional
\begin{eqnarray*}
\ ^{\shortmid }\delta \ _{h}^{\shortmid }\mathcal{V} &=&\ ^{\shortmid
}\delta \ ^{\shortmid }\mathcal{V[}\tau ,\ _{s}^{\shortmid }\Lambda (\tau
),\ _{s}^{\shortmid }\mathcal{K}(\tau ,\ ^{\shortmid }u^{\beta _{s}}(\
^{\shortmid }r,\theta ,\varphi ,\theta _{2},\theta _{3},p_{7}));\psi (\tau
,\ ^{\shortmid }r,\theta ,),\ \mathring{g}_{4}(\tau ,\ ^{\shortmid }r,\theta
,\varphi ),\ ^{\shortmid }\mathring{g}^{6}(\tau ,\ ^{\shortmid }r,\theta
,\varphi ,\theta _{2}), \\
&&\ ^{\shortmid }\mathring{g}^{7}(\tau ,\ ^{\shortmid }r,\theta ,\varphi
,\theta _{2},\theta _{3},p_{7});\eta _{4}(\tau ,\ ^{\shortmid }r,\theta
,\varphi ),\ ^{\shortmid }\eta ^{6}(\tau ,\ ^{\shortmid }r,\theta ,\varphi
,\theta _{2}),\ ^{\shortmid }\eta ^{8}(\tau ,\ ^{\shortmid }r,\theta
,\varphi ,\theta _{2},\theta _{3},p_{7})]
\end{eqnarray*}%
We can chose solutions with $\ _{s}^{\shortmid }\Lambda (\tau )=\
_{s}^{\shortmid }\Lambda _{0}=\Lambda _{0}$ and write the formulas for
thermodynamic variables (\ref{thvcann}) in the form
\begin{eqnarray}
\ _{s}^{\shortmid }\widehat{\mathcal{W}}_{[h\eta ]} &=&\int\nolimits_{\tau
^{\prime }}^{\tau _{0}}\frac{d\tau }{32(\pi \tau )^{4}}\int_{\
_{s}^{\shortmid }\widehat{\Xi }}\left( \tau \lbrack \Lambda
_{0}^{2}-1\right) \ ^{\shortmid }\delta \ _{h}^{\shortmid }\mathcal{V}(\tau
),\ \ _{s}^{\shortmid }\widehat{\mathcal{Z}}_{[h\eta ]}=\exp \left[
\int\nolimits_{\tau ^{\prime }}^{\tau _{0}}\frac{d\tau }{(2\pi \tau )^{4}}%
\int_{\ _{s}^{\shortmid }\widehat{\Xi }}\ ^{\shortmid }\delta \
_{h}^{\shortmid }\mathcal{V}(\tau )\right]  \label{thvcanntbh} \\
\ _{s}^{\shortmid }\widehat{\mathcal{E}}_{[h\eta ]} &=&-\int\nolimits_{\tau
^{\prime }}^{\tau _{0}}\frac{d\tau }{64(\pi )^{4}\tau ^{2}}\int_{\
_{s}^{\shortmid }\widehat{\Xi }}\left( [\Lambda _{0}-\frac{1}{\tau }\right)
\ ^{\shortmid }\delta \ _{h}^{\shortmid }\mathcal{V}(\tau ),\ \
_{s}^{\shortmid }\widehat{\mathcal{S}}_{[h\eta ]}=-\int\nolimits_{\tau
^{\prime }}^{\tau _{0}}\frac{d\tau }{32(\pi \tau )^{4}}\int_{\
_{s}^{\shortmid }\widehat{\Xi }}\left( \tau \Lambda _{0}-1\right) \
^{\shortmid }\delta \ _{h}^{\shortmid }\mathcal{V}(\tau ).  \notag
\end{eqnarray}%
The above quantities encode a nonassociative R-flux into $\ ^{\shortmid }\delta \
_{h}^{\shortmid }\mathcal{V}.$ Prescribing generating sources $\
_{s}^{\shortmid }\mathcal{K},$ we construct statistical thermodynamic models
for nonholonomic deformations of phase Thangerlini BHs. The thermodynamic
variables computed for $\ ^{\shortmid }\delta \ _{h}^{\shortmid }\mathcal{V}$
are functionals of generating functions $\eta _{4},\ ^{\shortmid }\eta ^{6}$
and $\ ^{\shortmid }\eta ^{8},$ which also define respective families of
parametric solutions with s-metrics (\ref{offtangfix}). The $\eta $%
-generating functions also determine symmetric, $\
_{\star }^{\shortparallel }\mathbf{\check{q}}_{\alpha _{s}\beta _{s}},$ and
antisymmetric, $\ _{\star }^{\shortparallel }\mathbf{a}_{\alpha _{s}\beta
_{s}},$ s-metrics (see formulas (\ref{aux40b}) and (\ref{aux40aa}))
characterizing the nonassociative properties of phase space MGTs.

\subsubsection{Phase space thermodynamics for quasi-stationary R-flux
deformed double 4-d BHs}

Here the prime s-metrics are taken in the form $\ \ ^{\shortmid }%
\mathring{g}_{\alpha _{s}}(\tau ,\ ^{\shortmid }u^{\beta _{s}})$ (\ref%
{pr2bhdata}) instead of $\ ^{\shortmid }\mathring{g}_{\alpha _{s}}(\tau ,\
^{\shortmid }u^{\beta _{s}})$ (\ref{primetangd}). We can compute the Perelman thermodynamic variables
if the nonassociative R-flux $\kappa $-deformations result in
target s-metrics and N-connection coefficients $\ _{\star }^{\shortmid }%
\mathbf{g}_{\tau _{s}\alpha _{s}}$ given by (\ref{2bhscoef}) and (\ref%
{2bhnscoef}). For $\chi $-polarization functions we have,
\begin{eqnarray}
\ _{2}\Phi (\tau ) &=&2\sqrt{|\ _{2}\Lambda (\tau )\ g_{4}(\tau )|}=\ 2\sqrt{%
|\ _{h}\Lambda _{0}\ \eta _{4}(\tau )\ \mathring{g}_{4}|}\simeq 2\sqrt{|\
_{h}\Lambda _{0}\ \zeta _{4}(\tau )\ \mathring{g}_{4}|}[1-\frac{\kappa }{2}%
\chi _{4}(\tau ,r,\theta ,\varphi )],  \label{genf2} \\
\ _{3}^{\shortmid }\Phi (\tau ) &=&2\sqrt{|\ _{3}^{\shortmid }\Lambda (\tau
)\ ^{\shortmid }g^{6}(\tau )|}=\ 2\sqrt{|\ _{c}\Lambda _{0}\ ^{\shortmid
}\eta ^{6}(\tau )\ \ ^{\shortmid }\mathring{g}^{6}|}\simeq 2\sqrt{|\
_{c}\Lambda _{0}\ ^{\shortmid }\zeta ^{6}(\tau )\ \ ^{\shortmid }\mathring{g}%
^{6}|}[1-\frac{\kappa }{2}\ ^{\shortmid }\chi ^{6}(\tau ,r,\theta ,\varphi
,\ _{\shortmid }^{p}r,\ _{\shortmid }^{p}\theta )],  \notag \\
\ _{4}^{\shortmid }\Phi (\tau ) &=&2\sqrt{|\ _{4}^{\shortmid }\Lambda (\tau
)\ ^{\shortmid }g^{7}(\tau )|}=\ 2\sqrt{|\ _{c}\Lambda \ ^{\shortmid }\eta
^{7}(\tau )\ \ ^{\shortmid }\mathring{g}^{7}|}\simeq  \notag \\
&&2\sqrt{|\ _{c}\Lambda _{0}~^{\shortmid }\zeta ^{7}(\tau )\ _{\circ \circ
}^{\shortmid }g^{7}|}[1-\frac{\kappa }{2}\ ^{\shortmid }\chi ^{7}(\tau
,r,\theta ,\varphi ;\ _{\shortmid }^{p}r,\ _{\shortmid }^{p}\theta ,\
_{\shortmid }^{p}\varphi )],  \notag
\end{eqnarray}%
where $\ _{1}\Lambda (\tau )=\ _{2}\Lambda (\tau )\simeq \ _{h}\Lambda
_{0}=const,$ $\ _{3}\Lambda (\tau )=\ _{4}\Lambda (\tau )\simeq \
_{c}\Lambda _{0}=const$ and \newline
$~\ ^{\shortmid }\mathring{g}_{\alpha _{s}}(\tau ,\ ^{\shortmid }u^{\beta
_{s}})$ $\simeq ~\ ^{\shortmid }\mathring{g}_{\alpha _{s}}(\ ^{\shortmid
}u^{\beta _{s}}(r,\theta ,\varphi ;\ _{\shortmid }^{p}r,\ _{\shortmid
}^{p}\theta ,\ _{\shortmid }^{p}\varphi ).$

The effective volume functional generated by (\ref{genf2}) is of the form $\
^{\shortmid }\delta \ ^{\shortmid }\mathcal{V=}\ ^{\shortmid }\delta \
^{\shortmid }\mathcal{V}_{0}+\kappa \ ^{\shortmid }\delta \ ^{\shortmid }%
\mathcal{V}_{1}$ (\ref{2bhscoef}), with
\begin{eqnarray}
&&\ ^{\shortmid }\delta \ _{0}^{\shortmid }\mathcal{V}=\ ^{\shortmid }\delta
\ _{0}^{\shortmid }\mathcal{V}[\tau ,\ _{h}\Lambda _{0},\ _{c}\Lambda _{0},\
_{s}^{\shortmid }\mathcal{K}(\tau ,\ ^{\shortmid }u^{\beta _{s}}(r,\theta
,\varphi ;\ _{\shortmid }^{p}r,\ _{\shortmid }^{p}\theta ,\ _{\shortmid
}^{p}\varphi ));\psi (\tau ,r,\theta ,),\ ^{\shortmid }\mathring{g}%
_{4}(r,\theta ,\varphi ),\ ^{\shortmid }\mathring{g}^{6}(r,\theta ,\varphi
;\ _{\shortmid }^{p}r,\ _{\shortmid }^{p}\theta ),  \notag \\
&&\ \ ^{\shortmid }\mathring{g}^{7}(r,\theta ,\varphi ;\ _{\shortmid
}^{p}r,\ _{\shortmid }^{p}\theta ,\ _{\shortmid }^{p}\varphi );\zeta
_{4}(\tau ,r,\theta ,\varphi ),\ ^{\shortmid }\zeta ^{6}(\tau ,r,\theta
,\varphi ;\ _{\shortmid }^{p}r,\ _{\shortmid }^{p}\theta ),\ ^{\shortmid
}\zeta ^{8}(\tau ,r,\theta ,\varphi ;\ _{\shortmid }^{p}r,\ _{\shortmid
}^{p}\theta ,\ _{\shortmid }^{p}\varphi )]  \notag \\
&&\mbox{ and }  \notag \\
&&\ ^{\shortmid }\delta \ _{1}^{\shortmid }\mathcal{V}=\ ^{\shortmid }\delta
\ _{1}^{\shortmid }\mathcal{V}[\tau ,\ _{h}\Lambda _{0},\ _{c}\Lambda _{0},\
_{s}^{\shortmid }\mathcal{K}(\tau ,\ ^{\shortmid }u^{\beta _{s}}(r,\theta
,\varphi ;\ _{\shortmid }^{p}r,\ _{\shortmid }^{p}\theta ,\ _{\shortmid
}^{p}\varphi ));\psi (\tau ,r,\theta ,),\ \mathring{g}_{4}(r,\theta ,\varphi
),\ \ ^{\shortmid }\mathring{g}^{6}(r,\theta ,\varphi ;\ _{\shortmid
}^{p}r,\ _{\shortmid }^{p}\theta ),  \notag \\
&&\ \ ^{\shortmid }\mathring{g}^{7}(r,\theta ,\varphi ;\ _{\shortmid
}^{p}r,\ _{\shortmid }^{p}\theta ,\ _{\shortmid }^{p}\varphi );\zeta
_{4}(\tau ,r,\theta ,\varphi ),\ ^{\shortmid }\zeta ^{6}(\tau ,r,\theta
,\varphi ;\ _{\shortmid }^{p}r,\ _{\shortmid }^{p}\theta ),\ ^{\shortmid
}\zeta ^{8}(\tau ,r,\theta ,\varphi ;\ _{\shortmid }^{p}r,\ _{\shortmid
}^{p}\theta ,\ _{\shortmid }^{p}\varphi );  \notag \\
&&\chi _{4}(\tau ,r,\theta ,\varphi ),\ ^{\shortmid }\chi ^{6}(\tau
,r,\theta ,\varphi ;\ _{\shortmid }^{p}r,\ _{\shortmid }^{p}\theta ),\
^{\shortmid }\chi ^{8}(\tau ,r,\theta ,\varphi ;\ _{\shortmid }^{p}r,\
_{\shortmid }^{p}\theta ,\ _{\shortmid }^{p}\varphi )].  \label{volec2bh}
\end{eqnarray}

Inserting the effective volume elements (\ref{volec2bh}) into (\ref%
{thvcann}), we compute the linear $\kappa $-decomposition of the thermodynamic variables (\ref{klinthvar}) as,%
\begin{eqnarray}
\ _{s}^{\shortmid }\widehat{\mathcal{W}}^{[dbh]}&=&\ _{s}^{\shortmid }%
\widehat{\mathcal{W}}_{0}^{[dbh]}+\kappa \ _{s}^{\shortmid }\widehat{%
\mathcal{W}}_{1}^{[dbh]}[\chi _{4},~^{\shortmid }\chi ^{6},~~^{\shortmid
}\chi ^{8}]  \label{thvcandbh} \\
&=&\int\nolimits_{\tau ^{\prime }}^{\tau _{0}}\frac{d\tau }{16(\pi \tau )^{4}%
}\int_{\ _{s}^{\shortmid }\widehat{\Xi }}\left( \tau \lbrack \ _{h}\Lambda
_{0}+\ _{c}\Lambda _{0}]^{2}-2\right) \ ^{\shortmid }\delta \
_{0}^{\shortmid }\mathcal{V}(\tau )  \notag \\
&&+\kappa \int\nolimits_{\tau ^{\prime }}^{\tau _{0}}\frac{d\tau }{16(\pi
\tau )^{4}}\int_{\ _{s}^{\shortmid }\widehat{\Xi }}\left( \tau \lbrack \
_{h}\Lambda _{0}+\ _{c}\Lambda _{0}]^{2}-2\right) \ ^{\shortmid }\delta \
_{1}^{\shortmid }\mathcal{V}(\tau ,\chi _{4},~^{\shortmid }\chi
^{6},~~^{\shortmid }\chi ^{8}),  \notag
\end{eqnarray}%
\begin{eqnarray*}
\ _{s}^{\shortmid }\widehat{\mathcal{Z}}^{[dbh]} &=&\ _{s}^{\shortmid }%
\widehat{\mathcal{Z}}_{0}^{[dbh]}\times \ _{s}^{\shortmid }\widehat{\mathcal{%
Z}}_{1}^{[dbh]} \\
&=&\exp [\int\nolimits_{\tau ^{\prime }}^{\tau _{0}}\frac{d\tau }{(2\pi \tau
)^{4}}\int_{\ _{s}^{\shortmid }\widehat{\Xi }}\ ^{\shortmid }\delta \
_{0}^{\shortmid }\mathcal{V}(\tau )]\times \exp [\kappa \int\nolimits_{\tau
^{\prime }}^{\tau _{0}}\frac{d\tau }{(2\pi \tau )^{4}}\int_{\
_{s}^{\shortmid }\widehat{\Xi }}\ ^{\shortmid }\delta \ _{1}^{\shortmid }%
\mathcal{V}(\tau ,\chi _{4},~^{\shortmid }\chi ^{6},~~^{\shortmid }\chi
^{8})],
\end{eqnarray*}%
\begin{eqnarray*}
\ _{s}^{\shortmid }\widehat{\mathcal{E}}^{[dbh]} &=&\ _{s}^{\shortmid }%
\widehat{\mathcal{E}}_{0}^{[dbh]}+\kappa \ _{s}^{\shortmid }\widehat{%
\mathcal{E}}_{1}^{[dbh]}[\chi _{4},~^{\shortmid }\chi ^{6},~~^{\shortmid
}\chi ^{8}] \\
&=&-\int\nolimits_{\tau ^{\prime }}^{\tau _{0}}\frac{d\tau }{128\pi ^{4}\tau
^{2}}\int_{\ _{s}^{\shortmid }\widehat{\Xi }}\left( \ _{h}\Lambda _{0}+\
_{c}\Lambda _{0}-\frac{2}{\tau }\right) \ ^{\shortmid }\delta \
_{0}^{\shortmid }\mathcal{V}(\tau ) \\
&&-\kappa \int\nolimits_{\tau ^{\prime }}^{\tau _{0}}\frac{d\tau }{128\pi
^{4}\tau ^{2}}\int_{\ _{s}^{\shortmid }\widehat{\Xi }}\left( \ _{h}\Lambda
_{0}+\ _{c}\Lambda _{0}-\frac{2}{\tau }\right) \ ^{\shortmid }\delta \
_{1}^{\shortmid }\mathcal{V}(\tau ,\chi _{4},~^{\shortmid }\chi
^{6},~~^{\shortmid }\chi ^{8}),
\end{eqnarray*}%
\begin{eqnarray*}
_{s}^{\shortmid }\widehat{\mathcal{S}}^{[dbh]} &=&\ _{s}^{\shortmid }%
\widehat{\mathcal{S}}_{0}^{[dbh]}+\kappa \ _{s}^{\shortmid }\widehat{%
\mathcal{S}}_{1}^{[dbh]}[\chi _{4},~^{\shortmid }\chi ^{6},~~^{\shortmid
}\chi ^{8}] \\
&=&-\int\nolimits_{\tau ^{\prime }}^{\tau _{0}}\frac{d\tau }{128(\pi \tau
)^{4}}\int_{\ _{s}^{\shortmid }\widehat{\Xi }}\left( \tau \lbrack \
_{h}\Lambda _{0}+\ _{c}\Lambda _{0}]-4\right) \ ^{\shortmid }\delta \
_{0}^{\shortmid }\mathcal{V}(\tau ) \\
&&-\kappa \int\nolimits_{\tau ^{\prime }}^{\tau _{0}}\frac{d\tau }{128\pi
^{4}\tau ^{2}}\int_{\ _{s}^{\shortmid }\widehat{\Xi }}\left( \tau \lbrack \
_{h}\Lambda _{0}+\ _{c}\Lambda _{0}]-4\right) \ ^{\shortmid }\delta \
_{1}^{\shortmid }\mathcal{V}(\tau ,\chi _{4},~^{\shortmid }\chi
^{6},~~^{\shortmid }\chi ^{8}).
\end{eqnarray*}

The thermodynamic variables in (\ref{thvcanntbh}) and (\ref{thvcandbh}) can be
computed in explicit form by integrating over the respective volume functionals $%
\ ^{\shortmid }\delta \ _{h}^{\shortmid }\mathcal{V}$ and $\ ^{\shortmid
}\delta \ _{0}^{\shortmid }\mathcal{V},\ ^{\shortmid }\delta \
_{1}^{\shortmid }\mathcal{V}.$ We do not provide such cumbersome formulas here. 
The geometric and thermodynamic values/variables provided above
show a very different dependence of the nonholonomic deformations of the phase
space Tangherlini and double BHs with respect to the temperature $\tau ,$ the effective
cosmological constants, and ion the possible terms comign from the generating
functions $\eta _{4},\ ^{\shortmid }\eta ^{6},\ ^{\shortmid }\eta ^{8}$
or linear $\kappa $-deformations by $\chi _{4},\ ^{\shortmid
}\chi ^{6},\ ^{\shortmid }\chi ^{8}.$ Using a nonholonomic, relativistic
generalization of Perelman's thermodynamic approach to geometric flows and
Ricci solitons, we can distinguish possible thermodynamic effects of various
type modifications of phase space higher dimension BHs or double BHs
solutions.

We now discuss some physical implications of (\ref{thvcandbh}) for nonassociative star product and R-flux quasi-stationary
deformations of double BH configurations in phase spaces. For very special classes of generating functions with ellipsoidal symmetry of $\left( \chi
_{4},~^{\shortmid }\chi ^{6},~^{\shortmid }\chi ^{8}\right) $, the primary double BH metrics transform into generic off-diagonal ones  for
double BEs as explained in section \ref{sss412}. In such cases, we can define certain polarized and $\varphi $-anisotropic versions of the
Bekenstein-Hawking temperature and entropy. Nevertheless, more general classes of nonassociative deformations involving momentum type variables and
generic off-diagonal metrics result in double quasi-stationary s-metrics. This leads to a change of the  Bekenstein-Hawking thermodynamic paradigm
into a more general Perelman, type even for small $\kappa $-deformations. The nonlinear symmetries of such solutions, given in formulas
(\ref{nonltransf}), allow one to introduce effective cosmological constants $\ _{h}\Lambda _{0}$ and$\ _{c}\Lambda _{0}.$ Nevertheless, we can not consider solutions like double Schwarzschild - (A)dS configurations if general classes of effective sources are used. So, nonassociative star product and R-flux deformations result in substantial modifications of the base spacetime Lorentz structure if the solutions of nonassociative vacuum gravitational equations contain dependencies on momentum coordinates. We can prescribe the nonholonomic phase space structure when the entropy of $\ _{s}^{\shortmid }\widehat{\mathcal{S}}_{0}^{[dbh]}$ and
$\kappa \ _{s}^{\shortmid }\widehat{\mathcal{S}}_{1}^{[dbh]}$ can be related to $S_{0}(\varphi )$ and
$\ _{\shortmid }S_{0}(\ _{\shortmid }^{p}\varphi )$ from section \ref{sss412}. However, this is true only for very special diagonalizable phase space metrics. Even in such cases, there will be a difference between the geometric flow temperature $\tau $ and the double BE temperature $T(\varphi ).$  Nonassociative deformations of double BH configurations are also characterized by geometric flow thermodynamic energies
$\ _{s}^{\shortmid }\widehat{\mathcal{E}}_{0}^{[dbh]}$ and $\kappa \ _{s}^{\shortmid }\widehat{\mathcal{E}}_{1}^{[dbh]}$ which do
not have analogs in the framework of the Bekenstein-Hawking thermodynamics.

\section{Discussion, Concluding Remarks and Open Questions}

\label{sec5} We begin with a summary and discussion of the main results of
this article. In this work we followed the approach to nonassociative
gravity with $\star $-product and R-flux deformations in string theory \cite%
{blumenhagen16,aschieri17}. This was generalized in nonholonomic form \cite%
{partner01,partner02} with the aim of constructing physically important,
exact solutions. The possibility to decouple and integrate the 
nonassociative vacuum gravitational equations for
quasi-stationary solutions, using the anholonomic frame and connection
deformation method (AFCDM) was given in \cite{partner02}. In another
work \cite{partner03}, we showed how to construct parametric
solutions on a base spacetime manifold defining BEs and encoding
nonassociative R-flux effects.   
In this article, in addition to gaining a more complete understanding of physical effects of
nonassociative R-flux deformations, we also studied 
certain new classes of 8-d parametric quasi-stationary solutions. This included the following new and original results:

\begin{enumerate}
\item In section \ref{sec2} and the Appendix, we outlined the
necessary formulas for the AFCDM which allowed us to construct exact 
8-d solutions, with a fixed energy parameter, in the conventional
co-fiber space of the phase space. This co-fiber space was modeled as a cotangent Lorentz bundle
nonholonomically deformed by a nonassociative star product determined by a
R-flux nongeometric structure. Such nonholonomic geometric constructions are
nonassociative versions of geometrical and physical models from \cite%
{bubuianu19}. The methods and solutions of thre presenrt work
are very different from \cite{bubuianu19} which foused on generalized
``rainbow" configurations with a dependence on a variable energy parameter $%
E $ and on a temperature parameter $\tau $. Here, we
consider fixed values $E_{0}$ and $\tau _{0}$ and study possible
relativistic momentum effects which via nonlinear symmetries (\ref{nonlinsym}) 
and also appendix \ref{apbssns}.

\item In section \ref{tangph} we extended the Tangherlini type 6-d BH solutions \cite{tangherlini63,pappas16} to 8-d phase space. These new solutions
are higher dimensional BHs with a conventional horizon and radius in
phase space and which, via nonlinear symmetries and effective cosmological
constants (see formulas (\ref{nonlinsym}) and appendix \ref{apbssns}),
encode nonassociative data.

\item Another class of solutions for the nonassociative, vacuum gravitational
equations is given in section \ref{prim2bh}. These describe double BHs - a 4-d BH 
on the base spacetime and another 4-d BH in a co-fiber (momentum) space.
These two 4-d BHs are not independent because of the nonlinear symmetries of the
generating functions and the effective sources.

\item In subsection \ref{ssqstfe} off-diagonal R-flux deformations of the prime 8-d metrics 
were given which resulted in quasi-stationary solutions 
with a fixed energy parameter. In
general, it is not clear what physical properties these solutions, which
encoded nonassociative R-flux data, may have.  It is
possible to prove certain stability conditions for such BE solutions.

\item  In section \ref{pthvarqs} we generalized the Bekenstein-Hawking 
entropy approach to phase space BH and BE configurations and speculated how the
corresponding temperature and associated thermodynamic values might encode
nonassociative data. We concluded that this approach was
limited solutions with conventional horizons. For more general configurations in nonassociative
gravity we need a new type of geometric
thermodynamic models, which was based on the concept of Perelman
thermodynamic variables \cite{perelman1}, as well as generalizations of Perelman's thermodynamic 
variables \cite{ibubuianu20,ibubuianu21,rajpoot17,bubuianu19}.  

\item Also in section \ref{pthvarqs}, the concept of W-entropy was generalized to
8-d phase spaces. We showed how a statistical analogy for nonholonomic phase
space Ricci flows can be formulated in order to compute
thermodynamic variables for quasi-stationary solutions of nonassociative
Ricci solitons with fixed temperature parameter $\tau _{0}$. It was
proved that Perelman-like thermodynamic variables for $\kappa $-dependent
solutions are determined by certain temperature and hypersurface
integrals with effective volume elements.

\item Section \ref{extherm} gave two explicit examples of how the Perelman
thermodynamic variables can be computed for general quasi-stationary R-flux
deformations of phase space Tangherlini BHs and deformed double 4-d BHs.
These two configurations are distinguished thermodynamically by different
dependencies on the effective cosmological constants and temperature. This
new geometric thermodynamic physics can not be studied using standard 
Bekenstein-Hawking thermodynamics.
\end{enumerate}

All prime and target phase space solutions constructed in this work are
characterized additionally by nonassociative metrics $\ _{\star
}^{\shortparallel }\mathbf{q}_{\alpha _{s}\beta _{s}}=\ _{\star
}^{\shortparallel }\mathbf{\check{q}}_{\alpha _{s}\beta _{s}}+\ _{\star
}^{\shortparallel }\mathbf{a}_{\alpha _{s}\beta _{s}}$ (\ref{aux40b}) with
symmetric and nonsymmetric components. For quasi-stationary
solutions, such generic off-diagonal metrics can be computed as induced
ones, determined by corresponding symmetric s-metrics. This was proven using
the AFCDM in \cite{partner02,partner03} -- see the respective discussions on
``nonassociative hair" BEs and BHs at the end of subsections
in \ref{ssqstfe}. In general, such target phase space solutions have
nontrivial, nonolonomically induced torsion but we can always extract zero
torsion, LC-solutions. 

But what about the legacy of using Perelman's thermodynamics for Ricci flows
to prove the Poincar\`{e}-Thorston conjecture for geometric flows of Riemannian metrics \cite%
{perelman1}? There are difficulties to proving
certain nonassociative/ noncommutative analogs of the Poincar\`{e}
conjecture in general forms -- see discussions in
\cite{ibubuianu20,ibubuianu21,rajpoot17,bubuianu19}. One could formulate a
number of such topological and geometric models which depend on the type of
nonassociative/noncommutative solutions studied, as in \cite%
{jordan32,jordan34,kurdgelaidze,drinf,okubo,castro1,alvarez06,luest10,vacaru08}.
Various nonassociative theories are different from those with $\star$%
-products and R-fluxes considered in this work. So, it is not clear how to
formulate a general mathematical framework involving fundamental topology
and geometric analysis, for all models of nonassociative
and noncommutative spaces. Nevertheless, self-consistent generalizations of
the statistic and geometric thermodynamics using Perelman's F- and
W-functionals are possible if one considers nonholonomic
deformations with functionals of type (\ref{wfperelmctl}). They encode
nonassociative quantities in $\kappa $-dependent form and result in a Ricci
soliton, \textit{i.e.} modified vacuum Einstein equations, of type (\ref%
{cannonsymparamc2}). Applying the AFCDM, this system of nonlinear PDEs can
be decoupled and integrated in very general form (as we proved in section \ref%
{ssecqs}, and with more details in \cite{partner02}). We can associate and compute for
such generic off-diagonal solutions respective Perelman-like thermodynamic
variables (\ref{gthermodvalues}). Thus, such modified gravity and
modified thermodynamic theories can be formulated in a self-consistent, relativistic
and physical form, even if there is not a rigorous
mathematical version of the nonassociative Poincar\`{e} hypothesis. It should be
noted also that if we can attribute physical significance to generalized
classes of  nonassociative solutions, then the concept of
Bekenstein-Hawking entropy is not applicable. However, the W-entropy and
related statistical Perelman thermodynamics can be applied for very general 
modified theories and their solutions.

The results of this paper support the \textbf{main hypothesis} from the
Introduction in two senses:\ (1) we can construct physically
interesting nonassociative,  nonholonomic phase
space  BE and BH solutions; and (2) such configurations are characterized by corresponding
statistical and geometric thermodynamic variables in a Bekenstein-Hawking
approach or, in more general cases, in a modified Perelman theory for
nonholonomic Ricci solitons. Nevertheless, there are number of important
open questions that should be investigated in the future which are laid out
in \cite{partner01,partner02}. Here we outline four of most important open questions:

\begin{itemize}
\item Q1: A full investigation of nonassociative gravity theories should
involve models with nontrivial sources and other types of nonassociative/
noncommutative structures which are not necessarily determined by R-flux
deformations but based on other types of star product or algebraic
structures. One of the next steps is to study models with star R-flux
nonholonomic deformations of Einstein-Yang-Mills-Higgs systems resulting in
nonassociative gravity and matter field theories. Such nonassociative models
should generalize the Einstein-Eisenhart-Moffat theories \cite%
{einstein25,einstein45,eisenhart51,eisenhart51,moffat95}. 

\item Q2: Nonassociative theories determined by a star product of type (\ref%
{starpn}) on tensor products of cotangent bundles, are a class of
nonassociative generalizations of the Finsler-Lagrange-Hamilton geometry. 
In (non) commutative/ supersymmetric variants,
such models were given in \cite%
{vacaru96b,vacaru09a,gheor14,bubuianu19}. The importance of Finsler-like
variables is that they can be reformulated as some equivalent almost
Kaehler/ complex variables. An even deeper
question is to formulate and study models of quantum gravity using
Lagrange-Hamilton geometric methods encoding nonassociative and
noncommutative structures.

\item Q3: Nonassociative gravity with star and R-flux structures was given
originally in \cite{blumenhagen16,aschieri17}. This was enough to develop a new direction of
the theory of nonassociative/ noncommutative Ricci flows. However, to study
possible physical implications we need to work with nonholonomic dyadic
shell structures as stated in \cite{partner01} and section \ref{sec4}. In a
formal nonassociative geometric form, we can generalize the Perelman F- and
W-functionals (\ref{wfperelmctl}), $\ _{s}^{\shortmid }\widehat{\mathcal{F}}%
(\tau )\rightarrow \ _{s}^{\shortmid }\widehat{\mathcal{F}}^{\star }(\tau )$
and $\ _{s}^{\shortmid }\widehat{\mathcal{W}}(\tau )\rightarrow \
_{s}^{\shortmid }\widehat{\mathcal{W}}^{\star }(\tau ),$ in terms of the
nonassociative phase space canonical Ricci s-tensor, $\ ^{\shortparallel}%
\widehat{\mathbf{R}}ic_{\alpha _{s}\beta _{s}}^{\star }$ and Ricci canonical
s-scalar, $\ _{s}^{\shortparallel }\widehat{\mathbf{R}}sc^{\star }$.
In this way, we can derive a $\star $-deformed Hamilton
nonassociative geometric flow equation
and define nonassociative versions of the Perelman thermodynamic generating
function (\ref{spf}) and variables (\ref{gthermodvalues}) {\it i.e.}, $\
_{s}^{\shortmid }\widehat{\mathcal{Z}}(\tau )\rightarrow \ _{s}^{\shortmid }%
\widehat{\mathcal{Z}}^{\star }(\tau )$ and $\ _{s}^{\shortmid }\widehat{%
\mathcal{E}}(\tau )\rightarrow \ _{s}^{\shortmid }\widehat{\mathcal{E}}%
^{\star }(\tau ), \ _{s}^{\shortmid }\widehat{\mathcal{S}}(\tau )\rightarrow
\ _{s}^{\shortmid }\widehat{\mathcal{S}}^{\star }(\tau ), \ _{s}^{\shortmid }%
\widehat{\sigma }(\tau)\rightarrow \ _{s}^{\shortmid }\widehat{\sigma }%
^{\star }(\tau )$. These constructions can be motivated by finding
exact/parametric solutions which for $\kappa $-dependent decompositions
will result in geometric thermodynamic models that are similar to those
considered in the previous section. Although it may be difficult to
attribute a physical significance to nonassociative geometric thermodynamics
we have shown that it is always possible to formulate well-defined
commutative kinetic, diffusion and thermodynamic models encoding almost
symplectic structures determined by R-flux deformations.

\item Q4: Finally we point to a promising avenue to extend the geometric
flow information theory \cite{ibubuianu20,ibubuianu21} to some versions with
nonassociative qubits and entanglement and conditional entropies. These new
directions in modern quantum field theory/gravity / computers have deep
roots in nonassociative quantum mechanics \cite%
{jordan32,jordan34,kurdgelaidze,castro1} and motivations from string and
M-theory \cite%
{vacaru96b,mylonas13,kupriyanov15,gunaydin,alvarez06,luest10,blumenhagen10,condeescu13,blumenhagen13,kupriyanov19a}%
.
\end{itemize}

We plan to report on progress to answers for questions Q1-Q4 in future works.

\vskip6pt

\textbf{Acknowledgments:} This work is elaborated in the framework of a
research program supported to a Fulbright senior fellowship of SV and hosted
by DS at physics department at California State University at Fresno, USA.
It develops for nonassociative geometry and gravity some former research
projects on geometry and physics supported by fellowships and grants at the
Perimeter Institute and Fields Institute (Ontario, Canada), CERN (Geneva,
Switzerland) and Max Planck Institut f\"{u}r Physik / Werner Heisenberg
Institut, M\"{u}nchen (Germany) and DAAD. The sections 3 and 4 are related also to research programs proposed for visiting at CASLMU M\"{u}nchen,  COST Action CA21109  and PNRR-Initiative 8 of the Romanian Ministry of RID.   SV is also grateful to professors
D. L\"{u}st, N. Mavromatos, J. Moffat, Yu. A. Seti, P. Stavrinos, M. V.
Tkach, and E. V. Veliev for respective hosting visits and collaborations.


\appendix
\setcounter{equation}{0} \renewcommand{\theequation}
{A.\arabic{equation}} \setcounter{subsection}{0}
\renewcommand{\thesubsection}
{A.\arabic{subsection}}

\section{Nonholonomic deformations and nonassociative parametric solutions}

\subsection{Conventions on (non) associative coordinates, and s-adapted
geometric objects}

\label{appendixa} We follow the conventions from \cite%
{partner01,partner02,partner03} on local spacetime and phase space
coordinates and respective labels with boldface and non-boldface symbols:
\begin{eqnarray}
&\mbox{on}&\mathbf{V}\mbox{ and }\ _{s}\mathbf{V}:\ x=\{x^{i}\}=\
_{s}x=\{x^{i_{s}}\}=(x^{i_{1}},x^{a_{2}}\rightarrow y^{a_{2}})=(x^{i_{2}}),%
\mbox{ with }x^{4}=t,  \label{coordshell} \\
&&\mbox{ where }i,j,...=1,2,3,4;\mbox{ shells}:\ s=1,\mbox{ when }%
i_{1},j_{1},...=1,2;\ s=2,a_{2},b_{2},...=3,4;  \notag
\end{eqnarray}
\begin{eqnarray}
&\mbox{on }&T\mathbf{V}\mbox{ and }T_{s}\mathbf{\mathbf{V}}:\
u=(x,y)=\{u^{\alpha }=(u^{k}=x^{k},\ u^{a}=y^{a})\}=  \notag \\
&&\ _{s}u=(~_{s}x,~_{s}y)=\{u^{\alpha _{s}}=(u^{k_{s}}=x^{k_{s}},\
u^{a_{s}}=y^{a_{s}})\}=(x^{i_{1}},x^{i_{2}},x^{a_{3}}\rightarrow
y^{a_{3}},x^{a_{4}}\rightarrow y^{a_{4}}),\mbox{ shells }  \notag \\
&&s=1,2,3,4,\mbox{ where }\alpha ,\beta ,...=1,2,...8;\ a,b,...=5,6,7,8;\
a_{3},b_{3},...=(5,6);\ a_{4},b_{4},..=(7,8);  \notag \\
&\mbox{on }&T^{\ast }\mathbf{V}\mbox{ and }T_{s}^{\ast }\mathbf{V}:\ \
^{\shortmid }u=(x,\ ^{\shortmid }p)=\{\ u^{\alpha }=(u^{k}=x^{k},\ \
^{\shortmid }p_{a}=p_{a})\}  \notag \\
&=&(\ _{3}^{\shortmid }x,\ _{4}^{\shortmid }p)=\{\ ^{\shortmid }u^{\alpha
}=(\ ^{\shortmid }u^{k_{3}}=\ ^{\shortmid }x^{k_{3}},\ ^{\shortmid
}p_{a_{4}}=p_{a_{4}})\}=  \notag \\
&&\ _{s}^{\shortmid }u=(\ _{s}x,\ _{s}^{\shortmid }p)=\{\ ^{\shortmid
}u^{\alpha _{s}}=(x^{k_{s}},\ ^{\shortmid
}p_{a_{s}}=p_{a_{s}})\}=(x^{i_{1}},x^{i_{2}},\ ^{\shortmid
}p_{a_{3}}=p_{a_{3}},\ \ ^{\shortmid }p_{a_{4}}=p_{a_{4}}),  \notag \\
&=&(\ _{3}^{\shortmid }u~=\ _{3}^{\shortmid }x,\ _{4}^{\shortmid }p)=\{\
^{\shortmid }u^{\alpha _{3}}=(x^{i_{1}},x^{i_{2}},\ \ ^{\shortmid
}x^{i_{3}}\rightarrow ~\ ^{\shortmid }p_{a_{3}}),\ \ ^{\shortmid
}p_{a_{4}}\},\mbox{ where }\ \ ^{\shortmid }x^{\alpha
_{3}}=(x^{i_{1}},x^{a_{2}},\ \ ^{\shortmid }p_{a_{3}}=p_{a_{3}})  \notag
\end{eqnarray}
\begin{eqnarray}
&\mbox{on}&T_{\shortparallel }^{\ast }\mathbf{V}\mbox{ and }%
T_{\shortparallel s}^{\ast }\mathbf{V}:\ ^{\shortparallel }u=(x,\
^{\shortparallel }p)=\{\ ^{\shortparallel }u^{\alpha }=(u^{k}=x^{k},\
^{\shortparallel }p_{a}=(i\hbar )^{-1}p_{a})\}  \notag \\
&=&(\ _{3}^{\shortparallel }x,\ _{4}^{\shortparallel }p)=\{\
^{\shortparallel }u^{\alpha }=(^{\shortparallel }u^{k_{3}}=\
^{\shortparallel }x^{k_{3}},\ ^{\shortparallel }p_{a_{4}}=(i\hbar
)^{-1}p_{a_{4}})\}=  \notag \\
&&\ _{s}^{\shortparallel }u=(\ _{s}x,\ _{s}^{\shortparallel }p)=\{\
^{\shortparallel }u^{\alpha _{s}}=(x^{k_{s}},\ ^{\shortparallel
}p_{a_{s}}=(i\hbar )^{-1}p_{a_{s}})\}=(x^{i_{1}},x^{i_{2}},\
^{\shortparallel }p_{a_{3}}=(i\hbar )^{-1}p_{a_{3}},\ ^{\shortparallel
}p_{a_{4}}=(i\hbar )^{-1}p_{a_{4}}),  \notag \\
&=&(\ _{3}^{\shortparallel }u~=\ _{3}^{\shortparallel }x,\
_{4}^{\shortparallel }p)=\{\ ^{\shortparallel }u^{\alpha
_{3}}=(x^{i_{1}},x^{i_{2}},\ ^{\shortparallel }x^{i_{3}}\rightarrow \
^{\shortparallel }p_{a_{3}}),\ ^{\shortparallel }p_{a_{4}}\},\mbox{ where }\
^{\shortparallel }x^{\alpha _{3}}=(x^{i_{1}},x^{i_{2}},\ ^{\shortparallel
}p_{a_{3}}=(i\hbar )^{-1}p_{a_{3}}).  \notag
\end{eqnarray}%
In the above formulas, the coordinate $x^{4}=y^{4}=t$ is time-like and $p_{8}=E$
is energy-like. Boldface indices are used for spaces and geometric objects
enabled with N-/s-connection structure $\ _{s}^{\shortparallel }%
\mathbf{N}$. An upper or lower left label "$\
^{\shortparallel }$" is used to distinguish coordinates
with ``complexified momenta" of real phase coordinates $\
^{\shortmid }u^{\alpha }=(x^{k},p_{a})$ on $T^{\ast }\mathbf{V}$. The
formalism of N- and s-adapted labels and abstract index/symbol or frame
coefficient notations is given in such a way that allows one to formulate a
unified ``symbolic" nonholonomic geometric calculus.

We can consider also $\kappa $-decompositions of a general nonsymmetric metric (\ref%
{nssdm}),
\begin{equation}
\ _{\star }^{\shortparallel }\mathbf{q}_{\alpha _{s}\beta _{s}}=\ _{\star
}^{\shortparallel }\mathbf{q}_{\alpha _{s}\beta _{s}}^{[0]}+\ _{\star
}^{\shortparallel }\mathbf{q}_{\alpha _{s}\beta _{s}}^{[1]}(\kappa )=\
_{\star }^{\shortparallel }\mathbf{\check{q}}_{\alpha _{s}\beta _{s}}+\
_{\star }^{\shortparallel }\mathbf{a}_{\alpha _{s}\beta _{s}},
\label{aux40bb}
\end{equation}%
where $\ _{\star }^{\shortparallel }\mathbf{\check{q}}_{\alpha _{s}\beta
_{s}}$ is the symmetric part and $\ _{\star }^{\shortparallel }\mathbf{a}%
_{\alpha _{s}\beta _{s}}$ is the anti-symmetric part of star and R-flux
deformations. In these formulas,
\begin{eqnarray}
\ _{\star }^{\shortparallel }\mathbf{\check{q}}_{\alpha _{s}\beta _{s}}:= &&%
\frac{1}{2}(\ _{\star }^{\shortparallel }\mathfrak{g}_{\alpha _{s}\beta
_{s}}+\ _{\star }^{\shortparallel }\mathfrak{g}_{\beta _{s}\alpha _{s}})=\
_{\star }^{\shortparallel }\mathbf{g}_{\alpha _{s}\beta _{s}}-\frac{i\kappa
}{2}\left( \overline{\mathcal{R}}_{\quad \beta _{s}}^{\tau _{s}\xi _{s}}\
\mathbf{^{\shortparallel }e}_{\xi _{s}}\ _{\star }^{\shortparallel }\mathbf{g%
}_{\tau _{s}\alpha _{s}}+\overline{\mathcal{R}}_{\quad \alpha _{s}}^{\tau
_{s}\xi _{s}}\ \mathbf{^{\shortparallel }e}_{\xi _{s}}\ _{\star
}^{\shortparallel }\mathbf{g}_{\beta _{s}\tau _{s}}\right)  \label{aux40b} \\
&=&\ _{\star }^{\shortparallel }\mathbf{\check{q}}_{\alpha _{s}\beta
_{s}}^{[0]}+\ _{\star }^{\shortparallel }\mathbf{\check{q}}_{\alpha
_{s}\beta _{s}}^{[1]}(\kappa ),  \notag \\
&&\mbox{ for }\ _{\star }^{\shortparallel }\mathbf{\check{q}}_{\alpha
_{s}\beta _{s}}^{[0]}=\ _{\star }^{\shortparallel }\mathbf{g}_{\alpha
_{s}\beta _{s}}\mbox{ and }\ _{\star }^{\shortparallel }\mathbf{\check{q}}%
_{\alpha _{s}\beta _{s}}^{[1]}(\kappa )=-\frac{i\kappa }{2}\left( \overline{%
\mathcal{R}}_{\quad \beta _{s}}^{\tau _{s}\xi _{s}}\ \mathbf{%
^{\shortparallel }e}_{\xi _{s}}\ _{\star }^{\shortparallel }\mathbf{g}_{\tau
_{s}\alpha _{s}}+\overline{\mathcal{R}}_{\quad \alpha _{s}}^{\tau _{s}\xi
_{s}}\ \mathbf{^{\shortparallel }e}_{\xi _{s}}\ _{\star }^{\shortparallel }%
\mathbf{g}_{\beta _{s}\tau _{s}}\right) ;  \notag \\
\ _{\star }^{\shortparallel }\mathbf{a}_{\alpha _{s}\beta _{s}}:= &&\frac{1}{%
2}(\ _{\star }^{\shortparallel }\mathbf{q}_{\alpha _{s}\beta _{s}}-\ _{\star
}^{\shortparallel }\mathbf{q}_{\beta _{s}\alpha _{s}})=\frac{i\kappa }{2}%
\left( \overline{\mathcal{R}}_{\quad \beta _{s}}^{\tau _{s}\xi _{s}}\
\mathbf{^{\shortparallel }e}_{\xi _{s}}\ _{\star }^{\shortparallel }\mathbf{g%
}_{\tau _{s}\alpha _{s}}-\overline{\mathcal{R}}_{\quad \alpha _{s}}^{\tau
_{s}\xi _{s}}\ \mathbf{^{\shortparallel }e}_{\xi _{s}}\ _{\star
}^{\shortparallel }\mathbf{g}_{\beta _{s}\tau _{s}}\right)  \notag \\
&=&\ _{\star }^{\shortparallel }\mathbf{a}_{\alpha _{s}\beta
_{s}}^{[1]}(\kappa )=\frac{1}{2}(\ _{\star }^{\shortparallel }\mathbf{q}%
_{\alpha _{s}\beta _{s}}^{[1]}(\kappa )-\ _{\star }^{\shortparallel }\mathbf{%
q}_{\beta _{s}\alpha _{s}}^{[1]}(\kappa )).  \label{aux40aa}
\end{eqnarray}%
The nonholonomic distributions can be prescribed in the form $\
_{\star }^{\shortparallel }\mathbf{a}_{\alpha _{s}\beta _{s}}^{[0]}=0$ for
nonassociative star deformations of commutative theories with symmetric
metrics. To compute inverse metrics and s-metrics for such nonassociative
geometric models, we have to apply a more sophisticated procedure -- see
details in \cite{aschieri17} and \cite{partner02}. Respective nonsymmetric
inverse s-metrics can be parameterized in the form
\begin{equation}
\ _{\star }^{\shortparallel }\mathbf{q}^{\alpha _{s}\beta _{s}}=\ _{\star
}^{\shortparallel }\mathbf{\check{q}}^{\alpha _{s}\beta _{s}}+\ _{\star
}^{\shortparallel }\mathbf{a}^{\alpha _{s}\beta _{s}},  \label{aux40cc}
\end{equation}%
when $\ _{\star }^{\shortparallel }\mathbf{\check{q}}^{\alpha _{s}\beta
_{s}} $ is not the inverse to $\ _{\star }^{\shortparallel }\mathbf{\check{q}%
}_{\alpha _{s}\beta _{s}}$ and $\ _{\star }^{\shortparallel }\mathbf{a}%
^{\alpha _{s}\beta _{s}}$ is not inverse to $\ _{\star }^{\shortparallel }%
\mathbf{a}_{\alpha _{s}\beta _{s}}$. After certain classes of nonholonomic
solutions have been constructed in explicit form, we can redefine the
constructions in terms of star deformed LC-configurations using canonical
s-distortions (\ref{candistrnas}) and imposing additional nonholonomic
constraints (\ref{lccondnonass}).

The nonassociative canonical Riemann s-tensor $\mathbf{\mathbf{\mathbf{%
\mathbf{\ ^{\shortparallel }}}}}\widehat{\mathcal{\Re }}_{\quad }^{\star }=\{%
\mathbf{\mathbf{\mathbf{\mathbf{\ ^{\shortparallel }}}}}\widehat{\mathcal{%
\Re }}_{\quad \alpha _{s}\beta _{s}\gamma _{s}}^{\star \mu _{s}}\}$ can be
defined and computed \ for the data $(\ _{\star s}^{\shortparallel }%
\mathbf{g}=\{\ _{\star }^{\shortparallel }\check{\mathbf{q}}_{\alpha
_{s}\beta _{s}}\},\ _{s}^{\shortparallel }\widehat{\mathbf{D}}^{\star }=\{\
^{\shortparallel }\widehat{\mathbf{\Gamma }}_{\star \alpha _{s}\beta
_{s}}^{\gamma _{s}}\}),$ [see formulas (\ref{dmss1}), (\ref{aux40bb}) and (%
\ref{candistrnas}); and details in \cite{partner02}], when
\begin{eqnarray}
\mathbf{\mathbf{\mathbf{\mathbf{\ ^{\shortparallel }}}}}\widehat{\mathcal{%
\Re }}_{\quad \alpha _{s}\beta _{s}\gamma _{s}}^{\star \mu _{s}} &=&\mathbf{%
\mathbf{\mathbf{\mathbf{\ _{1}^{\shortparallel }}}}}\widehat{\mathcal{\Re }}%
_{\quad \alpha _{s}\beta _{s}\gamma _{s}}^{\star \mu _{s}}+\mathbf{\mathbf{%
\mathbf{\mathbf{\ _{2}^{\shortparallel }}}}}\widehat{\mathcal{\Re }}_{\quad
\alpha _{s}\beta _{s}\gamma _{s}}^{\star \mu _{s}},\mbox{ where }
\label{nadriemhopfcan} \\
\mathbf{\mathbf{\mathbf{\mathbf{\ _{1}^{\shortparallel }}}}}\widehat{%
\mathcal{\Re }}_{\quad \alpha _{s}\beta _{s}\gamma _{s}}^{\star \mu _{s}}
&=&\ \mathbf{^{\shortparallel }e}_{\gamma _{s}}\mathbf{\ ^{\shortparallel }}%
\widehat{\Gamma }_{\star \alpha _{s}\beta _{s}}^{\mu _{s}}-\ \mathbf{%
^{\shortparallel }e}_{\beta _{s}}\mathbf{\ ^{\shortparallel }}\widehat{%
\Gamma }_{\star \alpha _{s}\gamma _{s}}^{\mu }+\mathbf{\ ^{\shortparallel }}%
\widehat{\Gamma }_{\star \nu _{s}\tau _{s}}^{\mu _{s}}\star _{s}(\delta _{\
\gamma _{s}}^{\tau _{s}}\mathbf{\ ^{\shortparallel }}\widehat{\Gamma }%
_{\star \alpha _{s}\beta _{s}}^{\nu _{s}}-\delta _{\ \beta _{s}}^{\tau _{s}}%
\mathbf{\ ^{\shortparallel }}\widehat{\Gamma }_{\star \alpha _{s}\gamma
_{s}}^{\nu _{s}})+\mathbf{\ ^{\shortparallel }}w_{\beta _{s}\gamma
_{s}}^{\tau _{s}}\star _{s}\mathbf{\ ^{\shortparallel }}\widehat{\Gamma }%
_{\star \alpha _{s}\tau _{s}}^{\mu _{s}},  \notag \\
\ _{2}^{\shortparallel }\widehat{\mathcal{\Re }}_{\quad \alpha _{s}\beta
_{s}\gamma _{s}}^{\star \mu _{s}} &=&i\kappa \ ^{\shortparallel }\widehat{%
\Gamma }_{\star \nu _{s}\tau _{s}}^{\mu _{s}}\star _{s}(\mathcal{R}_{\quad
\gamma _{s}}^{\tau _{s}\xi _{s}}\ \mathbf{^{\shortparallel }e}_{\xi _{s}}%
\mathbf{\ ^{\shortparallel }}\widehat{\Gamma }_{\star \alpha _{s}\beta
_{s}}^{\nu _{s}}-\mathcal{R}_{\quad \beta _{s}}^{\tau _{s}\xi _{s}}\ \mathbf{%
^{\shortparallel }e}_{\xi _{s}}\mathbf{\ ^{\shortparallel }}\widehat{\Gamma }%
_{\star \alpha _{s}\gamma _{s}}^{\nu _{s}}).  \notag
\end{eqnarray}%
Using parametric decompositions of the star canonical s-connection in (\ref%
{nadriemhopfcan}),
\begin{equation}
\ ^{\shortparallel }\widehat{\mathbf{\Gamma }}_{\star \alpha _{s}\beta
_{s}}^{\gamma _{s}}=\ _{[0]}^{\shortparallel }\widehat{\mathbf{\Gamma }}%
_{\star \alpha _{s}\beta _{s}}^{\nu _{s}}+i\kappa \ _{[1]}^{\shortparallel}%
\widehat{\mathbf{\Gamma}}_{\star \alpha _{s}\beta _{s}}^{\nu _{s}}= \
_{[00]}^{\shortparallel }\widehat{\Gamma}_{\ast \alpha _{s}\beta _{s}}^{\nu
_{s}}+ \ _{[01]}^{\shortparallel }\widehat{\Gamma}_{\ast \alpha _{s}\beta
_{s}}^{\nu _{s}}(\hbar )+ \ _{[10]}^{\shortparallel}\widehat{\Gamma}_{\ast
\alpha _{s}\beta _{s}}^{\nu _{s}}(\kappa )+\ _{[11]}^{\shortparallel }%
\widehat{\Gamma }_{\ast \alpha _{s}\beta _{s}}^{\nu _{s}}(\hbar \kappa)+
O(\hbar ^{2},\kappa ^{2}...),  \label{paramscon}
\end{equation}%
we can compute such parametric decompositions of the nonassociative
canonical curvature tensor,%
\begin{equation*}
\ ^{\shortparallel }\widehat{\mathcal{\Re }}_{\quad \alpha _{s}\beta
_{s}\gamma _{s}}^{\star \mu _{s}}=\mathbf{\mathbf{\mathbf{\mathbf{\ }}}}\
_{[00]}^{\shortparallel }\widehat{\mathcal{\Re }}_{\quad \alpha _{s}\beta
_{s}\gamma _{s}}^{\star \mu _{s}}+\mathbf{\mathbf{\mathbf{\mathbf{\ }}}}\
_{[01]}^{\shortparallel }\widehat{\mathcal{\Re }}_{\quad \alpha _{s}\beta
_{s}\gamma _{s}}^{\star \mu _{s}}(\hbar )+\mathbf{\mathbf{\mathbf{\mathbf{\ }%
}}}\ _{[10]}^{\shortparallel }\widehat{\mathcal{\Re }}_{\quad \alpha
_{s}\beta _{s}\gamma _{s}}^{\star \mu _{s}}(\kappa )+\
_{[11]}^{\shortparallel }\widehat{\mathcal{\Re }}_{\quad \alpha _{s}\beta
_{s}\gamma _{s}}^{\star \mu _{s}}(\hbar \kappa )+O(\hbar ^{2},\kappa
^{2},...).
\end{equation*}%
Further h1-v2-c3-c4 decompositions are also possible for such formulas:

The nonassociative canonical Ricci s-tensor,
\begin{eqnarray*}
\mathbf{\mathbf{\mathbf{\mathbf{\ _{s}^{\shortparallel }}}}}\widehat{%
\mathcal{\Re }}ic^{\star } &=&\mathbf{\mathbf{\mathbf{\mathbf{\
^{\shortparallel }}}}}\widehat{\mathbf{\mathbf{\mathbf{\mathbf{R}}}}}%
ic_{\alpha _{s}\beta _{s}}^{\star }\star _{s}(\ \mathbf{^{\shortparallel }e}%
^{\alpha _{s}}\otimes _{\star s}\ \mathbf{^{\shortparallel }e}^{\beta _{s}}),%
\mbox{ where } \\
&&\mathbf{\mathbf{\mathbf{\mathbf{\ ^{\shortparallel }}}}}\widehat{\mathbf{%
\mathbf{\mathbf{\mathbf{R}}}}}ic_{\alpha _{s}\beta _{s}}^{\star }:=\
_{s}^{\shortparallel }\widehat{\mathcal{\Re }}ic^{\star }(\mathbf{\ }\
^{\shortparallel }\mathbf{e}_{\alpha _{s}},\ ^{\shortparallel }\mathbf{e}%
_{\beta _{s}})=\mathbf{\langle }\ \mathbf{\mathbf{\mathbf{\mathbf{\
^{\shortparallel }}}}}\widehat{\mathbf{\mathbf{\mathbf{\mathbf{R}}}}}ic_{\mu
_{s}\nu _{s}}^{\star }\star _{s}(\ \mathbf{^{\shortparallel }e}^{\mu
_{s}}\otimes _{\star _{s}}\ \mathbf{^{\shortparallel }e}^{\nu _{s}}),\mathbf{%
\mathbf{\ }\ ^{\shortparallel }\mathbf{e}}_{\alpha _{s}}\mathbf{\otimes
_{\star s}\ ^{\shortparallel }\mathbf{e}}_{\beta _{s}}\mathbf{\rangle }%
_{\star _{s}}.
\end{eqnarray*}%
is defined by contracting the first and fourth indices of the
nonassociative curvature s-tensor,
\begin{eqnarray}
\ ^{\shortparallel }\widehat{\mathbf{R}}ic_{\alpha _{s}\beta _{s}}^{\star
}:= &&\mathbf{\mathbf{\mathbf{\mathbf{\ ^{\shortparallel }}}}}\widehat{%
\mathcal{\Re }}_{\quad \alpha _{s}\beta _{s}\mu _{s}}^{\star \mu _{s}}=\
_{[00]}^{\shortparallel }\widehat{\mathbf{\mathbf{\mathbf{\mathbf{R}}}}}%
ic_{\alpha _{s}\beta _{s}}^{\star }+\mathbf{\mathbf{\mathbf{\mathbf{\ \ }}}}%
_{[01]}^{\shortparallel }\widehat{\mathbf{\mathbf{\mathbf{\mathbf{R}}}}}%
ic_{\alpha _{s}\beta _{s}}^{\star }(\hbar )+\mathbf{\mathbf{\mathbf{\mathbf{%
\ }}}}_{[10]}^{\shortparallel }\widehat{\mathbf{\mathbf{\mathbf{\mathbf{R}}}}%
}ic_{\alpha _{s}\beta _{s}}^{\star }(\kappa )  \notag \\
&&+\mathbf{\mathbf{\mathbf{\mathbf{\ }}}}_{[11]}^{\shortparallel }\widehat{%
\mathbf{\mathbf{\mathbf{\mathbf{R}}}}}ic_{\alpha _{s}\beta _{s}}^{\star
}(\hbar \kappa )+O(\hbar ^{2},\kappa ^{2},...),\mbox{where}  \notag \\
&&\ _{[00]}^{\shortparallel }\widehat{\mathbf{R}}ic_{\alpha _{s}\beta
_{s}}^{\star }=\ _{[00]}^{\shortparallel }\widehat{\mathcal{\Re }}_{\quad
\alpha _{s}\beta _{s}\mu _{s}}^{\star \mu _{s}}\mathbf{\mathbf{\mathbf{%
\mathbf{\ ,}}}}\ _{[01]}^{\shortparallel }\mathbf{\mathbf{\mathbf{\mathbf{%
\widehat{\mathbf{\mathbf{\mathbf{\mathbf{R}}}}}}}}}ic_{\alpha _{s}\beta
_{s}}^{\star }=\ _{[01]}^{\shortparallel }\mathbf{\mathbf{\mathbf{\mathbf{%
\widehat{\mathcal{\Re }}}}}}_{\quad \alpha _{s}\beta _{s}\mu _{s}}^{\star
\mu _{s}},  \label{driccicanonstar1} \\
&&\ _{[10]}^{\shortparallel }\mathbf{\mathbf{\mathbf{\mathbf{\widehat{%
\mathbf{\mathbf{\mathbf{\mathbf{R}}}}}}}}}ic_{\alpha _{s}\beta _{s}}^{\star
}=\ _{[10]}^{\shortparallel }\mathbf{\mathbf{\mathbf{\mathbf{\widehat{%
\mathcal{\Re }}}}}}_{\quad \alpha _{s}\beta _{s}\mu _{s}}^{\star \mu _{s}},\
_{[11]}^{\shortparallel }\widehat{\mathbf{\mathbf{\mathbf{\mathbf{R}}}}}%
ic_{\alpha _{s}\beta _{s}}^{\star }=\ _{[11]}^{\shortparallel }\mathbf{%
\mathbf{\mathbf{\mathbf{\widehat{\mathcal{\Re }}}}}}_{\quad \alpha _{s}\beta
_{s}\mu _{s}}^{\star \mu _{s}}.  \notag
\end{eqnarray}%
Because of nonholonomic structures, the canonical Ricci s-tensors are not
symmetric for general (non) commutative and nonassociative cases. Using
contractions with the inverse nonassociative and nonsymmetric s-metric $\
_{\star }^{\shortparallel }\mathfrak{g}^{\mu _{s}\nu _{s}}$ (\ref{aux40cc}),
we can compute the nonassociative nonholonomic canonical Ricci scalar
curvature:%
\begin{eqnarray}
\ _{s}^{\shortparallel }\widehat{\mathbf{R}}sc^{\star }:= &&\ _{\star
}^{\shortparallel }\mathbf{q}^{\mu _{s}\nu _{s}}\mathbf{\mathbf{\mathbf{%
\mathbf{\ ^{\shortparallel }}}}}\widehat{\mathbf{\mathbf{\mathbf{\mathbf{R}}}%
}}ic_{\mu _{s}\nu _{s}}^{\star }=\left( \ _{\star }^{\shortparallel }\mathbf{%
\check{q}}^{\mu _{s}\nu _{s}}+\ _{\star }^{\shortparallel }\mathbf{a}^{\mu
_{s}\nu _{s}}\right) \left( \mathbf{\mathbf{\mathbf{\mathbf{\
^{\shortparallel }}}}}\widehat{\mathbf{\mathbf{\mathbf{\mathbf{R}}}}}%
ic_{(\mu _{s}\nu _{s})}^{\star }+\mathbf{\mathbf{\mathbf{\mathbf{\
^{\shortparallel }}}}}\widehat{\mathbf{\mathbf{\mathbf{\mathbf{R}}}}}%
ic_{[\mu _{s}\nu _{s}]}^{\star }\right) =\ _{s}^{\shortparallel }\widehat{%
\mathbf{\mathbf{\mathbf{\mathbf{R}}}}}ss^{\star }+\ _{s}^{\shortparallel }%
\widehat{\mathbf{\mathbf{\mathbf{\mathbf{R}}}}}sa^{\star },  \notag \\
&&\mbox{ where }\ _{s}^{\shortparallel }\widehat{\mathbf{\mathbf{\mathbf{%
\mathbf{R}}}}}ss^{\star }=:\ _{\star }^{\shortparallel }\mathbf{\check{q}}%
^{\mu _{s}\nu _{s}}\mathbf{\mathbf{\mathbf{\mathbf{\ ^{\shortparallel }}}}}%
\widehat{\mathbf{\mathbf{\mathbf{\mathbf{R}}}}}ic_{(\mu _{s}\nu
_{s})}^{\star }\mbox{ and }\ _{s}^{\shortparallel }\widehat{\mathbf{\mathbf{%
\mathbf{\mathbf{R}}}}}sa^{\star }:=\ _{\star }^{\shortparallel }\mathbf{a}%
^{\mu _{s}\nu _{s}}\mathbf{\mathbf{\mathbf{\mathbf{\ ^{\shortparallel }}}}}%
\widehat{\mathbf{\mathbf{\mathbf{\mathbf{R}}}}}ic_{[\mu _{s}\nu
_{s}]}^{\star }.  \label{ricciscsymnonsym}
\end{eqnarray}%
In these formulas, respective symmetric $\left( ...\right) $ and
anti-symmetric $\left[ ...\right] $ operators are defined using the multiple
$1/2,$ when, for instance, $\mathbf{\mathbf{\mathbf{\mathbf{\
^{\shortparallel }}}}}\widehat{\mathbf{\mathbf{\mathbf{\mathbf{R}}}}}ic_{\mu
_{s}\nu _{s}}^{\star }=\mathbf{\mathbf{\mathbf{\mathbf{\ ^{\shortparallel }}}%
}}\widehat{\mathbf{\mathbf{\mathbf{\mathbf{R}}}}}ic_{(\mu _{s}\nu
_{s})}^{\star }+\mathbf{\mathbf{\mathbf{\mathbf{\ ^{\shortparallel }}}}}%
\widehat{\mathbf{\mathbf{\mathbf{\mathbf{R}}}}}ic_{[\mu _{s}\nu
_{s}]}^{\star }.$

\subsection{Nonassociative nonlinear symmetries of generating functions and effective sources}

\label{apbssns}By straightforward computations (see details in section 5.4
of \cite{partner02}), we can check that any s-metric define the same class
of exact/parametric quasi-stationary solutions of nonassociative vacuum
gravitational equations if the transforms (\ref{nonlinsym}) are subjected to
such conditions: {\small
\begin{eqnarray}
s=2: &&\frac{[(\ _{2}\Psi )^{2}]^{\diamond }}{~_{2}^{\shortparallel }%
\mathcal{K}}=\frac{[(\ _{2}\Phi )^{2}]^{\diamond }}{\ _{2}\Lambda _{0}},
\label{nonltransf} \\
&&\mbox{ i.e. }(\ _{2}\Psi )^{2}=(\ _{2}\Lambda _{0})^{-1}\int
dx^{3}(~_{2}^{\shortparallel }\mathcal{K})[(\ _{2}\Phi )^{2}]^{\diamond }\ %
\mbox{ and/or }(\ _{2}\Phi )^{2}=\ _{2}\Lambda _{0}\int
dx^{3}(~_{2}^{\shortparallel }\mathcal{K})^{-1}[(\ _{2}\Psi )^{2}]^{\diamond
},  \notag \\
\lbrack (\ _{2}\Psi )^{2}]^{\diamond } &=&-\int dy^{3}(~_{2}^{\shortparallel
}\mathcal{K})g_{4}^{\diamond }\mbox{ and/or }(\ _{2}\Phi )^{2}=-4\
_{2}\Lambda _{0}g_{4},  \notag \\
\ [(\ _{2}\Psi )^{2}]^{\diamond } &=&-\int dy^{3}(~_{2}^{\shortparallel }%
\mathcal{K})(\ ^{\shortparallel }\eta _{4}\ \ \ \mathring{g}_{4})^{\diamond }%
\mbox{ and/or }(\ _{2}\Phi )^{2}=-4\ _{2}\Lambda _{0}\ ^{\shortparallel
}\eta _{4}\ \ \mathring{g}_{4},  \notag \\
\lbrack (\ _{2}\Psi )^{2}]^{\diamond } &=&-\int dy^{3}(~_{2}^{\shortparallel
}\mathcal{K})[\ ^{\shortparallel }\zeta _{4}(1+\kappa \ ^{\shortparallel
}\chi _{4})\ \mathring{g}_{4}]^{\diamond }\mbox{ and/or }(\ _{2}\Phi
)^{2}=-4\ _{2}\Lambda _{0}\ ^{\shortparallel }\zeta _{4}(1+\kappa \
^{\shortparallel }\chi _{4})\ \mathring{g}_{4};  \notag
\end{eqnarray}%
\begin{eqnarray*}
s=3: &&\frac{~^{\shortparallel }\partial ^{6}[(\ _{3}^{\shortparallel }\Psi
)^{2}]}{~_{3}^{\shortparallel }\mathcal{K}}=\frac{~^{\shortparallel
}\partial ^{6}[(\ _{3}^{\shortparallel }\Phi )^{2}]}{\ _{3}^{\shortparallel
}\Lambda _{0}}, \\
&&\mbox{ i.e. }(\ _{3}^{\shortparallel }\Psi )^{2}=(\ _{3}^{\shortparallel
}\Lambda _{0})^{-1}\int d~^{\shortparallel }p_{6}(~_{3}^{\shortparallel }%
\mathcal{K})[(\ _{3}^{\shortparallel }\Phi )^{2}]\mbox{ and/or }\ (\
_{3}^{\shortparallel }\Phi )^{2}=\ _{3}^{\shortparallel }\Lambda _{0}\int
d~^{\shortparallel }p_{6}(~_{3}^{\shortparallel }\mathcal{K})^{-1}[(\
_{3}^{\shortparallel }\Psi )^{2}] \\
~^{\shortparallel }\partial ^{6}[(\ _{3}^{\shortparallel }\Psi )^{2}]
&=&-\int d~^{\shortparallel }p_{6}(~_{3}^{\shortparallel }\mathcal{K}%
)~^{\shortparallel }\partial ^{6}~^{\shortparallel }g^{6}\mbox{ and/or }(\
_{3}^{\shortparallel }\Phi )^{2}=-4\ _{3}^{\shortparallel }\Lambda
_{0}~^{\shortparallel }g^{6},\  \\
~^{\shortparallel }\partial ^{6}[(\ _{3}^{\shortparallel }\Psi )^{2}]
&=&-\int d~^{\shortparallel }p_{6}(~_{3}^{\shortparallel }\mathcal{K}%
)~^{\shortparallel }\partial ^{6}(\ ^{\shortparallel }\eta
^{6}~^{\shortparallel }\mathring{g}^{6})\mbox{ and/or }(\
_{3}^{\shortparallel }\Phi )^{2}=-4\ _{3}^{\shortparallel }\Lambda _{0}\
^{\shortparallel }\eta ^{6}~^{\shortparallel }\mathring{g}^{6}, \\
~^{\shortparallel }\partial ^{6}[(\ _{3}^{\shortparallel }\Psi )^{2}]
&=&-\int d~^{\shortparallel }p_{6}(~_{3}^{\shortparallel }\mathcal{K}%
)~^{\shortparallel }\partial ^{6}[\ ^{\shortparallel }\zeta _{6}(1+\kappa \
^{\shortparallel }\chi _{6})\ \mathring{g}_{6}]\mbox{ and/or }(\
_{3}^{\shortparallel }\Phi )^{2}=-4\ _{3}^{\shortparallel }\Lambda _{0}\
^{\shortparallel }\zeta _{6}(1+\kappa \ ^{\shortparallel }\chi _{6})\
~^{\shortparallel }\mathring{g}_{6};
\end{eqnarray*}%
\begin{eqnarray*}
s=4: &&\frac{[(\ _{4}^{\shortparallel }\Psi )^{2}]^{\ast }}{%
~_{4}^{\shortparallel }\mathcal{K}}=\frac{[(\ _{4}^{\shortparallel }\Phi
)^{2}]^{\ast }}{\ _{4}^{\shortparallel }\Lambda }, \\
&&\mbox{ i.e. }(\ _{4}^{\shortparallel }\Psi )^{2}=(\ _{4}^{\shortparallel
}\Lambda )^{-1}\int d~^{\shortparallel }E(~_{4}^{\shortparallel }\mathcal{K}%
)[(\ _{4}^{\shortparallel }\Phi )^{2}]^{\ast }\mbox{ and/or }\ (\
_{4}^{\shortparallel }\Phi )^{2}=\ _{4}^{\shortparallel }\Lambda \int
d~^{\shortparallel }E(~_{4}^{\shortparallel }\mathcal{K})^{-1}[(\
_{4}^{\shortparallel }\Psi )^{2}]^{\ast } \\
\lbrack (\ _{4}^{\shortparallel }\Psi )^{2}]^{\ast } &=&-\int
d~^{\shortparallel }E(~_{4}^{\shortparallel }\mathcal{K})[~^{\shortparallel
}g^{7}]^{\ast }\mbox{ and/or }(\ _{4}^{\shortparallel }\Phi )^{2}=-4\
_{4}^{\shortparallel }\Lambda _{0}~^{\shortparallel }g^{7},\  \\
\lbrack (\ _{4}^{\shortparallel }\Psi )^{2}]^{\ast } &=&-\int
d~^{\shortparallel }E(~_{4}^{\shortparallel }\mathcal{K})[\ ^{\shortparallel
}\eta ^{7}~^{\shortparallel }\mathring{g}^{7}]^{\ast }\mbox{ and/or }(\
_{4}^{\shortparallel }\Phi )^{2}=-4\ _{4}^{\shortparallel }\Lambda _{0}\
^{\shortparallel }\eta ^{7}~^{\shortparallel }\mathring{g}^{7}, \\
\lbrack (\ _{4}^{\shortparallel }\Psi )^{2}]^{\ast } &=&-\int
d~^{\shortparallel }E(~_{4}^{\shortparallel }\mathcal{K})~[\
^{\shortparallel }\zeta ^{7}(1+\kappa \ ^{\shortparallel }\chi ^{7})\
\mathring{g}^{7}]^{\ast }\mbox{ and/or }(\ _{4}^{\shortparallel }\Phi
)^{2}=-4\ _{4}^{\shortparallel }\Lambda _{0}\ ^{\shortparallel }\zeta
^{7}(1+\kappa \ ^{\shortparallel }\chi ^{7})\ ~^{\shortparallel }\mathring{g}%
^{7};
\end{eqnarray*}%
or, for }$~^{\shortparallel }E=~^{\shortparallel }E_{0}=const,${\small \
\begin{eqnarray*}
s=4: &&\frac{~^{\shortparallel }\partial ^{7}[(\ _{4}^{\shortparallel }\Psi
)^{2}]}{~_{4}^{\shortparallel }\mathcal{K}}=\frac{~^{\shortparallel
}\partial ^{7}[(\ _{4}^{\shortparallel }\Phi )^{2}]}{\ _{4}^{\shortparallel
}\Lambda _{0}}, \\
&&\mbox{ i.e. }(\ _{4}^{\shortparallel }\Psi )^{2}=(\ _{4}^{\shortparallel
}\Lambda _{0})^{-1}\int d~^{\shortparallel }p_{7}(~_{4}^{\shortparallel }%
\mathcal{K})[(\ _{4}^{\shortparallel }\Phi )^{2}]\mbox{ and/or }\ (\
_{4}^{\shortparallel }\Phi )^{2}=\ _{4}^{\shortparallel }\Lambda _{0}\int
d~^{\shortparallel }p_{7}(~_{4}^{\shortparallel }\mathcal{K})^{-1}[(\
_{4}^{\shortparallel }\Psi )^{2}] \\
~^{\shortparallel }\partial ^{7}[(\ _{4}^{\shortparallel }\Psi )^{2}]
&=&-\int d~^{\shortparallel }p_{7}(~_{4}^{\shortparallel }\mathcal{K}%
)~^{\shortparallel }\partial ^{7}~^{\shortparallel }g^{8}\mbox{ and/or }(\
_{4}^{\shortparallel }\Phi )^{2}=-4\ _{4}^{\shortparallel }\Lambda
_{0}~^{\shortparallel }g^{8},\  \\
~^{\shortparallel }\partial ^{7}[(\ _{4}^{\shortparallel }\Psi )^{2}]
&=&-\int d~^{\shortparallel }p_{7}(~_{4}^{\shortparallel }\mathcal{K}%
)~^{\shortparallel }\partial ^{7}(\ ^{\shortparallel }\eta
^{8}~^{\shortparallel }\mathring{g}^{8})\mbox{ and/or }(\
_{4}^{\shortparallel }\Phi )^{2}=-4\ _{4}^{\shortparallel }\Lambda _{0}\
^{\shortparallel }\eta ^{8}~^{\shortparallel }\mathring{g}^{8}, \\
~^{\shortparallel }\partial ^{7}[(\ _{4}^{\shortparallel }\Psi )^{2}]
&=&-\int d~^{\shortparallel }p_{7}(~_{4}^{\shortparallel }\mathcal{K}%
)~^{\shortparallel }\partial ^{7}[\ ^{\shortparallel }\zeta ^{8}(1+\kappa \
^{\shortparallel }\chi ^{8})\ \mathring{g}^{8}]\mbox{ and/or }(\
_{4}^{\shortparallel }\Phi )^{2}=-4\ _{4}^{\shortparallel }\Lambda _{0}\
^{\shortparallel }\zeta ^{8}(1+\kappa \ ^{\shortparallel }\chi ^{8})\
~^{\shortparallel }\mathring{g}^{8}.
\end{eqnarray*}%
}

Nonlinear symmetries given by the above formulas allow us to construct
different classes of exact/ parametric solutions or to express a known
solution in other forms. Working with  $(\ _{s}\Psi ,\
_{s}^{\shortparallel }\mathcal{K}),$ we can generate nonassociative vacuum
s-metrics encoding star R-flux contributions via $\ _{s}^{\shortparallel}%
\mathcal{K}$ when $\ _{s}\Psi $ are prescribed in order to generate target
quasi-stationary s-metrics. For $(\
_{s}\Phi ,\ _{s}\Lambda _{0}),$ the sources are re-encoded in $\ _{s}\Phi $
but with some approximations to effective cosmological constants $\
_{s}\Lambda _{0}.$ Usually, the procedure for constructing solutions
generated by $\ _{s}\Phi $ is simpler but there are additional
considerations in order to chose generating and integration functions
describing physically important target configurations. Choosing some
coefficients of a s-metric as generating functions and working with
$(\ _{s}\widehat{\mathbf{g}},\ _{s}^{\shortparallel }%
\mathcal{K}),$ we can construct new classes of solutions with certain
prescribed properties and re-defined in some closed to ``real physical"
configurations. Finally, we note that for $(\
_{s}^{\shortparallel}\eta \ _{s}^{\shortparallel }\mathring{g} _{\alpha
_{s}}\sim \ _{s}^{\shortparallel }\zeta (1+\kappa \ _{s}^{\shortparallel
}\chi _{\alpha _{s}}) \ _{s}^{\shortparallel }\mathring{g}_{\alpha _{s}},\
_{s}\Lambda _{0}) $ it is possible to embed self-consistently certain
background configurations $\ _{s}^{\shortparallel }\mathring{g}_{\alpha
_{s}} $ into some classes of more general vacuum
solutions. For small $\kappa $-dependent deformations, we can determine
physical properties and compute nonassociative classical and quantum
physical effects.

\end{document}